\documentclass[sn-mathphys-num]{sn-jnl}

\pdfoutput=1

\usepackage{graphicx}%
\usepackage{multirow}%
\usepackage{amsmath,amssymb,amsfonts}%
\usepackage{amsthm}%
\usepackage{mathrsfs}%
\usepackage[title]{appendix}%
\usepackage{xcolor}%
\usepackage{textcomp}%
\usepackage{manyfoot}%
\usepackage{booktabs}%
\usepackage{algorithm}%
\usepackage{algorithmicx}%
\usepackage{algpseudocode}%
\usepackage{listings}%
\usepackage{microtype,xspace}
\usepackage[utf8]{inputenc}
\usepackage{bm,amssymb,amsmath,amsfonts,slashed,graphicx,multirow,soul,mathtools,xspace,array,comment,color}
\hypersetup{colorlinks=true, allcolors=blue}
\usepackage{adjustbox}
\usepackage{siunitx}

\usepackage{bm} % allows to use \bm{} to make bold math symbols
\usepackage{float}
\usepackage{feynmf}
\allowdisplaybreaks
\usepackage{ bbold }
\usepackage{subfigure}
\usepackage[normalem]{ulem} % for crossing out text
\usepackage{enumerate}
\usepackage{multirow}
\usepackage{booktabs}
\usepackage{xparse}
\usepackage{etoolbox}
\usepackage{physics}
\usepackage{hypcap}
\usepackage{siunitx}
\usepackage{xcolor}

\NewDocumentCommand{\lwc}{ m m O{} o }{
	L^{\ifblank{#3}{}{#3,}#2 }_{\IfNoValueTF{#4}{#1}{\substack{#1\\#4}}}
}

\def\be{\begin{equation}}
\def\ee{\end{equation}}
\newcommand{\TeV}{\text{TeV}}
\newcommand{\GeV}{\text{GeV}}

\theoremstyle{thmstyleone}%
\theoremstyle{thmstyletwo}%

\theoremstyle{thmstylethree}%

\begin{document}

\title[Article Title]{\boldmath{Implications of $B \to K \nu \bar{\nu}$ under Rank-One Flavor Violation hypothesis}}


\author[1]{\fnm{David} \sur{Marzocca}}\email{david.marzocca@ts.infn.it}

\author[2]{\fnm{Marco} \sur{Nardecchia}}\email{marco.nardecchia@roma1.infn.it}

\author[1,3]{\fnm{Alfredo} \sur{Stanzione}}\email{alfredo.stanzione@sissa.it}

\author[4]{\fnm{Claudio} \sur{Toni}}\email{claudio.toni@pd.infn.it}

\affil[1]{INFN, Sezione di Trieste, SISSA, Via Bonomea 265, 34136, Trieste, Italy}

\affil[2]{Universit\`a degli Studi di Roma la Sapienza and INFN Section of Roma 1, Piazzale Aldo Moro 5, 00185, Roma, Italy}

\affil[3]{SISSA International School for Advanced Studies, Via Bonomea 265, 34136, Trieste, Italy}

\affil[4]{Dipartimento di Fisica e Astronomia `G.~Galilei', Universit\`a di Padova and INFN Sezione di Padova, Via F. Marzolo 8, 35131 Padova, Italy}

%\affil[3]{\orgdiv{Department}, \orgname{Organization}, \orgaddress{\street{Street}, \city{City}, \postcode{610101}, \state{State}, \country{Country}}}


\abstract{We study the implications of the observed excess in $B^+ \to K^+ \nu \bar{\nu}$ under the assumption of Rank-One Flavour Violation, {\emph{i.e.}} that New Physics couples to a single specific direction in flavour space.
By varying this direction we perform analyses at the level of the low-energy EFT, the SMEFT, and with explicit mediators such as leptoquarks and colorless vectors ($Z^\prime$ and $V^\prime$).
We study correlations with other flavour, electroweak and collider observables, 
finding that the most interesting ones are with $K \to \pi \nu \bar{\nu}$, $B_s \to \mu^+ \mu^-$, meson mixing and the LHC searches in $\tau^+ \tau^-$ high-energy tails.
Among the various mediators, the scalar leptoquarks $\tilde{R}_2$ and $S_1$ offer the best fits of the Belle-II excess, while being consistent with the other bounds. On the other hand, colorless vectors are strongly constrained by meson mixing and resonance searches in $p p \to \tau^+ \tau^-$. In all cases we find that a flavour alignment close to the third generation is generically preferred.}

\maketitle

\section{Introduction}
Rare semileptonic neutral-current decays of mesons are powerful probes of New Physics (NP) due to their loop and CKM suppression in the Standard Model (SM). Within this class of processes, the so-called golden-channel decays to neutrinos stand out thanks to the precision of the SM predictions, a consequence of the neutrality of neutrinos under electromagnetic interactions, implying the absence of long-distance non-perturbative contributions to the decay rate \cite{Altmannshofer:2009ma,Buras:2014fpa,Buras:2015yca,Blake:2016olu,Parrott:2022zte,Becirevic:2023aov}.

Particularly interesting is the $b \to s \nu \bar{\nu}$ transition, currently studied via several decays of $B$ mesons into Kaons. It is useful to define the ratios with the SM prediction as
\begin{equation}
R_{K^{(*)}}^\nu=\frac{\mathcal{B}(B \to K^{(*)} \nu\bar{\nu})}{\mathcal{B}(B \to K^{(*)} \nu\bar{\nu})_{\mathrm{SM}}} \ ,
\end{equation}
where in the following we employ the latest SM predictions from Ref.~\cite{Becirevic:2023aov}, which combine results on the hadronic form factors from different lattice groups.
An upper limit from the combination of all $B \to K^* \nu\bar{\nu}$ channels has been reported by the Belle experiment: $\mathcal{B}(B \to K^* \nu\bar{\nu}) < 2.7 \times 10^{-5}$ at the 90\% CL \cite{Belle:2017oht}. Assuming the central value being the SM point (which is also motivated by the $\sim 2\sigma$ excess in the $K^{*+}$ channel) we obtain $R^\nu_{K^*} = 1.0 \pm 1.1$.
For the neutral channel Belle set the upper limit $\mathcal{B}(B^0 \to K^0_S \nu\bar{\nu}) < 1.3 \times 10^{-5}$ at the 90\% CL \cite{Belle:2017oht},\footnote{We note that in the PDG it is reported an upper limit of $2.6 \times 10^{-5}$, quoting that same Belle paper. We use the value reported by Belle.}
which we translate to $R^\nu_{K^0_S} = 1.0 \pm 3.3$. 
Recently, the Belle-II experiment announced the first $3.5\sigma$ evidence for the $B^+ \to K^+ \nu\bar{\nu}$ decay, showing a $2.7\sigma$ excess over the SM prediction \cite{Belle-II:2023esi}. When combined with previous upper limits on the same mode they obtain $\mathcal{B}(B^+ \to K^+ \nu\bar{\nu}) = (1.3 \pm 0.4 ) \times 10^{-5}$, which corresponds to $R^\nu_{K^+} = 2.93 \pm 0.90$, a $2.1\sigma$ deviation from the SM.\footnote{The long-distance contribution $B^+ \to \nu \tau^+ \to K^+ \nu \bar{\nu}$ is removed from the SM prediction, since it is also treated as background in the experimental analysis.}
With 5~(50)~ab$^{-1}$ of integrated luminosity, Belle-II expects to measure these rates with approximately 30\%~(11\%) precision \cite{Belle-II:2018jsg}.

It is likely that the mild deviation from the SM observed in $B^+ \to K^+ \nu \bar{\nu}$ is simply due to a statistical fluctuation. If the excess is instead due to a contribution of physics beyond the SM
one would generically expect correlated deviations also in other measurements.
Such correlations would depend on the specific assumptions on the type of NP and its flavour structure.
Our goal is to study possible correlations between this excess and other processes investigated at both low and high-energy experiments, under the assumption that NP is heavier than the electroweak scale. As a consequence, we describe NP contributions in the form of Effective Field Theory (EFT) operators written in terms of SM fields.
Similar investigations have already been performed by several groups~\cite{Bause:2023mfe,Allwicher:2023xba,Athron:2023hmz,Greljo:2023bix,Chen:2024jlj,DAlise:2024qmp}, while for the opposite but complementary case of light NP see Refs.~\cite{Felkl:2023ayn,He:2023bnk,Berezhnoy:2023rxx,Altmannshofer:2023hkn,McKeen:2023uzo,Fridell:2023ssf,Ho:2024cwk,Gabrielli:2024wys,Hou:2024vyw,He:2024iju,Bolton:2024egx,Datta:2023iln}.
Regarding the flavour structure of NP, our goal is to consider a scenario that is both specific enough to induce interesting correlations while being generic enough to allow us to explore different directions in flavour space. We therefore assume that the EFT
flavour structure follows the \emph{Rank-One Flavour Violation} (ROFV) hypothesis \cite{Gherardi:2019zil}, where the flavour matrix of EFT coefficients of semileptonic operators is of rank-1. 
This flavour structure is automatically realised in several New Physics scenarios, such as all leptoquark models coupled (mostly) to a single lepton flavour, or with single vector-like quarks that mix with SM quarks. In general, it is realised in all cases where a single linear combination of quarks is coupled to NP:
\be
    \mathcal{L} \supset \lambda_i \bar{q}^i \mathcal{O}_{\rm NP} + {\rm h.c.}~.
\ee
Such structure implies correlations between operators involving different families of quarks. In the specific case discussed here, the channels most obviously correlated to $b \to s \nu \bar{\nu}$ transitions are $s \to d \nu \bar{\nu}$ and $b \to d \nu \bar{\nu}$. 

The $s \to d \nu \bar{\nu}$ transition can be tested via $K^+ \to \pi^+ \nu \bar{\nu}$ and $K_L \to \pi^0 \nu \bar{\nu}$ decays, currently investigated at the NA62 \cite{NA62:2020fhy,NA62:2021zjw} and KOTO \cite{KOTO:2018dsc} experiments, respectively. While an upper limit is currently set on the neutral channel, $\mathcal{B}(K_L \to \pi^0 \nu\bar{\nu}) < 2.4 \times 10^{-9}$ at the 90\% CL \cite{KOTO:2018dsc},
the NA62 experiments found evidence for the charged mode with significance larger than $5\sigma$: $\mathcal{B}(K^+ \to \pi^+ \nu\bar{\nu}) = \left(13.6 \, (^{+3.0}_{-2.7})_{\rm stat} (^{+1.3}_{-1.2})_{\rm syst} \right) \times 10^{-11}$ \cite{Kpiseminar}.
%the NA62 experiments found a $3.4\sigma$ evidence for the charged mode with $\mathcal{B}(K^+ \to \pi^+ \nu\bar{\nu}) = (10.6^{+4.0}_{-3.4} \pm 0.9 ) \times 10^{-11}$.
A final sensitivity of about 15\% is now expected by NA62 \cite{Piccini:2020xgx,HIKE:2023ext}, while KOTO should reach a 95\%CL upper limit of $1.8 \times 10^{-10}$ by the end of stage-I \cite{Aoki:2021cqa}.\footnote{The proposed HIKE experiment at CERN could reach  $\mathcal{O}(5\%)$ precision on $\mathcal{B}(K^+ \to \pi^+ \nu\bar{\nu})$ \cite{HIKE:2023ext}, while a stage-II upgrade at KOTO would allow a precision of about $20 \%$ in the neutral channel \cite{Aoki:2021cqa}.}
The $B \to \pi \nu\bar{\nu}$ and $B \to \rho \nu \bar{\nu}$ decays, measured at Belle \cite{Belle:2017oht}, instead test the $b \to d \nu \bar{\nu}$ partonic transition.

By adding further assumptions on the gauge structure of the new physics, or by assuming specific heavy mediators, further correlations with other leptonic and semileptonic decays, involving also charged leptons, appear, as well as with electroweak or collider observables.

In Section~\ref{sec:LEFT} we briefly review the low-energy EFT (LEFT) fit of $b \to s \nu \bar{\nu}$ transitions, introduce the ROFV hypothesis and study the correlations one can derive in the LEFT.
In Section~\ref{sec:SMEFT} we study the correlations arising at the level of SMEFT.
In Section~\ref{sec:models} we focus on semileptonic four-fermion operators and review single-mediator simplified models.
Finally, we summarise our results and conclude in Section~\ref{sec:conclusions}.
The details of all observables discussed in our analysis are collected in App.~\ref{App:Observables}.
In App.~\ref{app:Fitsbsnunu} we collect the fits of EFT coefficients related to $b\to s \nu \bar{\nu}$ in the various scenarios. Finally, in App.~\ref{app:LNV} we study possible interpretations of the $R^\nu_K$ excess in terms of lepton-number violating operators.

%-----------------------------------------------------
\section{$B\to K^{(*)} \nu \bar{\nu}$ and rank-one correlations in LEFT}
\label{sec:LEFT}

\begin{figure}[t]
\centering
    \includegraphics[width=0.47\textwidth]{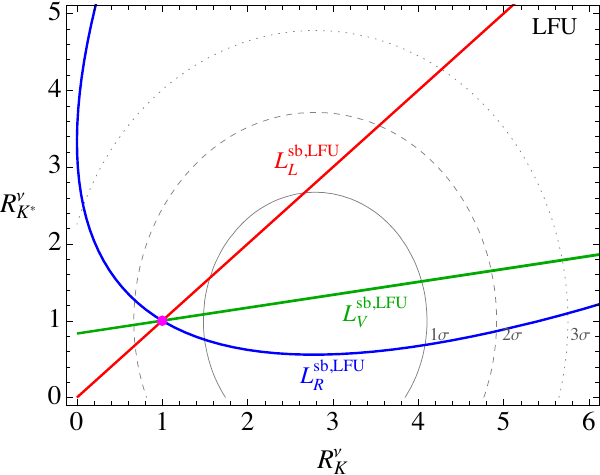}
    \hfill
    \includegraphics[width=0.47\textwidth]{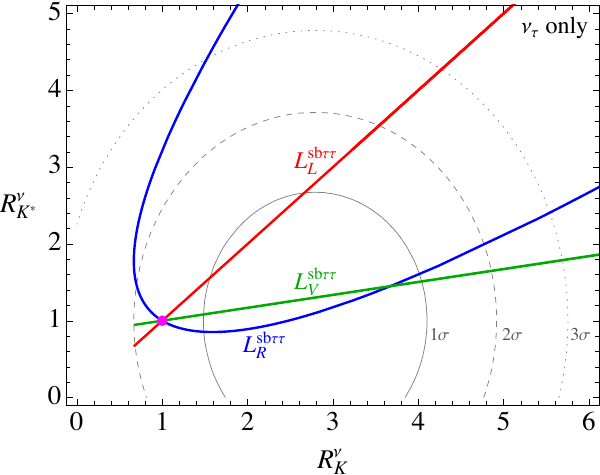} \\
    \includegraphics[width=0.48\textwidth]{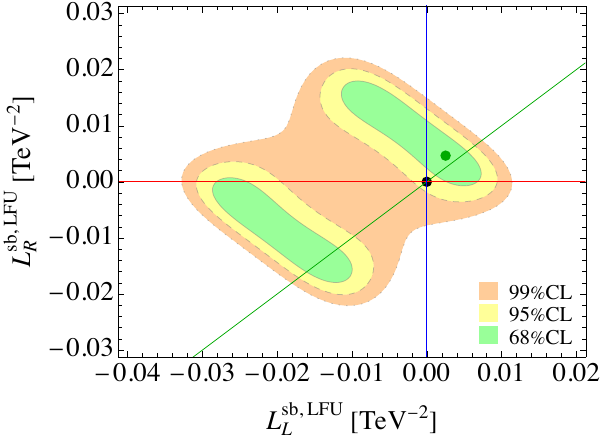}
    \hfill
    \includegraphics[width=0.48\textwidth]{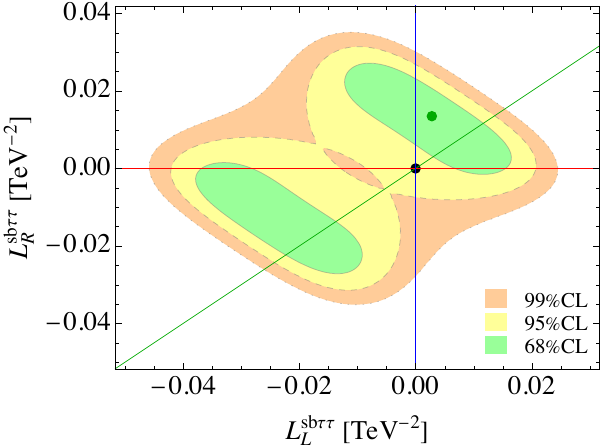}
    \caption{Top row: Fit of available data on all $B \to K^{(*)} \nu \bar\nu$ channels. Solid, dashed, and dotted gray lines represent $1\sigma$, $2\sigma$ and $3\sigma$ contours. The magenta point represents the SM.\\
    Bottom row: Fit in the $(L_L^{sb\alpha\beta}, \ L_R^{sb\alpha\beta})$ plane assuming lepton flavour universality (left) or contributions only in the tau neutrino (right). Red, blue, and green lines represents the EFT directions as in the plots above.}
    \label{fig:fit_BKnunu_data}
\end{figure}

Let us start by reviewing the fit of $B\to K^{(*)} \nu \bar{\nu}$ measurements in the EFT framework. The low energy effective theory relevant to the $b \to s\nu\bar{\nu}$
processes is described by the effective Lagrangian 
\cite{Altmannshofer:2009ma,Buras:2014fpa,Becirevic:2023aov}
\begin{equation}
\label{eqWETbs}
    \mathcal{L}_{\rm LEFT}^{bs\nu\nu} = 
    \sum_{\alpha\beta}\left[ (L_{L}^{sb\alpha\beta} + L_{L}^{sb, \text{ SM}} \delta^{\alpha \beta} ) 
   (\bar{s}_{L}\gamma_{\mu} b_{L}) (\bar\nu_L^\alpha \gamma^\mu \nu_L^\beta) +
    L_{R}^{sb\alpha\beta} (\bar{s}_{R}\gamma_{\mu} b_{R}) (\bar\nu_L^\alpha \gamma^\mu \nu_L^\beta) \right] \ ,
\end{equation}
where $L_{L}^{sb, \text{ SM}} = \frac{8 G_F V_{tb} V_{ts}^*}{\sqrt{2}} \frac{\alpha}{4\pi} C_L^{\rm SM}$, with $C_L^{\rm SM} \approx -6.32$ \cite{Becirevic:2023aov}, see App.~\ref{app:goldenmodes} for details.
We also define the lepton flavour universal (LFU) coefficients as $L_{L,R}^{sb \alpha \beta} = L_{L,R}^{sb, \text{ LFU}}\, \delta^{\alpha \beta}$, and the vector/axial combinations as $L_{V,A}^{sb \alpha\beta} \equiv L_R^{sb \alpha\beta} \pm L_L^{sb \alpha\beta}$.
In the top row of Fig.~\ref{fig:fit_BKnunu_data} we plot the correlation of $R_K^\nu$ and $R_{K^{*}}^\nu$ while varying the effective coefficients $L_{L,R,V}^{sb}$. In the bottom row we show the fit in the plane $L_L^{sb} - L_R^{sb}$, assuming for simplicity LFU or new physics coupled only to tau neutrinos. We observe a slight preference (less than $1\sigma$) for the right and vector currents, due to the absence of any excess in the $K^*$ channel. Future more precise measurements will clarify whether the Belle-II excess in the $K^+$ channel is something more than merely a statistical fluctuation and the interplay with the $K^*$ mode will provide more information on the chiral structure of the underlying new physics interactions. 
In the rest of the work we take the excess at face value and proceed with the new physics interpretation.

The two low-energy operators in Eq. \eqref{eqWETbs} are part of an effective Lagrangian involving all three quark families. Focusing on the new physics contribution one has
\begin{equation}
\label{eqLEFTlag}
\mathcal{L}_{\rm LEFT}^{\rm NP} = 
   \sum_{ij\alpha\beta}\left[ L_{L}^{ij\alpha\beta} 
   (\bar{d}^i_{L}\gamma_{\mu}d_{L}^j) (\bar\nu_L^\alpha \gamma^\mu \nu_L^\beta) +
    L_{R}^{ij\alpha\beta} (\bar{d}_{R}^i\gamma_{\mu}d_{R}^j) (\bar\nu_L^\alpha \gamma^\mu \nu_L^\beta) \right] \ .
\end{equation}
We study scenarios with NP coupled to left, right, or vector quark currents.
\begin{figure}
    \vspace{-0.5cm}
	\centering
	\includegraphics[scale=0.64]{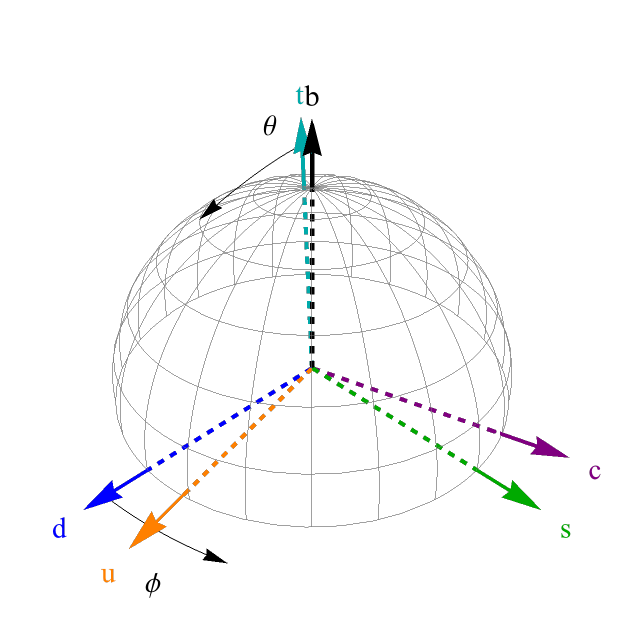}
	\caption{SM quark directions of the versor $\hat{n}$, shown in the semi-sphere described by the two angles $(\theta,\phi)$. Left-handed up quarks fit in this space by writing the $SU(2)_L$ quark doublet in the down mass basis: $q^i_L = (V_{ji}^* u_L^j, d^i_L)^t$. See Ref.~\cite{Gherardi:2019zil} for details.}
	\label{fig:ROFV}
\end{figure}
The key assumption of this work is that the NP sector responsible for the $R_K^\nu$ signal couples to a single direction in quark flavour space, inducing the ROFV structure \cite{Gherardi:2019zil}.
In the lepton side we consider instead flavour universality or coupling only to the tau flavour.
Under these assumptions, the Wilson coefficient matrices in Eq.~\eqref{eqLEFTlag} have the following structure:
\be
\label{eq:left_ROFV}
L_{L,R,V}^{ij\alpha\beta}= C_{L,R,V} \times \hat{n}_i \hat{n}_j^{*} \times 
\begin{cases}
\delta^{\alpha\beta} & \text{for LFU ,}\\
\delta^{\tau\alpha}\delta^{\tau\beta} & \text{for only tau flavour ,}
\end{cases}
\ee
where $C_{L,R,V} \in \mathbb{R}$ and $\hat{n}$ is a unitary vector in $U(3)_q$ flavour space. 
We can parametrize $\hat{n}$ as
\be
\hat{n}=
\begin{pmatrix}
e^{i\alpha_{db}}\sin\theta\cos\phi  \\
e^{i\alpha_{sb}}\sin\theta\sin\phi  \\
\cos\theta
\end{pmatrix}
\ee
where the angles and the phases can be chosen to lie in the following range
\be
\theta \in \left[0,\frac{\pi}{2}\right], \ \phi \in \left[0,2\pi\right) , \ \alpha_{sb}\in \left[-\frac{\pi}{2},\frac{\pi}{2}\right] , \ \alpha_{db}\in \left[-\frac{\pi}{2},\frac{\pi}{2}\right] \ .
\ee
The directions in flavour space associated to each SM quark can be shown in the semi-sphere described by the angles $(\theta,\phi)$ as in Fig.~\ref{fig:ROFV}.

%\begin{verbatim}
\begin{table}[t!]
\caption{Best-fit values of $sb$ LEFT coefficients from a fit to $B \to K^{(*)} \nu \bar{\nu}$ observables, as obtained in App.~\ref{appendix:fit_left_sb}.}
\label{tab:fit_left_sb}
%\begin{center}
\begin{tabular}{cc|cc|cc|}
%\begin{tabular}{@{}llllll@{}}
\multicolumn{2}{c|}{} & \multicolumn{2}{c|}{{\bf LFU}} & \multicolumn{2}{c|}{{\bf $\nu_\tau$ only}} \\
\midrule
\multicolumn{2}{c|}{{\bf RH}} & \multicolumn{2}{c|}{$L_R^{sb, \text{LFU}} \approx (11.5 \, \TeV)^{-2}$} & \multicolumn{2}{c|}{$L_R^{sb \tau\tau} \approx (7.7 \, \TeV)^{-2}$} \\
\midrule
\multicolumn{2}{c|}{{\bf LH}} &  \multicolumn{2}{c|}{$L_L^{sb, \text{LFU}} \approx (14.2 \, \TeV)^{-2}$} &  \multicolumn{2}{c|}{$L_L^{sb \tau\tau} \approx (9.2 \, \TeV)^{-2}$} \\
\midrule
\multicolumn{2}{c|}{{\bf V}} &  \multicolumn{2}{c|}{$L_V^{sb, \text{LFU}} \approx (11.6 \, \TeV)^{-2}$} &  \multicolumn{2}{c|}{$L_V^{sb \tau\tau} \approx (7.7 \, \TeV)^{-2}$} \\
\hline
\end{tabular}
%\end{center}
\end{table}
%\end{verbatim}

For any given $\theta$ and $\phi$, the overall coefficient $C_{L,R,V}$ and the phase $\alpha_{sb}$ can be univocally determined by the fit to $b\to s\nu\bar\nu$ observables. As described in detail in App.~\ref{appendix:fit_left_sb}, we find that we can set $L_{L,R,V}^{sb\alpha\beta}$ to be real and positive to be conservative. This yields $\alpha_{sb}=0$ and
\be
\label{eq:fixCLEFT}
C_{L,R,V} \cos\theta \sin\theta \sin\phi = \left. L^{sb}_{L,R,V}  \right|_{\rm best-fit}~,
\ee
where in the r.h.s we use the best-fit value resulting from the fit of all $B \to K^{(*)} \nu \bar{\nu}$ observables, including the Belle~II excess, which is reported in Table~\ref{tab:fit_left_sb} for the various cases. 

The ansatz in Eq.~\eqref{eq:left_ROFV} allows us to correlate the EFT coefficients for $b \to d \nu\bar{\nu}$ and $s\to d \nu \bar{\nu}$ transitions to the one of $b \to s \nu \bar{\nu}$, once the overall scale $C_{L,R,V}$ is fixed for any given $\theta$ and $\phi$ using Eq.~\eqref{eq:fixCLEFT}. For simplicity we set also $\alpha_{db}=0$.
This means that for any direction in the quark flavour space, {\emph{i.e.}} any $\theta$ and $\phi$, we get a precise prediction for the New Physics contribution to other observables. The EFT dependence of the relevant meson decays observables is discussed in App. \ref{App:Observables}.

\begin{figure}
	\centering
	\includegraphics[scale=0.4]{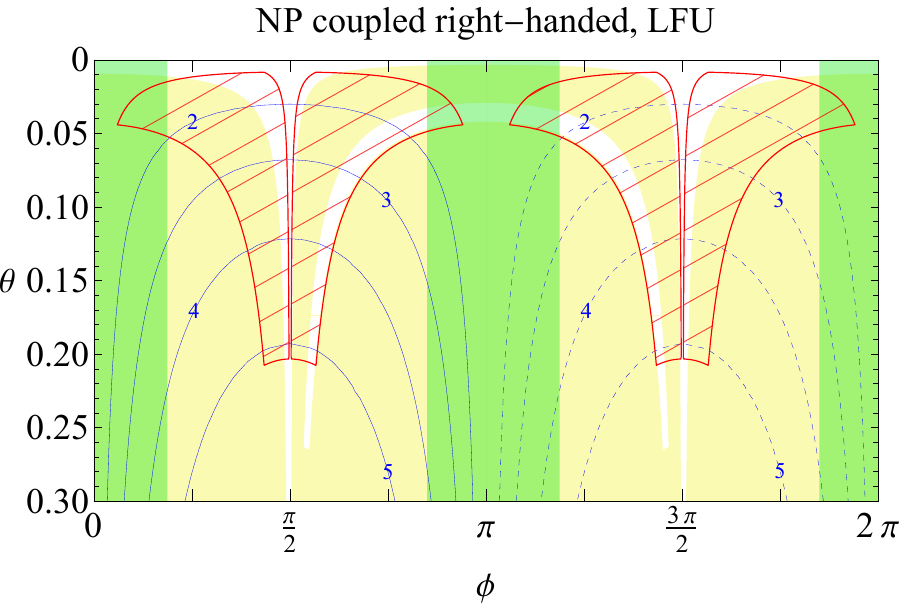}
	\quad
	\includegraphics[scale=0.4]{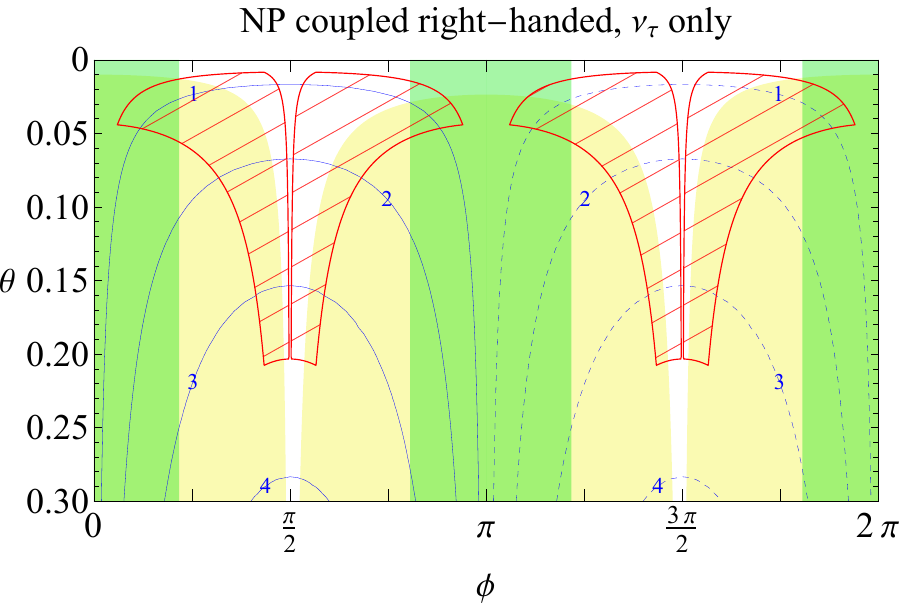}\\
    \vspace{4mm}
	\includegraphics[scale=0.4]{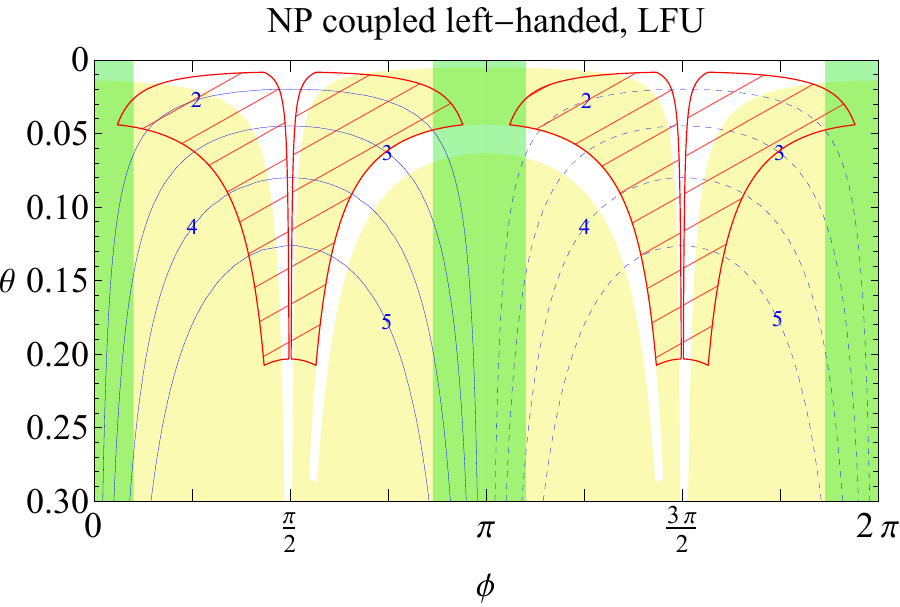}
	\quad
	\includegraphics[scale=0.4]{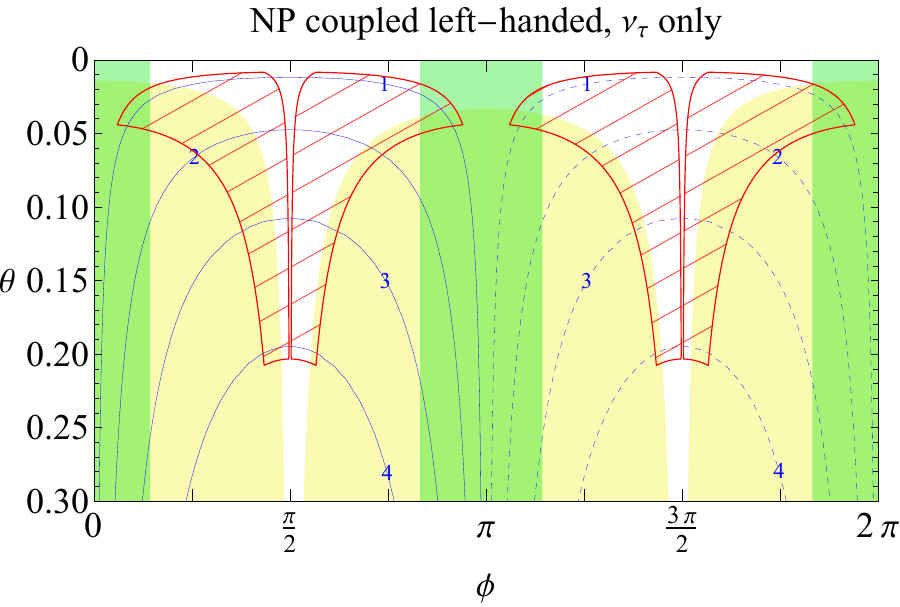}\\
	\includegraphics[scale=0.6]{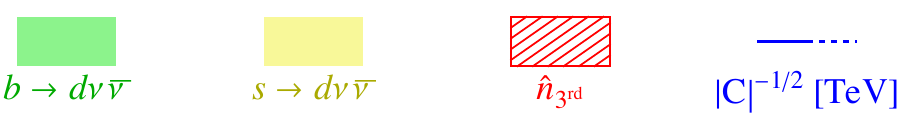}
	\caption{Limits in the $\phi-\theta$ plane for NP coupled to right-handed and left-handed quark currents assuming LFU (on the left) or coupling only to tau neutrinos (on the right). The case of vector-like current is very similar to the right-handed one and therefore not reported here.
 The solid (dashed) blue lines refer respectively to positive (negative) values of $C$, labelled as $|C|^{-1/2}$ [TeV]. The meshed red region correspond to the one aligned close to the third generation, Eq.~\eqref{eq:ROFV3rdgen} with $|a_{db,sb}| \in [0.2-5]$.}
	\label{fig:LEFT}
\end{figure}

The corresponding regions in the $\phi-\theta$ plane, excluded at the 95\% CL, are shown in Fig.~\ref{fig:LEFT} for the cases of interest. As we can see from the plots, the most severe bounds come from $K^+\to\pi^+\nu\bar\nu$, which requires $\hat{n}$ to be very close to the direction of the third family. The $B \to \pi \nu \bar{\nu}$ and $B \to \rho \nu\bar{\nu}$ limits, shown together in the picture, exclude instead values of $\phi \approx 0, \pi$, {\emph{i.e.}} directions of $\hat{n}$ too much aligned to the down quark $d$.

Among all the possible directions described by $\hat{n}$, a region that is particularly well motivated from the theoretical point of view is the one near the third generation:\\ \mbox{$\hat{n}_{3^{\rm rd}} = (\mathcal{O}(V_{td}), \mathcal{O}(V_{ts}), \mathcal{O}(1))$}. Indeed, NP is expected to couple more strongly to the third generation of quarks in several explicit models, both to evade the strong flavour and direct searches constraints, that involve light quarks, as well as to address the hierarchy problem of the electroweak scale or the SM flavour puzzle. This scenario can be parametrized by \cite{Gherardi:2019zil}
\be
    \hat{n}_{3^{\rm rd}} \propto ( a_{bd} e^{i \alpha_{bd}} |V_{td}|, a_{bs} e^{i \alpha_{bs}} |V_{ts}|, 1 ) \ ,
\label{eq:ROFV3rdgen}
\ee
where $a_{bd, bs}$ are $\mathcal{O}(1)$ real parameters. In Fig.~\ref{fig:LEFT} and in the following we show with a meshed-red area the region with $|a_{bd, bs}| \in [0.2 - 5]$.

In this section, we focused exclusively on operators that conserve total lepton number $L$. However, in App.~\ref{app:LNV} we also discuss a possible interpretation in terms of operators that violate $L$. There, we show that dimension-six operators with $|\Delta L| =2$ could potentially accommodate the excess. However, these operators are unlikely to fit within the ROFV framework since they involve both left- and right-handed quark chiralities, which in general couple independently to NP. As a result, these operators fall outside the scope of this paper and are not discussed further in the main text.

\section{Implications in the SMEFT}
\label{sec:SMEFT}

Assuming that the New Physics scale lies above the electroweak scale, it is natural to work within the Standard Model EFT (SMEFT) framework. The dim-6 operators made of quark currents and affecting the neutrino decay channels are
\begin{align}\label{eq:SMEFT}
\mathcal{O}_{lq}^{(1)\alpha\beta ij}=\left(\bar{l}_{L}^\alpha \gamma_\mu l^\beta_{L}\right)\left(\bar{q}^i_{L}\gamma^\mu q^j_{L}\right)
\ ,& \qquad
\mathcal{O}_{Hq}^{(1)ij}=\left(H^\dag{\overleftrightarrow D_\mu} H\right)\left(\bar{q}^i_{L}\gamma^\mu q^j_{L}\right) \ , \nonumber\\
\mathcal{O}_{lq}^{(3)\alpha\beta ij}=\left(\bar{l}^\alpha_{L}\gamma_\mu\sigma_a l^\beta_{L}\right)\left(\bar{q}^i_{L}\gamma^\mu\sigma_a q^j_{L}\right)
\ ,& \qquad
\mathcal{O}_{Hq}^{(3)ij}=\left(H^\dag\sigma_a{\overleftrightarrow D_\mu} H\right)\left(\bar{q}^i_{L}\gamma^\mu\sigma_a q^j_{L}\right) \ , \\
\mathcal{O}_{ld}^{\alpha\beta ij}=\left(\bar{l}^\alpha_{L}\gamma_\mu l^\beta_{L}\right)\left(\bar{d}^i_{R}\gamma^\mu d^j_{R}\right)
\ ,& \qquad
\mathcal{O}_{Hd}^{ij}=\left(H^\dag{\overleftrightarrow D_\mu} H\right)\left(\bar{d}^i_{R}\gamma^\mu d^j_{R}\right) \ . \nonumber
\end{align}
We choose to work in the down-type quark basis, so that the fermionic weak doublets in the operators above are $l^\alpha_{L}=(\nu_L^\alpha,\ell_L^\alpha)$ and $q^i_{L}=(V_{ji}^{*} u_L^j, d_L^i)$, where $V$ is the CKM matrix. Being aligned with the down-quark mass basis, the matching relations between the SMEFT operators in Eq.~\eqref{eq:SMEFT} and the LEFT ones in Eq.~\eqref{eqLEFTlag} are, at tree level:
\be
\label{smeft_to_left}
\begin{split}
&
L_{L}^{ij\alpha\beta}=C_{lq}^{(1)\alpha\beta ij}-C_{lq}^{(3)\alpha\beta ij}+C_{Hq}^{(1)ij}\delta_{\alpha\beta} + C_{Hq}^{(3)ij}\delta_{\alpha\beta}\,,\\
& L_{R}^{ij\alpha\beta}=C_{ld}^{\alpha\beta ij}+C_{Hd}^{ij}\delta_{\alpha\beta}\,.
\end{split}
\ee
These SMEFT operators can affect a broad set of low energy observables both via tree and loop level effects, getting sizable constraints on the flavour direction space. Therefore, we enlarge the set of observables under consideration,  whose correlation to di-neutrino modes is studied under the ROFV hypothesis. In detail, we include the rare decays $B_s(K_{L,S})\to \mu^+\mu^-$, meson-mixing constraints, the Higgs and electroweak fit from Ref.~\cite{Falkowski:2019hvp} (we use an updated version kindly provided by the authors),  and data from high-$p_T$ dilepton tails via the HighPT tool \cite{Allwicher:2022mcg}. Details on the EFT dependence of these observables are collected in App.~\ref{App:Observables}.

We assume that the operators are induced at the UV scale $\mu_{\mathrm{UV}}\equiv1$ TeV. The Wilson coefficients are evolved down to the EW scale $\mu_{EW}\sim M_Z$ with the RGEs from Refs.~\cite{Jenkins:2013wua,Jenkins:2013zja,Alonso:2013hga}. Going across the EW threshold, we match the SMEFT onto the LEFT at tree-level \cite{Jenkins:2017jig} and finally evolve the LEFT operators down to the low energy scales relevant for flavour observables \cite{Jenkins:2017dyc}. Following this procedure, we eventually build a global likelihood expressed in terms of the UV Wilson Coefficients. The RG equations are solved numerically using the DSixTools package \cite{Fuentes-Martin:2020zaz}, both above and below the EW scale.
In the following, unless otherwise specified, SMEFT coefficients are all understood to be evaluated at the UV scale $\mu_{\mathrm{UV}}$.

The analysis strategy for SMEFT operators follows the same steps outlined for the LEFT case. The only difference lies in the fit of
the excess from Belle~II,
now including also the $B_s \to \mu^+\mu^-$ decay, the
$R_{K^{(*)}}$ ratios and,
in some specific case,
the $\Delta M_{B_s}$ meson mass mixing, whose information would otherwise be lost
as they are mainly affected, trough the correlations coming from the SMEFT or the underlying UV structure, by the same SMEFT coefficients related to $R^{\nu}_{K^{(*)}}$.

Let us start by discussing the Higgs-quark SMEFT operators $O_{Hq}^{(1)}$, $O_{Hq}^{(3)}$ and $O_{Hd}$. The Belle~II excess, as well as the other neutrino channels, are sensitive to the $C_{Hq}^{(+)}\equiv C_{Hq}^{(1)}+ C_{Hq}^{(3)}$ and $C_{Hd}$ coefficients.
We find that the constraints set by $B_s\to\mu^+\mu^-$ and $\Delta M_{B_s}$ do not allow to accommodate the excess in $R^\nu_K$.
Indeed, an enhancement above $\sim25\%$ of the SM prediction for $B^+\to K^+\nu\bar\nu$ is highly disfavoured, in agreement with what was pointed out in \cite{Allwicher:2023xba}. We report a more detailed analysis of these cases in App.~\ref{appendix:fit_smeft_sb}.

In the rest of the paper we focus on four-fermion semileptonic SMEFT operators, specifically $O_{lq}^{(1)\alpha\beta ij}$, $O_{lq}^{(3)\alpha\beta ij}$, and $O_{ld}^{\alpha\beta ij}$.
We introduce the combinations $C_{lq}^{(\pm)}=C_{lq}^{(1)}\pm C_{lq}^{(3)}$, {\emph{i.e.}} the relevant degrees of freedom affecting the observables included in the analysis. The Belle~II excess is indeed sensitive to the $C_{lq}^{(-)}$ and $C_{ld}$ coefficients, while $B_s \to \mu^+ \mu^-$ provides bounds on the positive combination.
Since there could be various different possible  combinations of coefficients at the SMEFT level, it is useful to consider which operators are expected to be induced from UV models.
Given the EFT scales resulting from Fig.~\ref{fig:LEFT}, scenarios that induce the desired effect at the loop level are expected to have light and relatively strongly coupled states, disfavored by direct searches at the LHC.
Therefore, in the next Section we present a detailed study of simplified models of single-mediator scenarios, that generate different combinations of those semileptonic operators when the heavy state is integrated out at the tree-level.

\section{Simplified Models}
\label{sec:models}

The semileptonic operators in Eq.~\eqref{eq:SMEFT} can be induced by integrating out at the tree-level either leptoquarks or colorless heavy vectors.
The possible leptoquarks responsible for those operators are listed in Table~\ref{tab:LQ_mediators}, while colorless heavy vectors can either be $Z^\prime \sim ({\bf 1}, {\bf 1}, 0)$ or $V^\prime \sim ({\bf 1}, {\bf 3}, 0)$. An important feature of these UV models is the generation, at tree or loop level, of 4-quarks operators that affect $\Delta F=2$ processes, which are highly constrained by meson-antimeson mixing measurements \cite{UTfit,UTfit:2022hsi}
and whose bounds are so included as well in the following analysis. For details on the generation mechanisms of 4-quarks operators, see App.~\ref{appendix:four_quark_ops}.

As already explained in the previous Section, we first fit the $B\to K^{(*)}\nu\bar\nu$ decays together with the observables correlated trough the SMEFT or the underlying UV structure.
The details of the fit and the resulting plots for all the models we consider have been put in App.~\ref{appendix:fit_smeft_sb}.
In the second step we perform the global analysis in the $(\phi,\theta)$ plane including all the other observables discussed in the previous Section. For the constraints from high-$p_T$ tails we use the EFT approach, assuming that the mediator's mass is above a few TeV.

\subsection{Vector triplet $V^\prime$}

First we consider a colourless spin-1 mediator $V^\prime$ transforming as $({\bf 1},{\bf 3},0)$ under the SM group $G_{\rm SM}$, whose interaction to the SM fields is given by the Lagrangian
\be
\label{eq:lag:Vp}
\mathcal{L}\supset \sum_{a}\left[ \sum_{ij}g_q^{ij}(\bar{q}^i_{L}\gamma^\mu\sigma_a q^j_{L}) + \sum_{\alpha\beta}g_\ell^{\alpha\beta}\bar{l}^\alpha_{L}\gamma^\mu\sigma_a l^\beta_{L} \right]V_{a\mu}^\prime \ .
\ee
According to the ROFV hypothesis of this work, the rank of the quark coupling matrix $g_q^{ij}$ is assumed equal to one, {\emph{i.e.}} $g_q^{ij}=g_q\hat{n}_i\hat{n}_j^*$. A flavour structure such this could arise, for example, if the quark fields couple to $V^\prime$ only via mixing with a single heavy VLQ doublet $Q$ in the form
\be
    \mathcal{L}_{V^\prime} \supset g_{Q} V^\prime_\mu (\bar{Q} \gamma^\mu Q)  + M_Q \bar{Q} Q + ( M_i \bar{Q}q^i_{L} + h.c. \ ) \ ,
\ee
so that $\hat{n}_i\propto M_i$.
Once $V^\prime$ is integrated out, the matching with semileptonic SMEFT operators reads
\be
C_{lq}^{(3)\alpha\beta ij}=-\frac{1}{4}\frac{g_q^{ij}g_\ell^{\alpha\beta}}{M_{V^\prime}^2}, \quad C_{lq}^{(1)\alpha\beta ij}=C_{ld}^{\alpha\beta ij}=0 \ .
\ee
\begin{figure}
	\centering
	\includegraphics[scale=0.64]{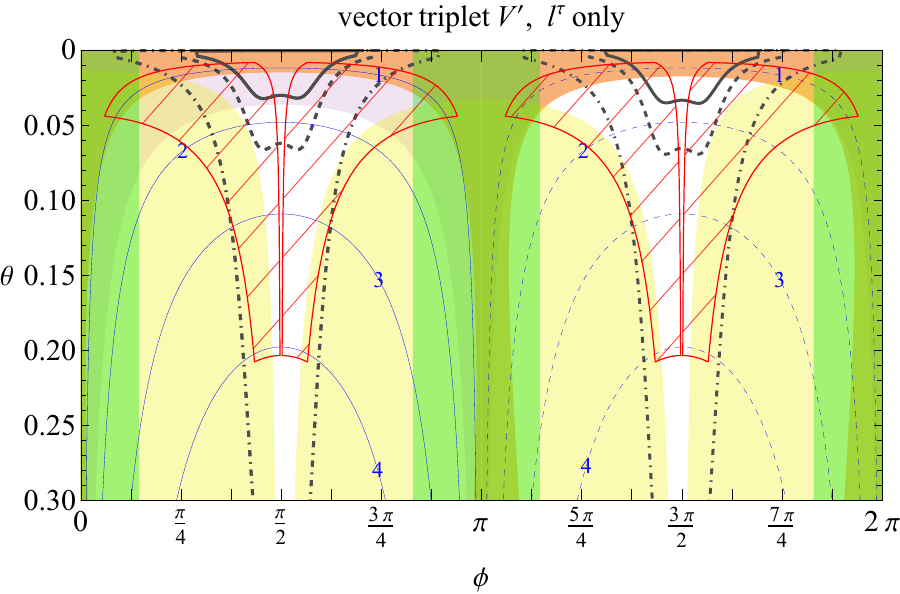}\\
	\includegraphics[scale=0.6]{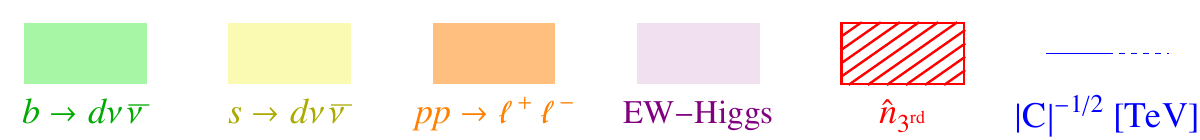}
	\caption{Most important limits on the $\phi-\theta$ plane for a $V^\prime$ mediator. Contours of the regions excluded by $\Delta F=2$ observables are shown for $|g_q/g_\ell|=0.1$(solid), $0.05$(dashed) and $0.01$(dot-dashed). The solid (dashed) blue lines refer respectively to positive (negative) values of $C=g_q g_\ell /4M_{V^\prime}^2$, labelled as $|C|^{-1/2}$ [TeV]. The meshed red region correspond to the to the third generation alignment, Eq.~\eqref{eq:ROFV3rdgen} with $|a_{db,sb}| \in [0.2-5]$.}
	\label{fig:SMEFT_Vp}
\end{figure}
On the lepton side, we consider either LFU or only tau flavour, namely
\be
\label{eq:g_l_Vp}
g_\ell^{\alpha\beta}=\begin{cases}
g_{\ell} \ \delta^{\alpha\beta} & \text{for LFU ,}\\
g_\ell \ \delta^{\tau\alpha}\delta^{\tau\beta} & \text{for only tau flavour .}
\end{cases}
\ee
Following this, we can write $C_{lq}^{(3)\alpha\beta ij} = C \ \hat{n}_i \hat{n}^*_j \ (g_\ell^{\alpha \beta}/g_\ell)$, where $C=g_q g_\ell /4M_{V^\prime}^2$.
From the 2D fits of $R_{K}^\nu$ in App.~\ref{appendix:fit_smeft_sb} (Fig.~\ref{fig:SMEFT_sb_Zp}),
we observe that $B_s\to\mu^+\mu^-$ highly disfavours any enhancement of $R_{K}^\nu$ in the LFU case.
On the other hand, a coupling to only tau flavour allows to accommodate the Belle~II result.
Assuming a real and positive EFT coefficient to be conservative, we find a best-fit value
\be
\label{eq:bestfitVp}
\left. C_{lq}^{(-)\tau\tau sb} \right|_{\rm best-fit} \approx(9.1 \, \TeV)^{-2} \ .
\ee
Setting this value, we then plot the excluded region at 95\% in the $\phi-\theta$ plane due to observables correlated trough the ROFV structure in Fig.~\ref{fig:SMEFT_Vp}.
We find again that $K\to\pi\nu\bar{\nu}$ decay is the strongest constraint, forcing $\phi \sim \frac{\pi}{2},\frac{3\pi}{2}$ to evade the bounds, but differently from the LEFT case, the  small $\theta$ region is now forbidden by high-$p_T$ tails and EW data.
With these parameters, the contribution to $b\to c \tau \nu$ transition is small and the effect in $R(D^{(*)})$ is limited to $\sim4$ percent or less.
Furthermore, $\Delta F = 2$ operators are induced at the tree-level, proportionally to $g_q^2 / M_{V^\prime}^2$, see Eq.~\eqref{eq:DeltaF2Vprime}. Compared to $C_{lq}^{(-)\tau\tau ij}$ they depend on the ratio $|g_q/g_\ell|$, which is an independent variable: smaller values suppress $\Delta F = 2$ effects.
Contours of the region excluded by meson-antimeson mixing at 95\% are shown for $|g_q/g_\ell|=0.1$, $0.05$ and $0.01$.
It is clear that we need a certain hierarchy between the coupling to quarks and tau, approximately $|g_q/g_\ell|\lesssim 0.05$,
in order to have a viable scenario.
Such maximal allowed value, combined with the constraint $g_\ell \lesssim 2.3$ coming from the tool of perturbative unitary (PU) \cite{Barducci:2023lqx} and the fixed best-fit value in Eq.~\eqref{eq:bestfitVp}, imply an upper limit for the vector mass:
\begin{align}
\label{eq:MVpUpperLim}
M_{V^\prime} \lesssim & \ 1391 \ \GeV \ \left( \frac{|g_q/g_\ell|^{\rm max}}{0.05} \right)^{1/2} \ \left|\sin\theta\cos\theta\sin\phi\right|^{\frac{1}{2}}  \\
\approx & \ 762 \ \GeV \ \left( \frac{|g_q/g_\ell|^{\rm max}}{0.05} \right)^{1/2} \ \left| \frac{\theta}{0.3} \right|^{1/2} \ , \nonumber
\end{align}

where in the last step we approximate $\sin\theta\sim\theta$ and $\cos\theta\sim\sin\phi\sim1$ as found from our analysis.
Such a light mass invalidates our use of the limits from high-$p_T$ tails, which employ the EFT approximation for heavy mediators. Nevertheless, we can consider the resonant channel $p p \to V^{\prime 0} \to \tau\tau$, and the corresponding search by ATLAS \cite{ATLAS:2015rbx,ATLAS:2020zms}. In our setup, since $|g_q / g_\ell| \ll 1$, the branching ratio to the tau pair will be $\sim 1/2 $ and given the preference for having $\hat{n}$ aligned to the third generation (small $\theta$) the production cross section will be dominated by $b-\bar{b}$ fusion (for completeness we include contributions from all quarks). Notwithstanding the large value of $g_\ell$, the width of the vector is approximately $\Gamma_{\rm tot} / M_{V^\prime} \approx 14\%$ so we use the narrow width approximation to obtain the total cross section:
\be
    \sigma(p p \to V^{\prime 0} \to \tau^+ \tau^-) \approx \frac{4 \pi^2}{3} \mathcal{B}(V^{\prime 0} \to \tau^+ \tau^-) \sum_{i,j = u,d,s,c,b} \frac{\Gamma(V^{\prime 0} \to q^i \bar{q}^j)}{M_{V^\prime}}  \frac{2}{s_0} \mathcal{L}_{q^i \bar{q}^j}(M_{V^{\prime}})~,
    \label{eq:xsecVptautau}
\ee
where $s_0 = (13 \ \TeV)^2$, $\mathcal{L}_{q^i \bar{q}^j}(E)$ is the $q^i-\bar{q}^j$ parton luminosity and $\Gamma(V^{\prime 0} \to q^i \bar{q}^j) = \frac{M_{V^\prime} N_c}{24 \pi} |g_q^{ij}|^2$. We include all neutral-current $(i,j)$ combinations, including flavour-violating ones.
For each value of $\theta-\phi$ we fix $M_{V^\prime}$ at the upper limit from Eq.~\eqref{eq:MVpUpperLim}, set $g_\ell = g_\ell^{\rm max} \approx 2.3$, and then fix $g_q$ by imposing the best-fit value for $C_{lq}^{(-)\tau\tau s b}$. Doing so we have all parameters required to compute the cross section and we can compare with the upper limit from ATLAS \cite{ATLAS:2015rbx,ATLAS:2020zms}.  We find that the whole region allowed by the global fit (white region in Fig.~\ref{fig:SMEFT_Vp}) is excluded for values $\theta \lesssim 0.22$, see Fig.~\ref{fig:SMEFT_Mmax_collider}.

In addition to this, another possible direct-search channel for such vectors is electroweak pair-production, \emph{i.e.} $p p \to V^+ V^-$ or $V^\pm V^0$, with subsequent decays into taus and neutrinos. The analysis of Ref.~\cite{ATLAS:2017qwn} looked for pair-produced charginos decaying into taus and missing energy, obtaining an upper limit of about 630~GeV for light LSP. This analysis could be recasted to derive a limit on the $V^\prime$ mass, however the coupling to EW gauge bosons would introduce further model dependence via the presence of a possible non-minimal coupling. This analysis goes well beyond the purpose of our work, so we do not pursue it further.

\begin{figure}
	\centering
	\includegraphics[scale=0.54]{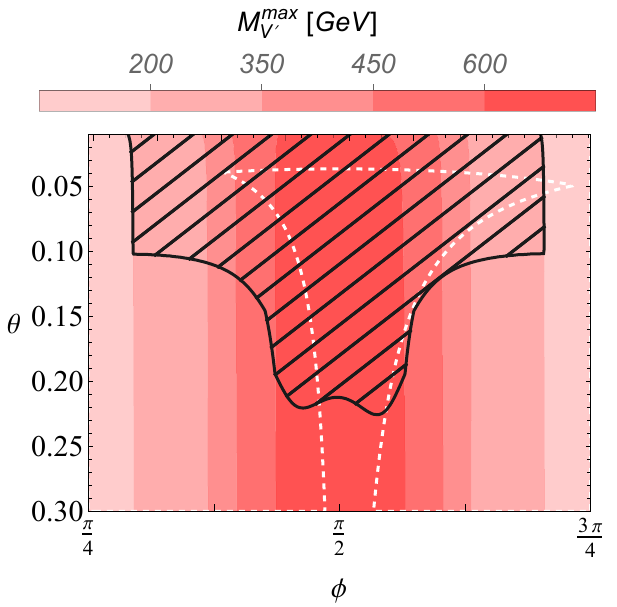}
	\includegraphics[scale=0.54]{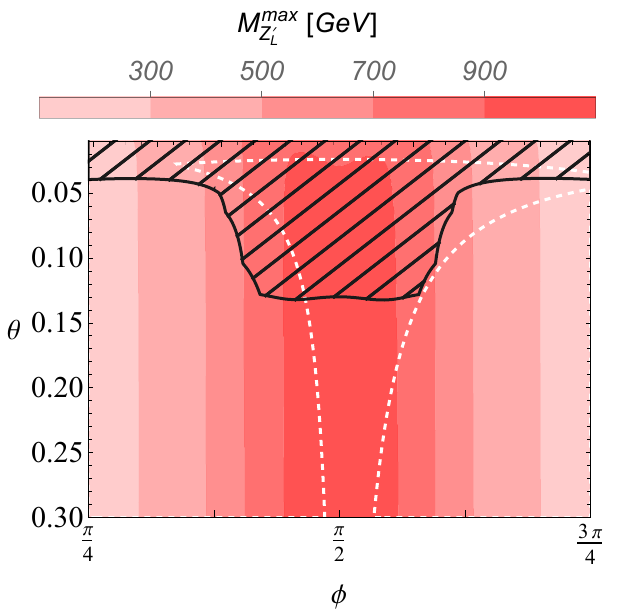}
	\includegraphics[scale=0.54]{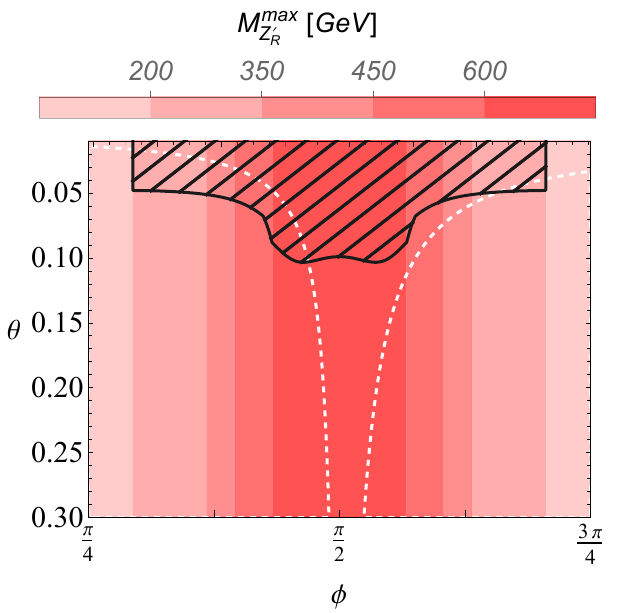}
	\caption{Maximal allowed values of the mediator mass in the $\phi-\theta$ around $\phi\sim\pi/2$ for $V^\prime$ and $Z_{L,R}^\prime$ with couplings to only tau leptons. The meshed black region is excluded by ATLAS trough the direct search of $pp \to X \to \tau^+\tau^-$~\cite{ATLAS:2020zms}. The region outside the dashed white line is excluded at 95\% by the same observables of Figs.~\ref{fig:SMEFT_Vp} and \ref{fig:SMEFT_Zp} except for the high-$p_T$ dilepton tails whose constraint does not hold at such small values of the mediator mass. The results for the region around $\phi\sim3\pi/2$ are qualitatively similar to these.}
	\label{fig:SMEFT_Mmax_collider}
\end{figure}

\subsection{Vector singlet $Z^\prime$}

We turn now to a colourless spin-1 mediator $Z^\prime$ transforming as $({\bf 1},{\bf 1},0)$ under $G_{\rm SM}$, with couplings
\be
\mathcal{L}\supset \left[ \sum_{ij}g_L^{ij}(\bar{q}^i_{L}\gamma^\mu q^j_{L}) + \sum_{ij}g_R^{ij}(\bar{d}^i_{R}\gamma^\mu d^j_{R}) + \sum_{\alpha\beta}g_\ell^{\alpha\beta}\bar{l}^\alpha_{L}\gamma^\mu l^\beta_{L} \right]Z_{\mu}^\prime \ .
\ee
We are interested in scenarios where the mediator couples to a quark current with only left ($g_R^{ij}=0$), right ($g_L^{ij}=0$) or vector ($g_L^{ij}=g_R^{ij}$) couplings, denoted respectively as $Z_{L,R,V}^\prime$.

However, we immediately discard the vector combination scenario $Z_V^\prime$ as it is non trivial to justify a ROFV structure for it and furthermore
it is highly disfavoured by meson-antimeson mixing constraints, as they would require a larger hierarchy among the lepton and quark couplings, namely $|g_q/g_\ell|\lesssim\mathcal{O}(10^{-3})$, with respect to the case with $Z_{L,R}^\prime$ mediator.
So we will not discuss further this option and hence focus on $Z_{L,R}^\prime$. We then assume the quark coupling matrix for both these mediator to have the ROFV structure
$g_{L,R}^{ij}= g_{q}\hat{n}_i \hat{n}_j^{*}$. New $Z_{L,R}^{\prime}$s with such coupling structure are justified by UV scenarios with heavy vector-like quarks, as already pointed out for $V^{\prime}$.
The tree-level matching with the relevant semileptonic SMEFT operators reads
\begin{gather}
\label{eq:lag:Zp}
C_{lq}^{(1)\alpha\beta ij}=-\frac{g_L^{ij}g_\ell^{\alpha\beta}}{M_{Z^\prime}^2}, \quad C_{lq}^{(3)\alpha\beta ij}=0, \quad C_{ld}^{\alpha\beta ij}=-\frac{g_R^{ij}g_\ell^{\alpha\beta}}{M_{Z^\prime}^2} \ ,
\end{gather}
where, on the lepton side we consider the same cases as done for the vector triplet, Eq.~\eqref{eq:g_l_Vp}.

\begin{figure}
	\centering
	\includegraphics[scale=0.4]{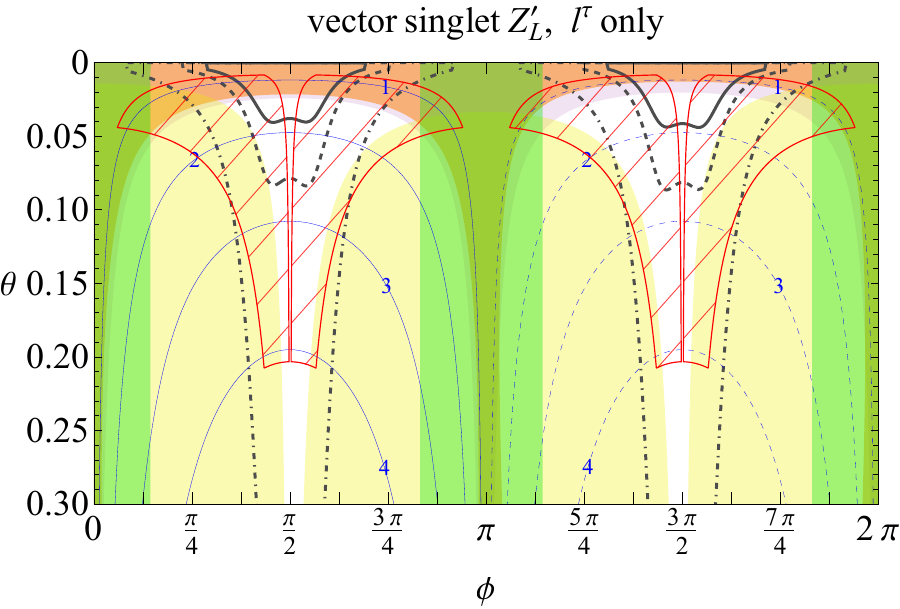}
 \quad \includegraphics[scale=0.4]{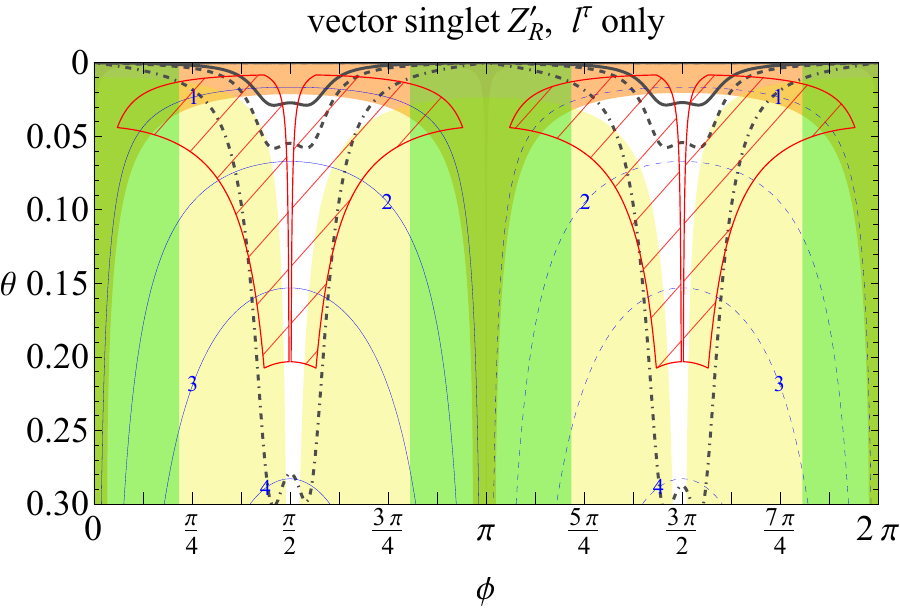}\\
	\includegraphics[scale=0.56]{plots/legend_smeft_Vp_or_Zp.pdf}
	\caption{Limits in the $\phi-\theta$ plane for $Z_L^\prime$ and $Z_R^\prime$ mediators. Contours of the regions excluded by $\Delta F=2$ observables are shown for $|g_q/g_\ell|=0.1$(solid), $0.05$(dashed) and $0.01$(dot-dashed). The solid (dashed) blue lines refer respectively to positive (negative) values of $C=-g_q g_\ell /M_{Z^\prime}^2$, labelled as $|C|^{-1/2}$ [TeV]. The meshed red region correspond to the to the third generation alignment, Eq.~\eqref{eq:ROFV3rdgen} with $|a_{db,sb}| \in [0.2-5]$.}
	\label{fig:SMEFT_Zp}
\end{figure}

The usual 2D fits of the Belle~II result can be found in App.~\ref{appendix:fit_smeft_sb}. In general, we find that we can accommodate the excess in all the scenarios, expect for the case of a $Z_R^\prime$ with LFU couplings.
However, we discard $Z^{\prime}_{L}$ boson with LFU coupling as well as it is highly disfavoured by the $\Delta F =2$ constraints, which would require a larger hierarchy among the lepton and quark couplings, namely $|g_q/g_\ell|\lesssim\mathcal{O}(10^{-2})$, respect to the case with lepton coupling only to tau flavour.
Assuming real coefficients, we find the best-fit values 
\be
\left. C_{lq}^{(-)\tau\tau sb} \right|_{Z_L, \rm best-fit} \approx(9.2 \, \TeV)^{-2} \, , \quad 
\left. C_{ld}^{\tau\tau sb} \right|_{Z_R, \rm best-fit} \approx(7.7 \, \TeV)^{-2} \, .
    \label{eq:ZpfixRnuK}
\ee
Using these values, the correlation with other observables under the ROFV hypothesis is displayed in Fig.\,~\ref{fig:SMEFT_Zp}. We can notice that the most severe bounds come again from $K\to\pi\nu\bar\nu$ that requires $\hat{n}$ to be very close to the direction of the third family while, as already discussed for the $V^{\prime}$ case, the small $\theta$ region is excluded at the SMEFT level by $\textrm{high}-p_T$ tails or by EW data.
Contours of the region excluded by meson-antimeson mixing at 95\% are shown for $|g_q/g_\ell|=0.1$, $0.05$ and $0.01$.
As in the vector triplet case, a hierarchy between lepton and quark couplings, namely $|g_q/g_\ell|\lesssim 0.05$, is required.

The discussion on the upper limit on the $Z^\prime$ mass and corresponding direct searches follows analogously to the previous section, with the main difference that the $Z^\prime$ cannot be pair-produced via electroweak interactions so $\tau^+ \tau^-$ resonant production is the dominant search channel.
From the maximal value of  $|g_q/g_\ell|$ allowed by meson mixing we can derive an upper limit for the vector mass after setting $g_\ell$ at the maximal value from perturbative unitarity, $g_{\ell}^{\rm max} \approx 1.65$. In this case the branching ratio into taus is expected to be equal to $1/2$ with good accuracy and we can compute the signal cross section as in Eq.~\eqref{eq:xsecVptautau} after solving for $g_q$ by imposing the best-fit values in Eq.~\eqref{eq:ZpfixRnuK}. In the case of $Z^\prime_R$ we do not include the contribution from $u$ and $c$ quarks, since it couples only to right-handed down quarks. The excluded region in the two models can be seen in Fig.~\ref{fig:SMEFT_Mmax_collider}, corresponding to a lower limit $\theta \gtrsim 0.13~(0.10)$ for $Z^\prime_L$ ($Z^\prime_R$).

\subsection{Leptoquarks}

Leptoquarks (LQs) are hypothetical BSM states that mediate interactions between leptons and quarks.
Since LQs interacting with lepton fields of different flavour would generate lepton flavour violating couplings once integrated out, we assume couplings only to the tau fields for these type of single-mediator simplified models.
In this case they naturally yield a Rank-One structure, where the matching of the coefficient $C$ of Eq.~\eqref{eq:fixCLEFT} to the LQ coupling and mass is shown in Table \ref{tab:LQ_mediators} while $\hat{n}_i$ is proportional to the LQ coupling to the SM fields as $\lambda_{i\tau}\equiv\lambda \ \hat{n}_i$.

%\begin{verbatim}
\begin{table}[t!]
%\begin{center}
\begin{tabular}{c|c|cc|cc|cc|cc|}
{\bf LQ} & {\bf Spin} & \multicolumn{2}{c|}{$G_{\rm SM}$} & \multicolumn{2}{c|}{{\bf Interaction term}} & \multicolumn{2}{c|}{{\bf SMEFT coeff.}} & \multicolumn{2}{c|}{$C$} \\
\midrule
$S_1$ & 0 & \multicolumn{2}{c|}{$(\bar{{\bf 3}},{\bf 1},1/3)$} & \multicolumn{2}{c|}{$\lambda_{i\tau}^* (\overline{q_{L}^{i,c}}\epsilon l^\tau_{L})S_1$} & \multicolumn{2}{c|}{$C_{lq}^{(1)}=-C_{lq}^{(3)}$} & \multicolumn{2}{c|}{$\frac{1}{2}\frac{|\lambda|^2}{m_{LQ}^2}$} \\
\midrule
$S_3$ & 0 & \multicolumn{2}{c|}{$(\bar{{\bf 3}},{\bf 3},1/3)$} & \multicolumn{2}{c|}{$\lambda_{i\tau}^* (\overline{q_{L}^{i,c}}\epsilon\sigma_a l^\tau_{L})(S_3)_a$} & \multicolumn{2}{c|}{$C_{lq}^{(1)}=3C_{lq}^{(3)}$} & \multicolumn{2}{c|}{$\frac{1}{2}\frac{|\lambda|^2}{m_{LQ}^2}$} \\
\midrule
$U_3$ & 1 & \multicolumn{2}{c|}{$({\bf 3},{\bf 3},2/3)$} & \multicolumn{2}{c|}{$\lambda_{i\tau} (\overline{q^i_{L}}\gamma_\mu\sigma_a l^\tau_{L})(U_3^\mu)_a$} & \multicolumn{2}{c|}{$C_{lq}^{(1)}=-3C_{lq}^{(3)}$} & \multicolumn{2}{c|}{$-\frac{1}{2}\frac{|\lambda|^2}{m_{LQ}^2}$} \\
\midrule
$\tilde{R}_2$ & 0 & \multicolumn{2}{c|}{$({\bf 3},{\bf 2},1/6)$} & \multicolumn{2}{c|}{$\lambda_{i\tau} \overline{d^i_{R}}( l^\tau_{L}\epsilon\tilde{R}_2)$} & \multicolumn{2}{c|}{$C_{ld}$} & \multicolumn{2}{c|}{$-\frac{1}{2}\frac{|\lambda|^2}{m_{LQ}^2}$} \\
\midrule
$V_2$ & 1 & \multicolumn{2}{c|}{$(\bar{{\bf 3}},{\bf 2},5/6)$} & \multicolumn{2}{c|}{$\lambda_{i\tau}^* \overline{d_{R}^{i,c}}\gamma_\mu(l^\tau_{L}\epsilon V_2^\mu)$} & \multicolumn{2}{c|}{$C_{ld}$} & \multicolumn{2}{c|}{$\frac{|\lambda|^2}{m_{LQ}^2}$} \\
\hline
\end{tabular}
%\end{center}
\caption{List of the leptoquark fields relevant for this work. Here $\epsilon\equiv i\sigma_2$ acts on the isospin indices.}
\label{tab:LQ_mediators}
\end{table}
%\end{verbatim}

\begin{figure}
	\centering
	\includegraphics[scale=0.45]{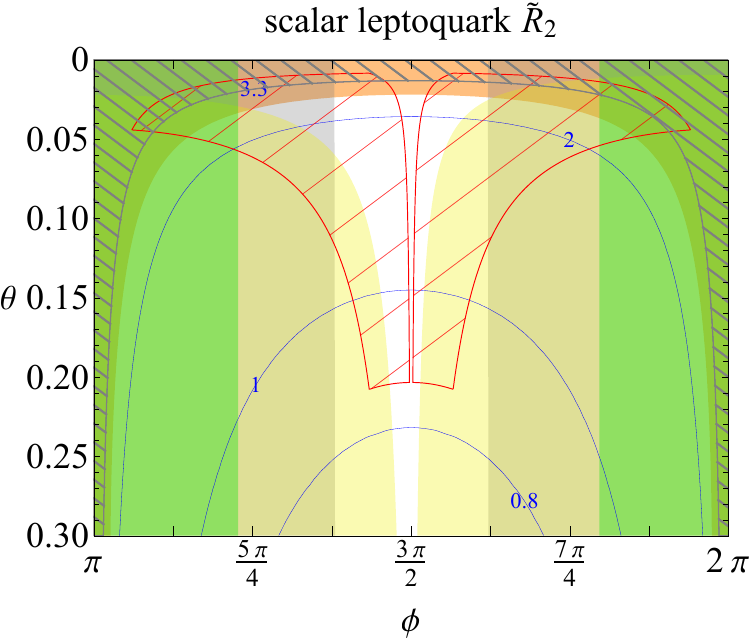}
	\quad
	\includegraphics[scale=0.45]{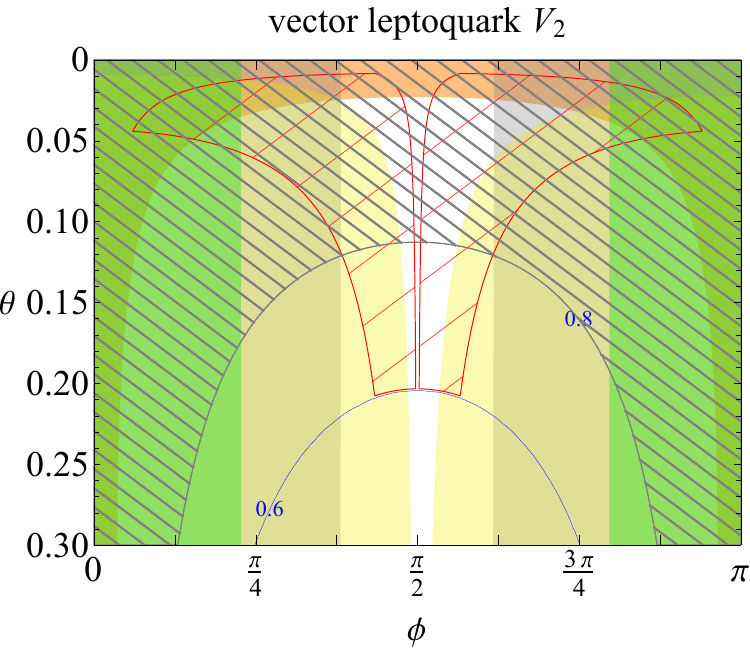}\\
    \includegraphics[scale=0.45]{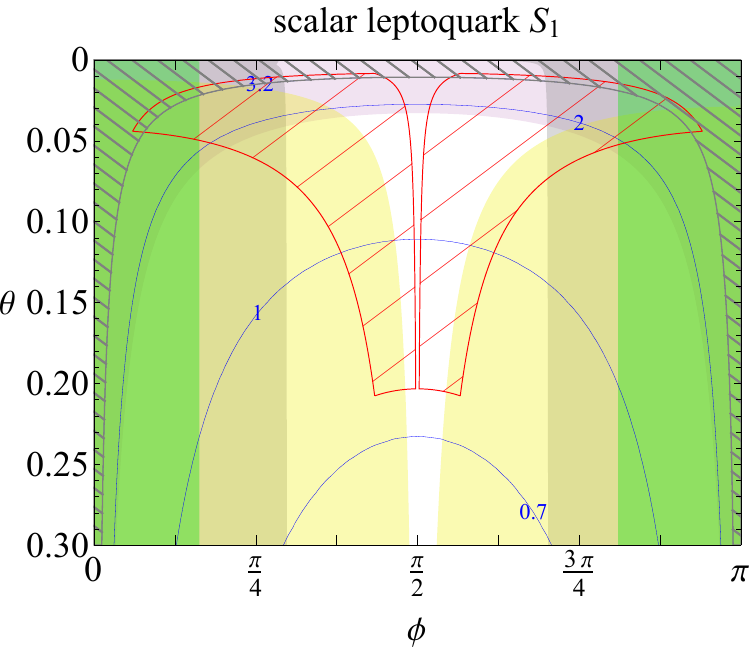}\\
    \includegraphics[scale=0.56]{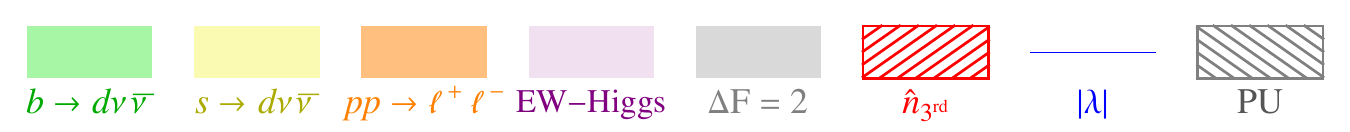}
	\caption{Limits in the $\phi-\theta$ plane for $S_1$, $\tilde{R}_2$ and $V_2$ leptoquarks with mass $m_{\rm LQ} = 2$~TeV. The solid blue lines refer to values of the LQ coupling $|\lambda|$. The meshed red region correspond to the one aligned to the third generation of quarks up to $\mathcal{O}({\rm CKM})$ rotations, Eq.~\eqref{eq:ROFV3rdgen}, with $|a_{db,sb}| \in [0.2-5]$. The dashed gray region is exclude by the argument of perturbative unitary.}
	\label{fig:LQ}
\end{figure}

In our analysis we also consider meson-antimeson mixing effects coming from loop-induced four-quark operators (see App.~\ref{appendix:four_quark_ops} for details). To do so, we set $m_{LQ}=2$ TeV as benchmark point, which is beyond the lower limit of 1.3 (1.5) TeV obtained by direct searches for scalar (minimally coupled vector) LQ coupled to third generation SM fermions \cite{ATLAS:2021jyv}.

The resulting 2D fits of the Belle~II excess for all the LQs listed in Table~\ref{tab:LQ_mediators} can be found in App.~\ref{appendix:fit_smeft_sb}.
Among the various leptoquarks we focus on $\tilde{R}_2$ and $V_2$, which give the best fit to the $R^\nu_K$ excess as shown in Fig.~\ref{fig:SMEFT_sb_LQ}, and on $S_1$, whose fit is slightly worse than the former, as an example of LQ coupled to left-handed quarks. This is understood by the fact that they induce a right-handed quark current, preferred by data over the left-handed one (see Fig.~\ref{fig:fit_BKnunu_data}).
The scalar triplet $S_3$ performs slightly worse than $S_1$ due to the larger contribution to meson mixing and the vector triplet $U_3$ is expected to induce an even larger effect in $\Delta F=2$ observables, although for a precise computation it would be necessary to introduce a UV-completion for the vector, resulting in a model-dependent analysis.
As usual, we find that we can set $C_{lq}^{(-)\tau\tau sb}$ or $C_{ld}^{\tau\tau sb}$ to be real and positive to be conservative.
This yields $\alpha_{sb}=0$ while the best-fit values of $C_{lq}^{(-)\tau\tau sb}$ or $C_{ld}^{\tau\tau sb}$ are respectively
\be\begin{split}
\left. C_{lq}^{(-)\tau\tau sb} \right|_{S_1, \rm best-fit} &\approx(8.5 \, \TeV)^{-2} \, , \\
\left. C_{ld}^{\tau\tau sb} \right|_{\tilde{R}_2, \rm best-fit} &\approx(7.5 \, \TeV)^{-2} \, , \\
\left. C_{ld}^{\tau\tau sb} \right|_{V_2, \rm best-fit} &\approx(7.5 \, \TeV)^{-2} \, .
\end{split}\ee

Once the overall coefficient is determined from $b\to s \nu\bar\nu$ data, we can constrain the direction in the quark flavour space, {\emph{i.e.}} $\theta$ and $\phi$, using the other observables correlated to $R_K^\nu$ in the ROFV hypothesis. Again, for simplicity we set $\alpha_{db}=0$.
The corresponding constrained regions in the $\phi-\theta$ plane are shown in Fig.~\ref{fig:LQ} for the cases of our interest.
Since $C_{lq}^{(-)\tau\tau sb}$ or equivalently $C_{ld}^{\tau\tau sb}$ is set to be positive, the values of $\phi$ are restricted to $[0,\pi]$ or to $[\pi,2\pi]$ depending on if the LQ induces respectively a positive or negative overall coefficient $C$.
From the plot we can read that for the $S_1$ case, the small $\theta$ region gets constrained by EW data, specifically the measurement of the $Z \to \tau \tau$ coupling, due to the mixing along the RG evolution of $C_{Hl}^{(+)}$ with $C_{lq}^{(-)}$ (see Eq.~\eqref{eq:HiggsBasis}), but it is not bounded by dilepton tails as the positive combination $C_{lq}^{(+)}$ vanishes for $S_1$. Viceversa, $\tilde{R}_2$ and $V_2$ get the strongest bound at small $\theta$ from high-$p_T$ di-tau tails.
These constraints are expected to improve substantially from future measurements at (HL-)LHC. Using the projections from the HighPT tool \cite{Allwicher:2022mcg}, we verified that the bounds on the Wilson coefficients will improve by a factor of approximately 2. Similar improvements are expected also for the colorless vectors.
As for $V^\prime$, also for the leptoquarks the contribution to $R(D^{(*)})$ is always negligible and cannot address the observed deviation from the SM prediction.
Different values of the LQ coupling $|\lambda|$ are outlined by blue contours in Fig.~\ref{fig:LQ}.
Furthermore, a PU bound on $\lambda$ can be computed by considering all the $2\to2$ scattering in the model as discussed in Refs.~\cite{Allwicher:2021rtd,Barducci:2023lqx}. By direct calculation one gets
\be
|\lambda|\lesssim\begin{cases}
3.2 & \text{for $S_1$ ,}\\
3.3 & \text{for $\tilde{R}_2$ ,}\\
0.8 & \text{for $V_2$ .}
\end{cases}
\label{eq:LQpert}
\ee
Such bound actually excludes a relevant portion of the parameter space in the $V_2$ plot, while we are safely below the PU threshold for $S_1$ and $\tilde{R}_2$.

\begin{figure}[t]
	\centering
	\includegraphics[scale=0.45]{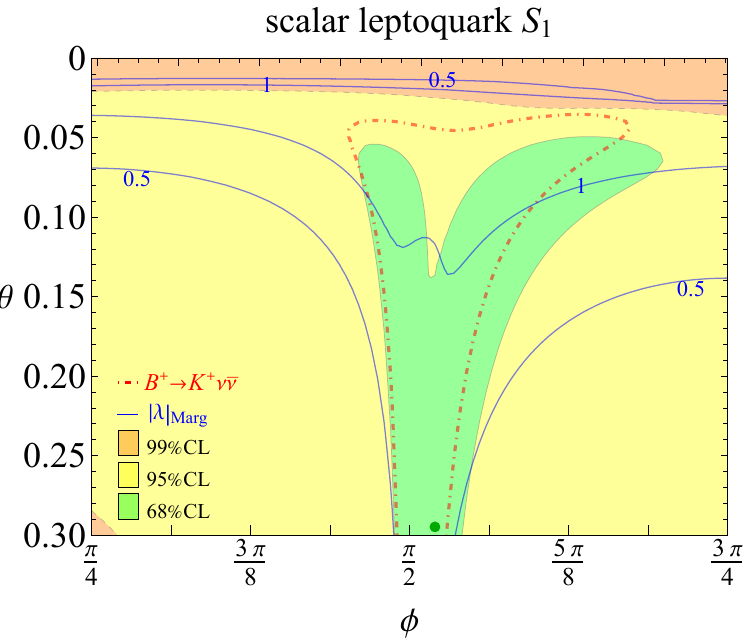}
	\quad
	\includegraphics[scale=0.45]{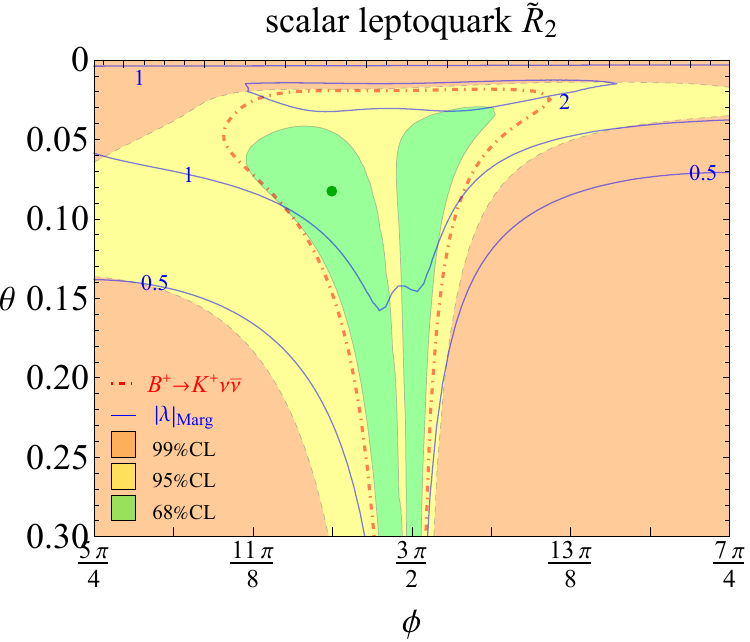}\\
    \vspace{4mm}
    \includegraphics[scale=0.45]{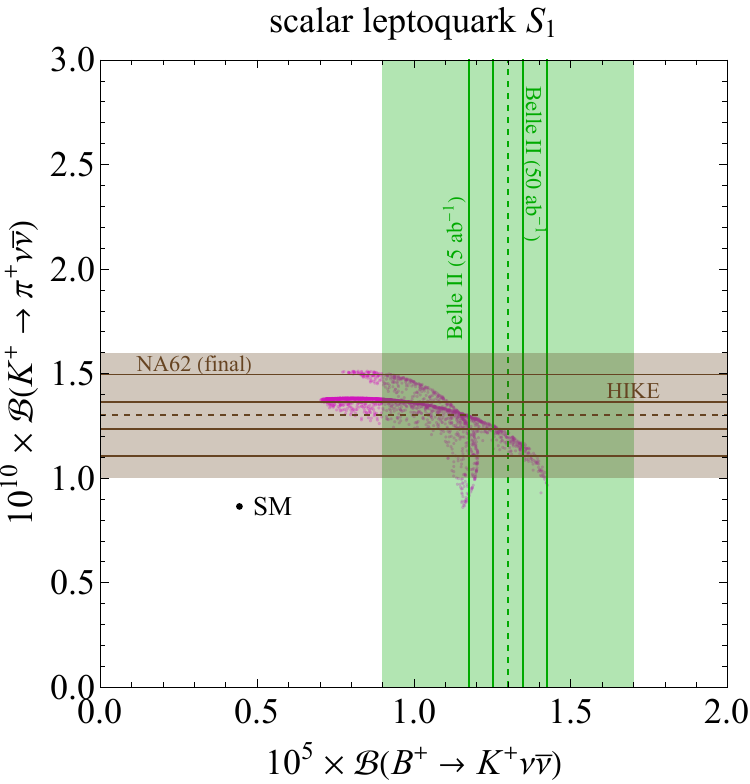}
    \quad
    \includegraphics[scale=0.45]{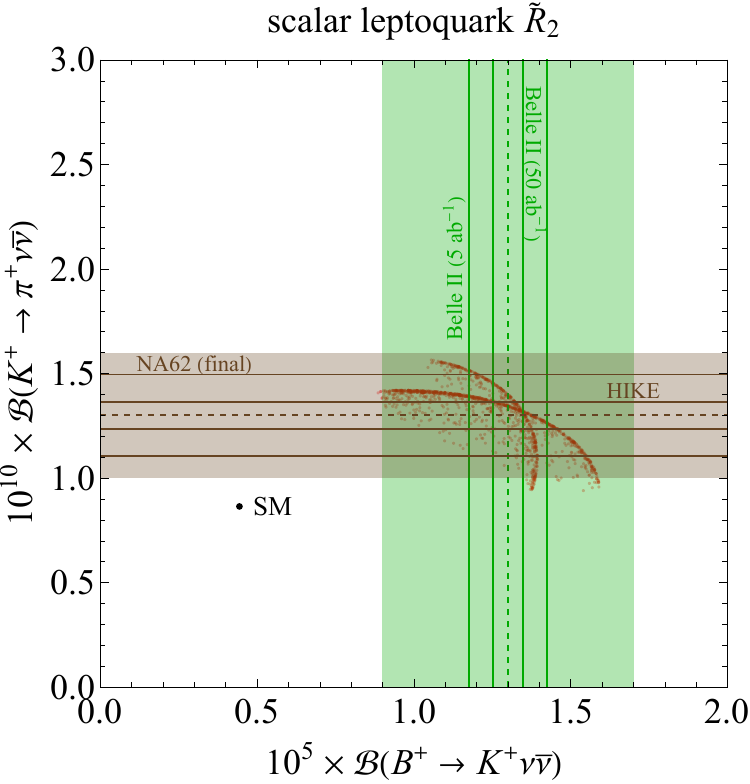}
	\caption{Top row: preferred region in the $\phi-\theta$ plane by the global fit for the $S_1$ and $\tilde{R}_2$ leptoquarks, where the value of $|\lambda|$ has been marginalised. In blue we show iso-lines of the value of $|\lambda|$ that marginalizes the fit.
    Bottom row: we show points taken from a random scan inside of the $1\sigma$ region of the global fit (green region in the plots above), mapped in the plane of $B^+ \to K^+ \nu \bar\nu$ and $K^+ \to \pi^+ \nu \bar\nu$. We overlay the present $1\sigma$ constraints as well as future prospects.}
	\label{fig:LQ_prof_obs}
\end{figure}

We note that $\tilde{R}_2$ and $V_2$ generate the same semileptonic coefficient $C_{ld}$, albeit with opposite overall sign, and same four-quark operator (see App.~\ref{appendix:four_quark_ops}), so that they exhibit a similar behaviour. 
However, most of the allowed region for $V_2$ is excluded by the perturbative unitarity bound, Eq.~\eqref{eq:LQpert}, as clear from Fig.~\ref{fig:LQ}.

When compared to $V^\prime$ and $Z^\prime$ vectors, that require a large hierarchy between the couplings to quarks and to leptons in order to pass $\Delta F=2$ constraints, leptoquarks enjoy an automatic loop suppression of flavor mixing.
We therefore identify $S_1$ and $\tilde{R}_2$ as the most favoured single-mediator simplified models to accommodate the observed enhancement in Belle~II measurement of $B^+\to K^+\nu\bar\nu$ decay.

Motivated by these results, we proceed with a further investigation of the region at small $\theta$ and $\phi \sim \frac{\pi}{2}$ or $\frac{3\pi}{2}$ through the following procedure. We consider the complete $\chi^2$ of our analysis (including so both $b \to s$ transitions and all the possible correlated observables) that depends on the overall SMEFT coefficient $C$ and on the angle variables $\theta,\,\phi$.\footnote{We set $\alpha_{sb}=\alpha_{db}=0$ for simplicity.}
We profile the $\chi^2$ over the $C$ parameter, {\emph{i.e.}} over $|\lambda|$ since we fix $m_{LQ}=2$~TeV, namely at each $(\theta,\phi)$ point we substitute the value of $|\lambda|$ that minimise the chi-square. The resulting $\chi^2_{\textrm{profiled}}(\theta, \phi)$ is used to derive the $1\sigma$, $2\sigma$ and $3\sigma$ regions shown in the top row of Fig.~\ref{fig:LQ_prof_obs}. The solid blue lines correspond to the value of $|\lambda|$ that minimised the $\chi^2$. The dot-dashed red line represents the $1\sigma$ region preferred by the excess in $B^+ \to K^+ \nu \bar\nu$.

Finally, we map the $1\sigma$ (green) regions of the top row in Fig.~\ref{fig:LQ_prof_obs} into regions in the $\mathcal{B}(B^+ \to K^+ \nu \bar{\nu}) - \mathcal{B}(K^+ \to \pi^+ \nu \bar{\nu})$ plane, in order to show the predicted values of these di-neutrino modes for the two simplified models under examination. We overlay the present $1\sigma$ constraints by Belle-II and NA62, as well as future prospects by the same experiments \cite{Belle-II:2018jsg} as well as by the proposed HIKE experiment at CERN \cite{HIKE:2023ext}. It is clear that future measurements will have a strong power to confirm or disprove definitely the present excess observed by Belle-II, and to resolve the specific direction in flavour space for the New Physics couplings under the ROFV hypothesis.

%\newpage
%************************************************************
\section{Conclusions} 
\label{sec:conclusions}
Numerous groundbreaking advancements in particle physics have been achieved through the measurement of low-energy observables.
It is also likely that a future discovery of physics beyond the SM will come from such experiments. In this case, correlations among these observables are crucial for establishing New Physics and identifying the appropriate UV extension of the Standard Model.

However, obtaining correlations requires making theoretical assumptions, particularly regarding the flavor structure of the New Physics.
A generic model-independent parametrization results in a complex analysis with numerous couplings, potentially obscuring the underlying UV physics and yielding generic correlations. Conversely, a model-dependent approach typically yields strong correlations but is limited in validity to the specific model.

In our study, we propose a simple and motivated ansatz for the New Physics that offers a balance between predictivity and generality: the Rank One Flavor Violation (ROFV) hypothesis \cite{Gherardi:2019zil}. In this scenario, New Physics couples to a specific direction in flavor space, which can be varied enabling a comprehensive exploration of this hypothesis using only a few parameters.

We specifically examine the recent hint of a potential deviation in the $B \to K \nu \bar\nu$ decays as observed by the Belle II collaboration.\footnote{Contrary to Refs.~\cite{Bause:2023mfe,Allwicher:2023xba} we conservatively use the combination of Belle II~\cite{Belle-II:2023esi} and Belle~\cite{Belle:2017oht}, resulting in a reduced value for the central value of $R^{\nu}_K$ respect to the cited works.} Possible correlations arise from  Lorentz symmetry, SM gauge symmetry, and flavor structure. While we consider the most general operators for Lorentz and gauge symmetries, we employ our ROFV hypothesis for the flavor aspect of the analysis.

At the LEFT analysis level, the minimal correlation under scrutiny is that with the decay $B \to K^* \nu\bar\nu$. Our primary findings are depicted in Fig.~\ref{fig:fit_BKnunu_data}. We find that right-handed and vector currents for quarks are favored over the left-handed counterpart, as the latter yields $R^{\nu}_{K^*}=R^{\nu}_{K}$, and currently, there is a good agreement between the observed value of $B \to K^* \nu\bar\nu$ and the Standard Model (SM) prediction. Our analysis using ROFV indicates that most important observable correlating with $B \to K \nu\bar\nu$ is the transition $K \to \pi \nu\bar\nu$, while from a more theoretical standpoint, the preference lies within the region aligned with the bottom quark.

Regarding the SMEFT, we have identified six potential operators that could contribute to $b \to s \nu \bar\nu$ at the tree level. Among them, three involve the Higgs doublet but are not considered phenomenologically viable due to constraints from $B_s \to \mu^+ \mu^-$ and $\Delta M_{B_s}$, confirming the finding of \cite{Allwicher:2023xba}. The remaining three are semileptonic four-fermion operators. Instead of conducting an exhaustive examination, we focused on simplified models of single-mediator scenarios, which yield fixed combinations of these semileptonic operators when the heavy state is integrated out at the tree level. Loop level mediator are not considered in our work because they have to be very light and strongly coupled to reproduce the anomalous effect and this scenario seems disfavoured by the present data.  

We discuss our findings on all the six potential mediators, ranking them from the most to the least effective.
For all the cases we assume coupling to tau lepton doublet only and for leptoquarks we fix the mass to the benchmark value of 2 TeV to avoid constraints from direct searches. 

\begin{enumerate}
\item $\tilde{R}_2 \sim ({\bf 3},{\bf 2},1/6)$. 
The optimal scenario involves the scalar leptoquark, which interacts with right-handed quarks. Leptoquarks are preferred due to their contribution to $\Delta F=2$ at the one-loop level, a feature lacking in the vector triplet $V^\prime$ and singlet $Z^\prime$. Presently, the data favors right-handed chirality for quarks.

\item $S_1\sim (\bar{{\bf 3}},{\bf 1},1/3)$ and $S_3\sim (\bar{{\bf 3}},{\bf 3},1/3)$. Less favorable than the previous scenario due to its coupling to left-handed quarks. A more precise measurement of $R^{\nu}_{K^*}$ could differentiate between this case and the preceding one.

\item $V_2 \sim (\bar{{\bf 3}},{\bf 2},5/6)$. 
This scenario suggests low mass and significant couplings, raising concerns regarding perturbative unitarity. Additionally, spin one states necessitate a UV completion to address various issues such as calculability of some observables.

\item $Z^\prime\sim ({\bf 1},{\bf 1},0)$. 
In this instance, $\Delta F=2$ effects emerge at the tree level. This necessitates the coupling to quarks to be substantially suppressed respect to the leptons one. An upper limit for the mass is deduced from indirect searches in every direction within the ROFV sphere. Although direct searches hold significant importance, they are largely contingent on the model. The only inevitable production mechanism is Drell-Yann, which is directly associated with our ROFV hypothesis. $Z^\prime$ is predicted to decay predominately in tau and neutrino pairs. A substantial portion of the parameter space is excluded, leaving only a small region where the mass of the $Z_L^\prime$ ($Z_R^\prime$) is not larger than approximately 900 (600) GeV.

\item $V^\prime\sim ({\bf 1},{\bf 3},0)$. Similar considerations apply as in the $Z^\prime$ case, with the distinction that there exists another production mechanism in direct searches due to pure gauge interactions. A comprehensive reassessment of the experimental findings is necessary; however, we anticipate the parameter space of this scenario to be more restricted, if not entirely ruled out.

\item $U_3 \sim ({\bf 3},{\bf 3},2/3) $.  
Compared to $V_2$, the situation worsens considerably due to its couplings to left-handed quarks and its large contribution to meson mixing, which does not allow to fit $R^\nu_K$ within $1\sigma$.
\end{enumerate}

Our primary phenomenological finding utilizing the ROFV hypothesis suggests the potential for significant effects in $K \to \pi \nu\bar\nu$. For this reason, and particularly for the most promising candidates $S_1$ and $\tilde{R}_2$, we conducted an analysis providing a more detailed understanding of the correlation between this observable and $R^{\nu}_{K}$ (see Fig.~\ref{fig:LQ_prof_obs}). We emphasize the importance of obtaining better measurements for both observables and present future prospects from experiments such as Belle II and NA62.\footnote{We note that a proposed experiment like HIKE \cite{HIKE:2023ext} would have been advantageous in elucidating this specific physics case.}

On the theoretical front, we emphasize that New Physics is favored to primarily couple to the third generation, in both the quark and charged lepton flavour spaces. This strengthens the case of a potential connection between the anomalous data and the Yukawa structure of the Standard Model couplings.

%\section*{Acknowledgments}

%The work of MN and CT is supported by the Italian Ministry of University and Research (MUR) via the PRIN 2022 project n.~2022K4B58X -- AxionOrigins. DM acknowledges support by the MUR grant PRIN 20224JR28W.

\backmatter

\bmhead{Acknowledgements}

The work of MN and CT is supported by the Italian Ministry of University and Research (MUR) via the PRIN 2022 project n.~2022K4B58X -- AxionOrigins. DM acknowledges support by the MUR grant PRIN 20224JR28W.

\begin{appendices}

\section{Observables}
\label{App:Observables}

%---------------------------------------
\subsection{Dineutrino meson decays}
\label{app:goldenmodes}

Here we recollect the dineutrino decay channels of meson, with the BSM contribution to these processes described by the Hamiltonian:
\begin{equation}
\begin{split}
& \mathcal{H}_{eff}^{BSM}=-\frac{4 G_F}{\sqrt{2}}\frac{\alpha}{4\pi}\sum_{i,j,\alpha,\beta} V_{ti}^*V_{tj}\left( C_L^{ij\alpha \beta}\mathcal{O}_{L}^{ij\alpha \beta} + C_R^{ij\alpha \beta}\mathcal{O}_{R}^{ij\alpha \beta} \right)\,, \\
& \mathcal{O}_{L}^{ij\alpha \beta}=(\bar{d}^i_{L}\gamma_{\mu}d^j_{L})(\bar{  \nu}^{\alpha}\gamma^{\mu}(1-\gamma_5)\nu^{\beta})\,,\quad \mathcal{O}_{R}^{ij\alpha \beta}=(\bar{d}^i_{R}\gamma_{\mu}d^j_{R})(\bar{  \nu}^{\alpha}\gamma^{\mu}(1-\gamma_5)\nu^{\beta})\, .
\end{split}
\end{equation}

\subsubsection{$B \to K^{(*)}\nu\bar{\nu}$}

The short-distance EFT coefficients in the Standard Model are given by $C_L^{sb\alpha\beta}=C_{L,\mathrm{SM}}^{sb}\delta_{\alpha\beta}$ and $C_{R}^{sb\alpha\beta}=0$, with $C_{L,\mathrm{SM}}^{sb}=-6.32(7)$.
Theoretical predictions of these modes have been updated with latest lattice results in Ref.~\cite{Becirevic:2023aov}, providing $\mathcal{B}(B^+ \to K^+ \nu \bar\nu)_{\rm SM} = (4.44 \pm 0.30) \times 10^{-6}$, $\mathcal{B}(B^+ \to K^{*+} \nu \bar\nu)_{\rm SM} = (9.8 \pm 1.4) \times 10^{-6}$, $\mathcal{B}(B^0 \to K_S \nu \bar\nu)_{\rm SM} = (2.05 \pm 0.14) \times 10^{-6}$, where the tree-level contribution has been removed.
It is common to define the ratios
\begin{equation}
R_{K^{(*)}}^{\nu}\equiv\frac{\mathcal{B}(B\to K^{(*)}\nu \bar\nu)}{\mathcal{B}(B\to K^{(*)}\nu \bar\nu)_{\mathrm{SM}}}=1+\Delta{R_{K^{(*)}}^{\nu}}\,,
\end{equation}
where the deviation from the SM expectation induced by New Physics can be parametrized as
\begin{equation}
    \begin{split}
    \Delta R_{K^{(*)}}^{\nu} & =\sum_{\alpha}\frac{2\mathrm{Re}[C_{L,\mathrm{SM}}^{sb}(C_L^{sb\alpha \alpha} +  C_R^{sb\alpha \alpha})]}{3 |C_{L,\mathrm{SM}}^{sb}|^2}+\sum_{\alpha,\beta}\frac{|C_L^{sb\alpha\beta}+ C_R^{sb\alpha \beta}|^{2}}{3|C_{L,\mathrm{SM}}^{sb}|^2}- \\
    & - \eta_{K^{(*)}}\sum_{\alpha,\beta}\frac{\mathrm{Re}[ C_{R}^{sb\alpha \beta}(C_{L,\mathrm{SM}}^{sb}+ C_L^{sb\alpha \beta})]}{3|C_{L,\mathrm{SM}}^{sb}|^2}\,.
    \end{split}
\end{equation}
In the formula above $\eta_K=0$ and $\eta_{K^{*}}=3.33(7)$, so that the decay mode to the $K$ meson only depends on the vector current, {\emph{i.e.}} on the combination $C_V= C_L +  C_R$.

\subsubsection{$K \to \pi\nu\bar{\nu}$}

Decays of $K_L$ and $K_0$ to pions and neutrinos have been discussed in Refs.~\cite{Marzocca:2021miv,Buras:2015qea,Buras:2015yca}. The branching ratios read:
\begin{equation}
\begin{split}
\mathcal{B}(K^+\to\pi^+\nu\bar{\nu})_{th} &= \mathcal{B}(K^+\to\pi^+\nu_e\bar{\nu}_e)_{\textrm{SM}}\sum_{\alpha,\beta=1,2}\left|\delta_{\alpha\beta}+\frac{C_V^{ds\alpha\beta}}{C_{L,\textrm{SM}}^{ds11}}\right|^2\\
&+\mathcal{B}(K^+\to\pi^+\nu_\tau\bar{\nu}_\tau)_{\textrm{SM}}\left[\left|1+\frac{C_V^{ds33}}{C_{L,\textrm{SM}}^{ds33}}\right|^2+\sum_{\alpha=1,2}\left(\left|\frac{C_V^{ds\alpha 3}}{C_{L,\textrm{SM}}^{ds33}}\right|^2+\left|\frac{C_V^{ds3\alpha}}{C_{L,\textrm{SM}}^{ds33}}\right|^2\right)\right]\,,
\end{split}
\end{equation}
\begin{equation}
\begin{split}
\mathcal{B}(K_L\to\pi^0\nu\bar{\nu})_{th} &= \frac{1}{3}\mathcal{B}(K_L\to\pi^0\nu\bar{\nu})_{\textrm{SM}}\Bigg[\sum_{\alpha,\beta=1,2}\left(\delta_{\alpha\beta}+\frac{\textrm{Im}[N_{ds}C_V^{ds\alpha\beta}]}{\textrm{Im}[N_{ds}C_{L,\textrm{SM}}^{ds11}]}\right)^2\\
&+\left(1+\frac{\textrm{Im}[N_{ds}C_V^{ds33}]}{\textrm{Im}[N_{ds}C_{L,\textrm{SM}}^{ds33}]}\right)+\sum_{\alpha=1,2}\left(\left(\frac{\textrm{Im}[N_{ds}C_V^{ds\alpha 3}]}{\textrm{Im}[N_{ds}C_{L,\textrm{SM}}^{ds33}]}\right)+\left(\frac{\textrm{Im}[N_{ds}C_V^{ds3\alpha}]}{\textrm{Im}[N_{ds}C_{L,\textrm{SM}}^{ds33}]}\right)\right)\Bigg]\,,
\end{split}
\end{equation}
where we defined $N_{ds}=(G_F\alpha V_{td}^*V_{ts}/\sqrt{2}\pi)$, $C_V=C_L+C_R$ and the SM values for the branching ratios and Wilson coefficients are 
\begin{equation}
\begin{split}
\mathcal{B}(K^+\to\pi^+\nu_e\bar{\nu}_e)_{\textrm{SM}} &= 3.06\times 10^{-11}\,,\\
\mathcal{B}(K^+\to\pi^+\nu_\tau\bar{\nu}_\tau)_{\textrm{SM}}&= 2.52\times 10^{-11}\,,\\
\mathcal{B}(K_L\to\pi^0\nu\bar{\nu})_{\textrm{SM}}&= 3.4\times 10^{-11}\,,
\end{split}
\end{equation}
\begin{equation}
C_{L,\textrm{SM}}^{ds\alpha\beta} = -\frac{1}{s_W^2}\left(X_t+\frac{V_{cd}^*V_{cs}}{V_{td}^*V_{ts}}X^\alpha_c\right)\delta_{\alpha\beta}\,,\quad C_{R,\textrm{SM}}^{ds\alpha\beta}=0\,,
\end{equation}
with $X_t =1.481$, $X^e_c=X^\mu_c = 1.053\times 10^{-3}$ and $X^{\tau}_c = 0.711 \times 10^{-3}$.

\subsubsection{$B \to \pi \nu\bar{\nu},\rho \nu\bar{\nu}$}

Differential and total branching ratios of $B \to \pi \nu \bar{\nu}$ and $B \to \rho \nu \bar{\nu}$ are studied, e.g., in Ref.~\cite{Bause:2021cna}. The general formula for this processes is:
\begin{equation}\label{eq:BtoF}
    \mathcal{B}(B\to F_d \nu \bar{\nu})=A^{BF_d}_+ x_{bd}^+ + A^{BF_d}_- x_{bd}^-\,,
\end{equation}
with
\begin{equation}
    x_{bd}^{\pm}=\sum_{\alpha,\beta}\left|2 V_{td}^{*}V_{tb}\left(C_{L,\mathrm{SM}}^{db}\delta_{\alpha\beta}+C_L^{db\alpha\beta}\pm C_R^{db\alpha\beta}\right)\right|^2\,,
\end{equation}
and where we have defined the coefficients in such a way that $C_{L,\mathrm{SM}}^{db}=-6.32$. The coefficients $A_{-}^{BP}$ are zero for pseudoscalar mesons while, in general, $A_{-}^{BV} \gg A_{+}^{BV}$ for vectors. They are given in Table~\ref{tab:ABFpm}.
%
%\begin{verbatim}
\begin{table}[!h]
%\begin{center}
\begin{tabular}{|c c c|} 
\hline
 $B \to F_d$ & $A_{+}^{BF}\,[10^{-8}]$ & $A_{-}^{BF}\,[10^{-8}]$ \\
 \hline
 \hline
  $B^+\to \pi^+$ & $332\pm34$ & $0$ \\ 
 \hline
  $B^0\to \pi^0$ & $154\pm16$ & $0$ \\
 \hline
  $B^+\to \rho^+$ & $126\pm26$ & $1236\pm502$ \\
 \hline
 $B^0\to \rho^0$ & $59\pm12$ & $573\pm233$ \\
 \hline
\end{tabular}
%\end{center}
\caption{Coefficients $A^{BF_d}_{\pm}$ as in Eq.~\eqref{eq:BtoF}.}
\label{tab:ABFpm}
\end{table}
%\end{verbatim}
%
%
\subsection{Leptonic meson decays}
\label{appendix:dilepton_decays}

The effective Hamiltonian describing leptonic meson decays $M \to \mu^+ \mu^-$, where $M=B,K_L,K_S$, is:
\begin{equation}\label{eq:C9C10WETLEFT}
\mathcal{H}_{eff}=-\frac{4G_F}{\sqrt{2}}\frac{\alpha}{4\pi}V_{ti}^{*}V_{tj}\sum_{k=10,10'}C_k^{ij\alpha\beta} \mathcal{O}_k^{ij\alpha\beta}.
\end{equation}
where
\begin{equation}\label{eq:C9C10}
O_{10}^{ij\alpha\beta}=(\bar{d}^i\gamma_{\mu}P_{L}d^j)(\bar{\ell}^\alpha \gamma^{\mu}\gamma_5\ell^\beta)\,,\quad O_{10'}^{ij\alpha\beta}=(\bar{d}^i\gamma_{\mu}P_{R}d^j)(\bar{\ell}^\alpha\gamma^{\mu}\gamma_5\ell^\beta)\,.
\end{equation}
The relation between these coefficients and the LEFT ones is:
\begin{equation}
    C_{10}^{ij\alpha\beta} = \frac{\sqrt{2}\pi}{G_F\alpha V_{ti}^*V_{tj}}\left( 
    \lwc{de}{LR}[V][ij\alpha\beta] - \lwc{ed}{LL}[V][\alpha\beta ij]
    \right)
    \,\quad
    C_{10'}^{ij\alpha\beta} = \frac{\sqrt{2}\pi}{G_F\alpha V_{ti}^*V_{tj}}\left( 
    \lwc{ed}{RR}[V][ij\alpha\beta] - \lwc{ed}{LR}[V][\alpha\beta ij]
    \right)
\end{equation}
\subsubsection{$B_s \to \mu^+\mu^-$}
The branching ratios for leptonic decays of $B_s$ are discussed, for example, in Refs.~\cite{Buras:1998raa,Becirevic:2016zri}. In case muon pair in the final state: 
\begin{equation}
    \mathcal{B}(B_s\to \mu^+ \mu^-)_{\mathrm{th}}=\frac{\alpha^2 G_F^2  \tau_{B_s}f_{B_s}^2 m_{B_s}m_{\mu}^2}{16 \pi^3} 
    \sqrt{1-\frac{4 m_{\mu}^2}{m_{B_s}^2}}|V_{tb}^*V_{ts}||C_{10,\textrm{SM}}^{sb\mu\mu}+C_{10}^{sb\mu\mu}-C_{10}^{\prime\,sb\mu\mu}|^2.
\label{eq:BrBll}
\end{equation}
When comparing the theoretical prediction of $B_s$ decays to untagged experimental data, the sizeable decay width differences in the $B_{s}^0-\bar{B}_{s}^0$ system must be taken into account. This is done by using an effective lifetime. To a good approximation one has \cite{DeBruyn:2012wj,DeBruyn:2012wk} 
\be
    \mathcal{B}(B_s\to \mu^+ \mu^-)_{\rm eff}\simeq\frac{1}{1-y_s}\mathcal{B}(B_s\to \mu^- \mu^+)_{\rm th} \, ,
\ee
with $y_s=\Delta\Gamma_{B_s}/(2\Gamma_{B_s})=0.064(4)$, according to the current PDG and HFLAV average \cite{Workman:2022ynf}. The experimental measurement of this branching ratio is \cite{Workman:2022ynf}:
\be
\mathcal{B}(B_s\to \mu^- \mu^+)=3.01 \pm 0.35 \cdot 10^{-9}\,.
\ee
\subsubsection{$K_{L,S}\to \mu^+\mu^-$}
Neglecting indirect CP-violation and assuming $f_{K^0}=f_{K^+}$, the branching ratios for kaon decays read:
\begin{equation}
\begin{split}
& \mathcal{B}(K_L \to \mu^+ \mu^-)_{\mathrm{SD}}=\frac{\alpha^2 G_F^2  \tau_{K}f_{K}^2 m_{K}m_{\mu}^2}{8 \pi^3} 
    \sqrt{1-\frac{4 m_{\mu}^2}{m_{K}^2}}|V_{ts}^*V_{td}|\left(\mathrm{Re}[C_{10,\textrm{SM}}^{sd\mu\mu}+C_{10}^{sd\mu\mu}-C_{10}^{\prime\,sd\mu\mu}]\right)^2\,,\\
& \mathcal{B}(K_S \to \mu^+ \mu^-)_{\mathrm{SD}}=\frac{\alpha^2 G_F^2  \tau_{K}f_{K}^2 m_{K}m_{\mu}^2}{8 \pi^3} 
    \sqrt{1-\frac{4 m_{\mu}^2}{m_{K}^2}}|V_{ts}^*V_{td}|\left(\mathrm{Im}[C_{10,\textrm{SM}}^{sd\mu\mu}+C_{10}^{sd\mu\mu}-C_{10}^{\prime\,sd\mu\mu}]\right)^2\,.
\end{split}
\end{equation}
The standard model contribution to SD physics is discussed in \cite{DAmbrosio:2022kvb}. For $K_S$, whose SM contribution is dominated by long distance effects, we use the estimate of Ref.~\cite{Isidori:2003ts}: $\mathcal{B}(K_S \to \mu^+ \mu^-)_{\mathrm{LD}}\approx 4.99 \cdot 10^{-12}$, which is orders of magnitude below present sensitivity.
The bounds on these observables are:
\be
\begin{split}
\mathcal{B}(K_L \to \mu^+ \mu^-)_{\mathrm{SD}} & < 2.5\cdot10^{-9} \ \text{\cite{Isidori:2003ts}} \, , \\
\mathcal{B}(K_S \to \mu^+ \mu^-) & < 2.1\cdot10^{-10} \,(90\% {\rm CL}) \ \text{\cite{LHCb:2020ycd}}
\end{split}
\ee

\subsection{$R(D^{(*)})$}
\label{app:RDst}
We consider the Standard Model predictions and the experimental results reported in \cite{hflav}, namely
\be
\begin{split}
&R_{D}^{\textrm{SM}}=0.298\pm0.004\,, \quad R_{D^{*}}^{\textrm{SM}}=0.254\pm0.005 \ , \\
&R_{D}=0.356\pm0.029\, , \quad R_{D^{*}}=0.284\pm0.013~, \quad \rho = -0.37~,
\end{split}
\ee
where $\rho$ is the rescaled correlation. Within our simplified models, the only 4-quarks operators contributing to these charged current processes are the current-current operators made of left-handed fermions. We define the low energy effective Hamiltonian:
\be
\mathcal{H}_{\textrm{eff}}=\frac{4 G_F}{\sqrt{2}}V_{cb}(1+C_{V_L})\left[(\bar{c}\gamma^{\mu}P_L b)(\bar{\tau}\gamma_{\mu}P_L\nu_{\tau})\right]\,,
\ee
so that we simply parametrize the EFT prediction normalized to the SM one as
\be
\frac{R_{D^{(*)}}}{R_{D^{(*)}}^{\textrm{SM}}}=\abs{1+C_{V_L}}^2\,.
\ee

\subsection{Electroweak precision data}
\label{app:EWPT}

In order to reduce the correlation between directions mainly constrained by EW precision data or Higgs data, the fit of Ref.~\cite{Falkowski:2019hvp} employed the so called \emph{Higgs basis}, i.e. specific combinations of the usual Warsaw basis coefficient that we defined in the main text. We use an updated version of the results of that analysis, kindly provided by the authors.
Since the WCs relevant to our analysis get mostly constrained by the vertex corrections of the $Z$ boson coupling to leptons, we report here the expression of the $\delta g_{Z \psi}$ parameters as a function of the Warsaw basis coefficients and refer the reader to the original paper for more details.
\be\begin{split}
    \label{eq:HiggsBasis}
    \delta g^{W \ell}_L & =   C^{(3)}_{H l} + f(1/2,0) - f(-1/2,-1), \\
    \delta g^{Z\ell}_L & =    - \frac{1}{2} C^{(3)}_{H l} - \frac{1}{2} C_{H l}^{(1)}+   f(-1/2, -1) , \\ 
    \delta g^{Z\ell }_R & =  - {1\over 2} C_{H e}^{(1)}   +  f(0, -1) ,\\ 
\end{split}\ee
where
\be
    f(T^3,Q)  \equiv  \bigg \{  
-   Q  {g_L  g_Y \over  g_L^2 -  g_Y^2} C_{H WB} 
-  {\bf 1}  \left ( {1 \over 4} C_{H D}  
+  {1 \over 2 } \Delta_{G_F}  \right )  \left ( T^3 + Q { g_Y^2 \over  g_L^2 -   g_Y^2} \right )  \bigg \} {\bf 1} ~,
\ee
and $\Delta_{G_F} = [C_{Hl}^{(3)}]_{11}+[C_{Hl}^{(3)}]_{22}-\frac{1}{2} [C_{ll}]_{1221}$.
This set of parameters is constrained mainly by electroweak data from LEP.

\subsection{Four-quark operators and $\Delta F=2$ observables}
\label{appendix:four_quark_ops}

In this Appendix we recollect the list of four-quark operators generated at tree or loop level by the mediators considered in Sect.~\ref{sec:models}.
The relevant operators are
\begin{gather}
\mathcal{O}_{qq}^{(1)ijkl}=\left(\bar{q}^i_{L}\gamma_\mu q^j_{L}\right)\left(\bar{q}^k_{L}\gamma^\mu q^l_{L}\right)
\ , \nonumber\\
\label{eq:four_quark_ops}
\mathcal{O}_{qq}^{(3)ijkl}=\left(\bar{q}^i_{L}\gamma_\mu\sigma_a q^j_{L}\right)\left(\bar{q}^k_{L}\gamma^\mu\sigma_a q^l_{L}\right)
\ , \nonumber \\
\mathcal{O}_{dd}^{ijkl}=\left(\bar{d}^i_{R}\gamma_\mu d^j_{R}\right)\left(\bar{d}^k_{R}\gamma^\mu d^l_{R}\right) \ , \\
\mathcal{O}_{qd}^{(1) ijkl}=\left(\bar{q}^i_{L}\gamma_\mu q^j_{L}\right)\left(\bar{d}^k_{R}\gamma^\mu d^l_{R}\right)
\ , \nonumber
\end{gather}
and are highly constrained by $\Delta F=2$ observables like meson-antimeson mixing.
These operators are generated at tree level by the exchange of colourless vector mediator like $V^\prime$ or $Z^\prime$. The matching of the non-vanishing corresponding SMEFT coefficients reads
\be
C_{qq}^{(3)ijkl}=-\frac{1}{8}\frac{g_q^{ij}g_q^{kl}}{M_{V^\prime}^2}
    \label{eq:DeltaF2Vprime}
\ee
for the Lagrangian \eqref{eq:lag:Vp} and
\be
C_{qq}^{(1)ijkl}=-\frac{1}{2}\frac{g_L^{ij}g_L^{kl}}{M_{Z^\prime}^2}, \quad C_{qd}^{(1) ijkl}=-\frac{g_L^{ij}g_R^{kl}}{M_{Z^\prime}^2}, \quad C_{dd}^{ijkl}=-\frac{1}{2}\frac{g_R^{ij}g_R^{kl}}{M_{Z^\prime}^2} \ ,
\label{eq:DeltaF2Zprime}
\ee
for the Lagrangian \eqref{eq:lag:Zp}.
Leptoquarks instead generate four-quark operators only at loop level, arising from the same interaction terms shown in Table~\ref{tab:LQ_mediators}.
One can easily compute the non vanishing SMEFT coefficient of the loop-induced four-quark operators for scalar leptoquarks, getting by direct calculation \cite{Dorsner:2016wpm,Gherardi:2020det,Bobeth:2017ecx}
\begin{gather}
C_{qq}^{(1)ijkl}=C_{qq}^{(3)ijkl}=-\frac{\lambda_{i\tau}\lambda_{j\tau}^*\lambda_{k\tau}\lambda_{l\tau}^*}{256\pi^2m_{LQ}^2} \quad \text{for $S_1$}
\ , \\
C_{qq}^{(1)ijkl}=9C_{qq}^{(3)ijkl}=-\frac{9\lambda_{i\tau}\lambda_{j\tau}^*\lambda_{k\tau}\lambda_{l\tau}^*}{256\pi^2m_{LQ}^2} \quad \text{for $S_3$}
\ , \\
C_{dd}^{ijkl}=-\frac{\lambda_{i\tau}\lambda_{j\tau}^*\lambda_{k\tau}\lambda_{l\tau}^*}{64\pi^2m_{LQ}^2} \quad \text{for $\tilde{R}_2$}
\ , \nonumber
\end{gather}
while the loop diagram with vector LQ is UV divergent and requires a complete UV model to be calculable.
However, in order to evaluate the impact of the four-quark operators on the phenomenological analysis we aim to perform, we estimate the loop-induced SMEFT coefficients by considering just the $g_{\mu\nu}$-term of the vector LQ propagator inside the loop, which leads to a finite, although incomplete, result (the same procedure has been followed for example in Ref.~\cite{Davidson:1993qk}). The matching of the non-vanishing corresponding SMEFT coefficients thus evaluated then reads
\begin{gather}
C_{qq}^{(1)ijkl}=9C_{qq}^{(3)ijkl}=-\frac{9\lambda_{i\tau}\lambda_{j\tau}^*\lambda_{k\tau}\lambda_{l\tau}^*}{64\pi^2m_{LQ}^2} \quad \text{for $U_3$}
\ , \nonumber \\
C_{dd}^{ijkl}=-\frac{\lambda_{i\tau}\lambda_{j\tau}^*\lambda_{k\tau}\lambda_{l\tau}^*}{16\pi^2m_{LQ}^2} \quad \text{for $V_2$} \ . 
\end{gather}

We employ the expressions of NP contributions to meson mixing observables from Ref.~\cite{Aebischer:2020dsw}. Notice that they are provided in terms of WCs evolved up to a scale $\mu=5$~TeV, while we set our NP scale at 1 TeV. We neglect this difference, hiding the small effect in the uncertainty of working with a leading log resummation in the RG evolution. For the numerical constraints we take the results of the fit, updated in summer 2023, by the UTfit collaboration \cite{UTfit,UTfit:2022hsi}.

\begin{figure}[t]
	\centering
	\includegraphics[scale=0.4]{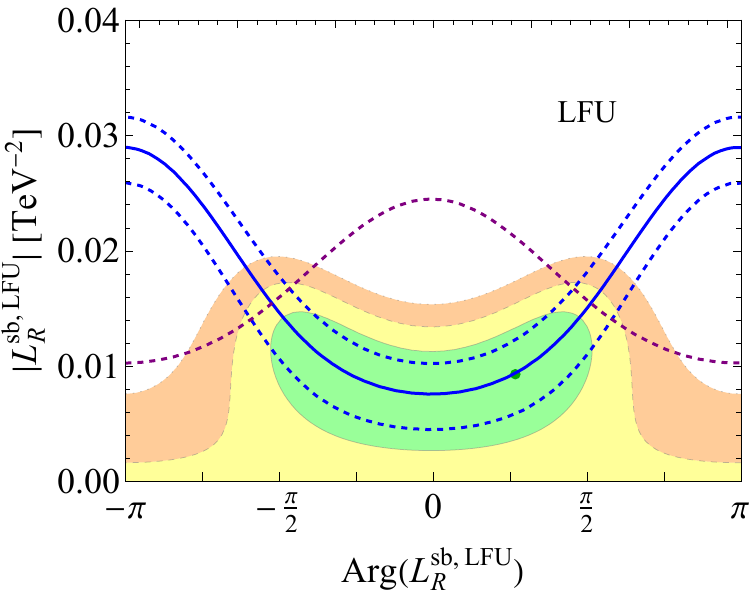}
	\quad
	\includegraphics[scale=0.4]{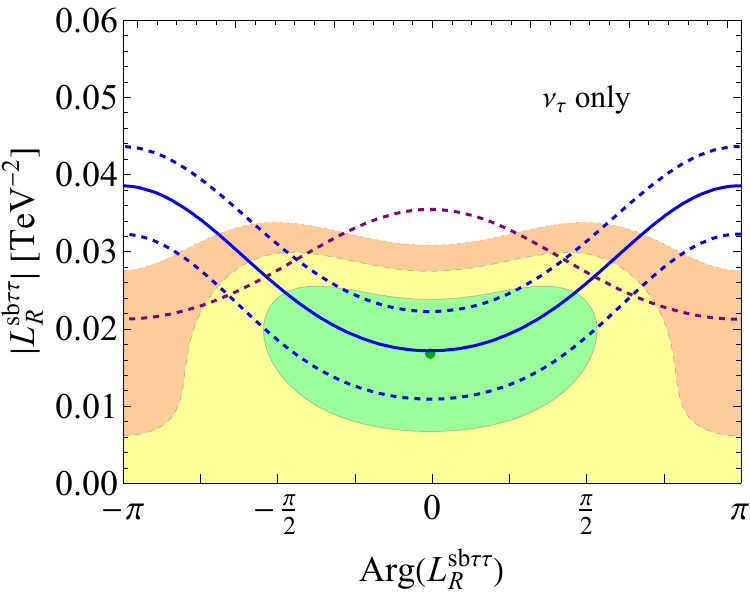}\\
    \vspace{2mm}
	\includegraphics[scale=0.4]{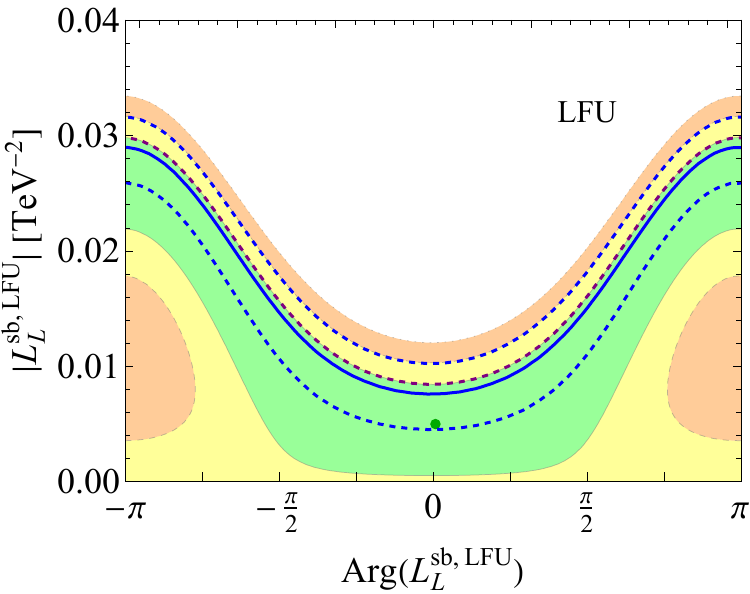}
	\quad
	\includegraphics[scale=0.4]{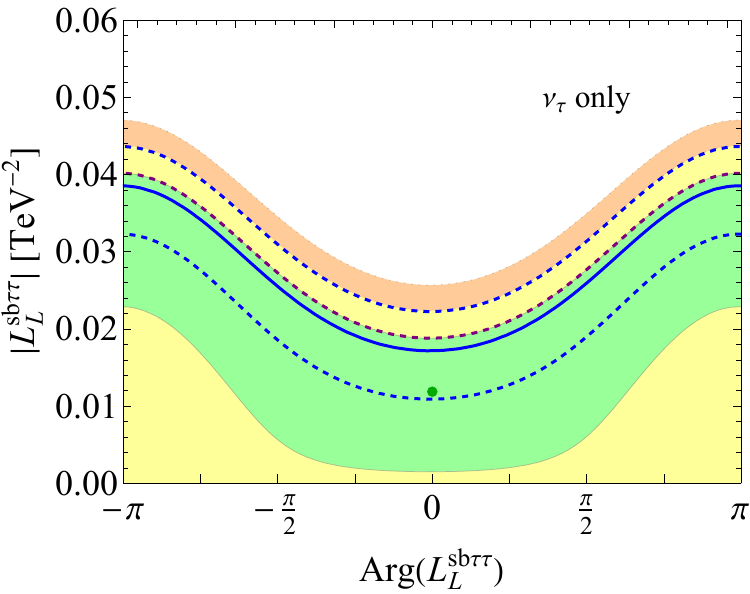}\\
    \vspace{2mm}
	\includegraphics[scale=0.4]{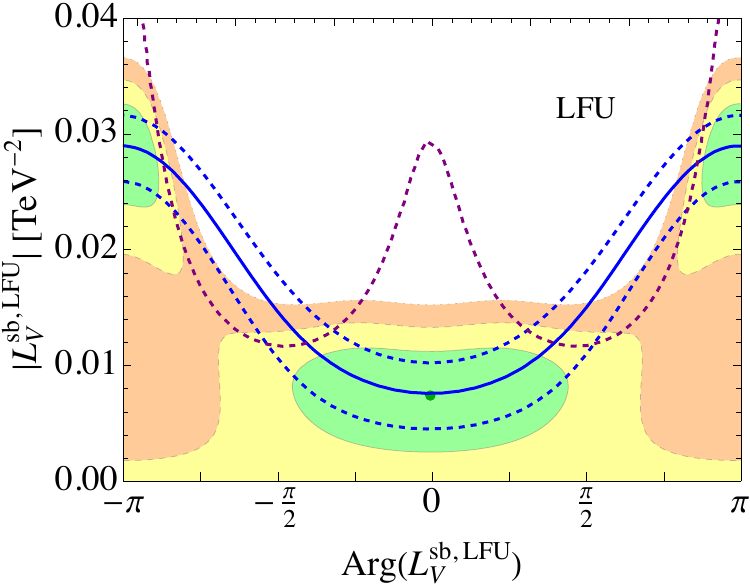}
	\quad
	\includegraphics[scale=0.4]{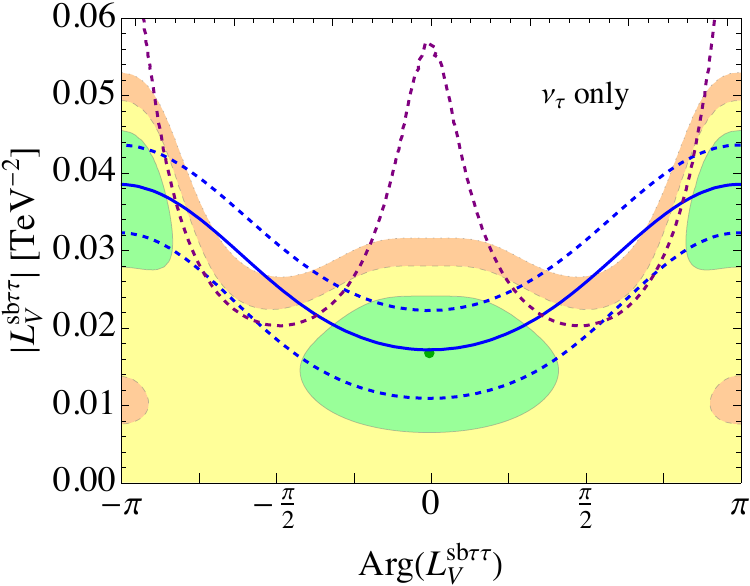}\\
    \vspace{2mm}
    \includegraphics[scale=0.6]{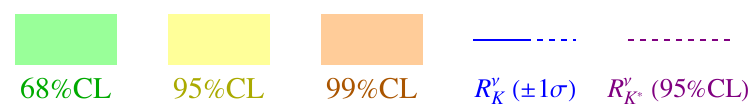}
	\caption{2D fit in the $\text{Arg}(L^{sb\alpha\beta})-|L^{sb\alpha\beta}|$ for NP coupled to right-handed, left-handed, vector quark current while assuming LFU (on the left) or coupling only to tau neutrinos (on the right). The solid (dashed) blue line refers to the central value ($\pm1\sigma$ value) of $R_{K^+}^\nu$ observable. The dashed purple line refers to the upper limit at 95\% of $R_{K^*}^\nu$ observable.}
	\label{fig:LEFT_sb}
\end{figure}

%%%%%%%%%%%%%%%%%%%%%%%%%%%%%%%%%%%%%%%%%%%%%%%%%%%%%%%%%%%%%%%
\section{Fits of $b\to s\nu\bar\nu$}
\label{app:Fitsbsnunu}

In this Appendix we discuss the 2D fits of the EFT coefficients related to the $b\to s\nu\bar\nu$ transition, independently on the ROFV assumption. 
While, at the level of LEFT the EFT coefficient responsible for $b\to s\nu\bar\nu$ transitions does not impact other classes of observables, in case of the SMEFT it does. We discuss each case individually in the following.

\subsection{LEFT}
\label{appendix:fit_left_sb}

At the level of the LEFT, non-trivial correlations among the coefficients of different operators are absent, so we have to consider only $B\to K^{(*)}\nu\bar\nu$ observables.
We study scenarios of NP coupled to right, left or vector quark current while assuming either LFU or only tau flavour on the lepton side.
We plot in Fig.~\ref{fig:LEFT_sb} the regions at $1\sigma$, $2\sigma$ and 3$\sigma$ in the plane $\text{Arg}(L^{sb\alpha\beta})-|L^{sb\alpha\beta}|$. The coefficient explicitly reads
\be
L_{L,R,V}^{sb\alpha\beta}=C_{L,R,V} \ e^{i\alpha_{sb}} \ \cos\theta\sin\theta\sin\phi \times\begin{cases}
\delta^{\alpha\beta} & \text{for LFU ,}\\
\delta^{\tau\alpha}\delta^{\tau\beta} & \text{for only tau flavour ,}
\end{cases}
\ee
which means that, for any given $\theta$ and $\phi$, the overall coefficient $C$ and the phase $\alpha_{sb}$ are univocally determined.
We note that in all the cases we consider the fits favour a real and positive value of the LEFT coefficient. We thus evaluate the best-fit value at vanishing complex phase and report the obtained values in the main text.

\subsection{SMEFT and simplified models}
\label{appendix:fit_smeft_sb}

In this Appendix we discuss the 2D fits of the coefficient related to the $b\to s\nu\bar\nu$ transition in the SMEFT.
Among the observables of the fit, additionally to the LEFT case, we must include also
the $B_s \to \mu^+\mu^-$ decay, the $R_{K^{(*)}}$ ratios and,
where explicitly specified,
the $\Delta M_{B_s}$ meson mass mixing, whose information would otherwise be lost as they are mainly affected, trough the correlations coming from the SMEFT or the underlying UV structure, by the same SMEFT coefficients related to $R^{\nu}_{K^{(*)}}$.

\subsection*{Vector-like quarks}

\begin{figure}[t]
	\centering
	\includegraphics[scale=0.45]{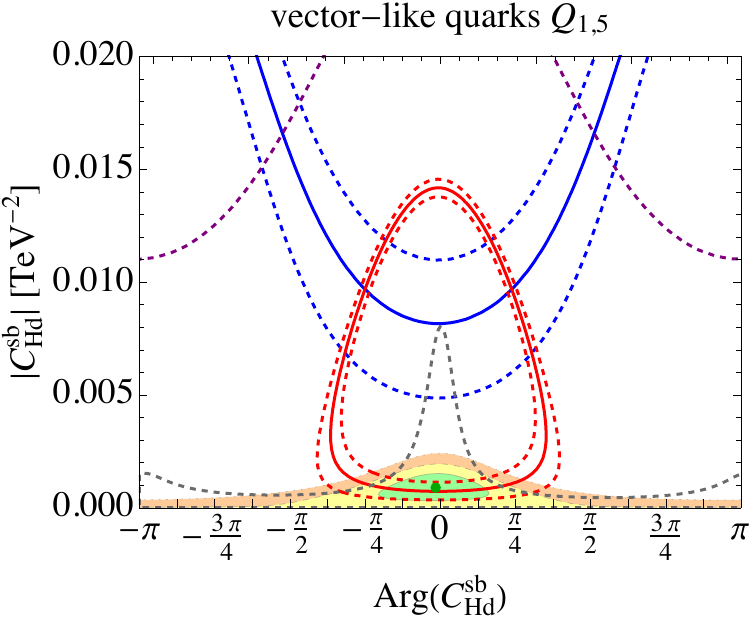}
	\quad
	\includegraphics[scale=0.45]{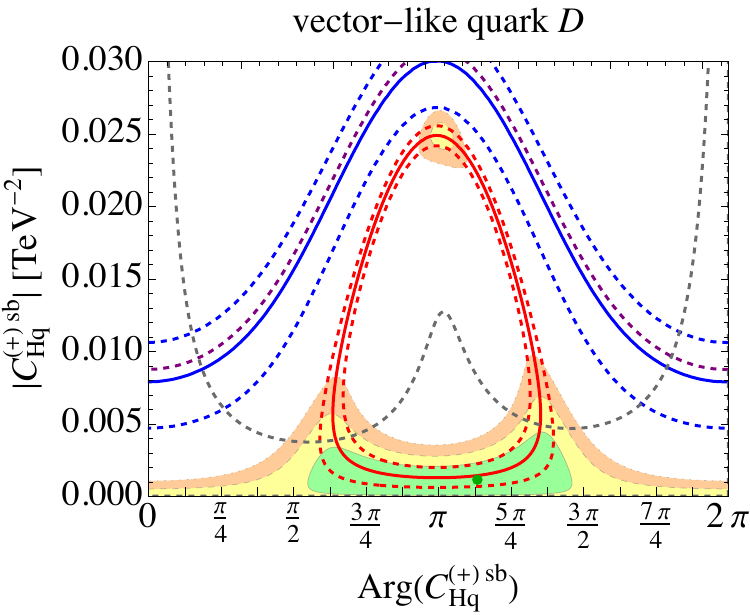}\\
    \vspace{2mm}
	\includegraphics[scale=0.45]{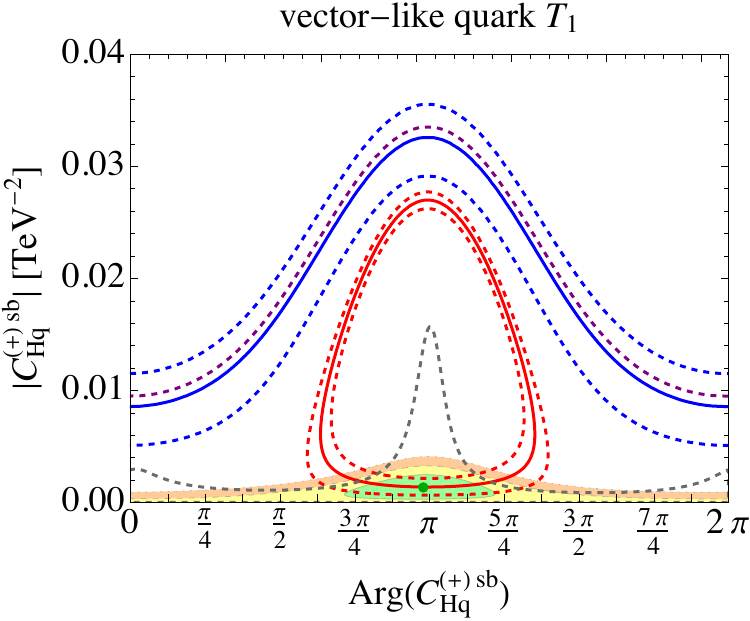}
	\quad
	\includegraphics[scale=0.45]{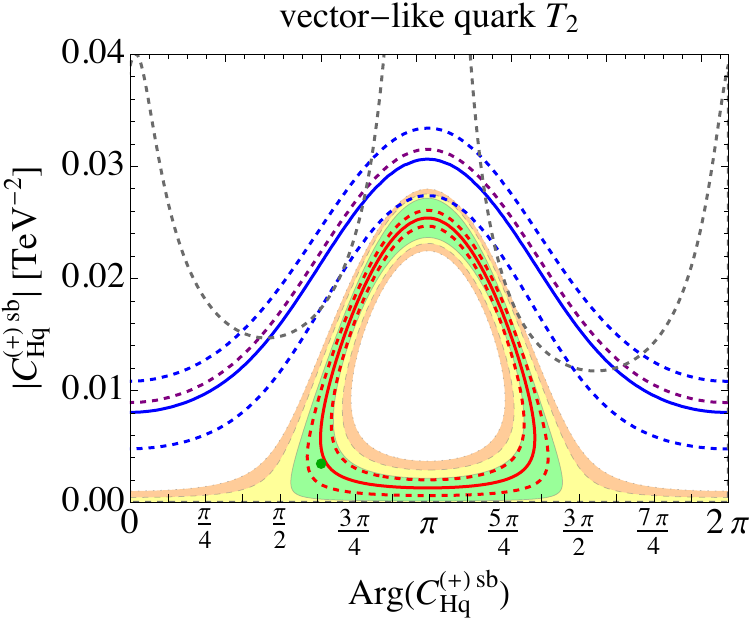}\\
    \includegraphics[scale=0.54]{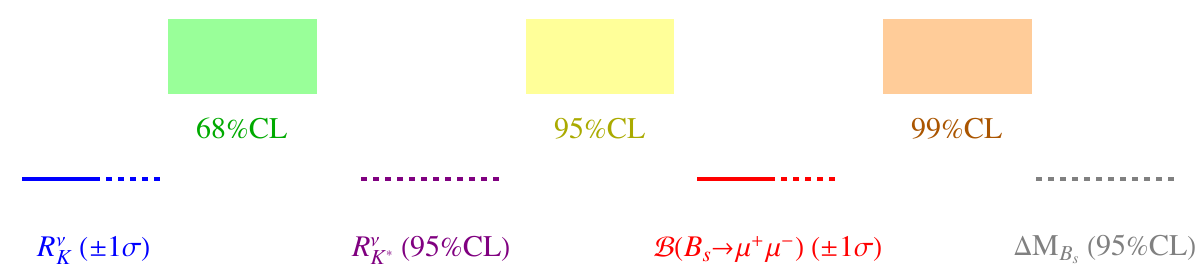}
	\caption{2D fit in the $|C_{Hq}^{(+)sb}(C_{Hd}^{sb})|-\text{Arg}(C_{Hq}^{(+)sb}(C_{Hd}^{sb}))$ plane for several vector-like quarks models. The solid (dashed) blue lines refer to the central value ($\pm1\sigma$ contours) of the $R_{K^+}^\nu$ observable. The solid (dashed) red lines refer to the central value ($\pm1\sigma$ contours) of the $B_s\to\mu^+\mu^-$ observable. The region above the dashed gray lines is excluded at 95\% from the $\Delta M_{B_s}$ observable. The dashed purple line refers to the upper limit at 95\% of $R_{K^*}^\nu$ observable.}
    \vspace{-4mm}
	\label{fig:SMEFT_higgsquark_sb}
\end{figure}

We list in Table~\ref{tab:higgsquark_mediators} some examples of heavy BSM states, such as vector-like quarks, that can generate Higgs-quark coefficients in the EFT when integrated out.
The 2D fits in the $\text{Arg}(C_{Hq}^{(+)sb}(C_{Hd}^{sb}))-|C_{Hq}^{(+)sb}(C_{Hd}^{sb})|$ plane are illustrated in Fig.~\ref{fig:SMEFT_higgsquark_sb}, displaying also the bounds from $B_s\to\mu^+\mu^-$ and $\Delta M_{B_s}$.
As can be seen, none of these models allows to fit the measured value of $R_\nu^K$.

%\begin{verbatim}
\begin{table}[t!]
%\begin{center}
\begin{tabular}{cc|c|cc|ccc}
\multicolumn{2}{c|}{{\bf Simplified model}} & {\bf Spin} & \multicolumn{2}{c|}{{\bf SM irrep}} & \multicolumn{3}{c}{{\bf SMEFT couplings}} \\
\midrule
\multicolumn{2}{c|}{$D$} & 1/2 & \multicolumn{2}{c|}{$({\bf 3},{\bf 1},-1/3)$} & \multicolumn{3}{c}{$C_{Hq}^{(1)}=C_{Hq}^{(3)}$} \\
\midrule
\multicolumn{2}{c|}{$T_1$} & 1/2 & \multicolumn{2}{c|}{$({\bf 3},{\bf 3},-1/3)$} & \multicolumn{3}{c}{$C_{Hq}^{(1)}=-3C_{Hq}^{(3)}$} \\
\midrule
\multicolumn{2}{c|}{$T_2$} & 1/2 & \multicolumn{2}{c|}{$({\bf 3},{\bf 3},2/6)$} & \multicolumn{3}{c}{$C_{Hq}^{(1)}=3C_{Hq}^{(3)}$} \\
\midrule
\multicolumn{2}{c|}{$Q_1$} & 1/2 & \multicolumn{2}{c|}{$({\bf 3},{\bf 2},1/6)$} & \multicolumn{3}{c}{$C_{Hd}$} \\
\midrule
\multicolumn{2}{c|}{$Q_5$} & 1/2 & \multicolumn{2}{c|}{$({\bf 3},{\bf 2},-5/6)$} & \multicolumn{3}{c}{$C_{Hd}$} \\
\hline
\end{tabular}
%\end{center}
\caption{List of vector-like quark mediators and the corresponding Higgs-quark couplings generated when when integrating them out at the tree level.}
\label{tab:higgsquark_mediators}
\end{table}
%\end{verbatim}

\subsection*{Vector triplet $V^\prime$ and vector singlet $Z^\prime$}

\begin{figure}
	\centering
    \includegraphics[scale=0.45]{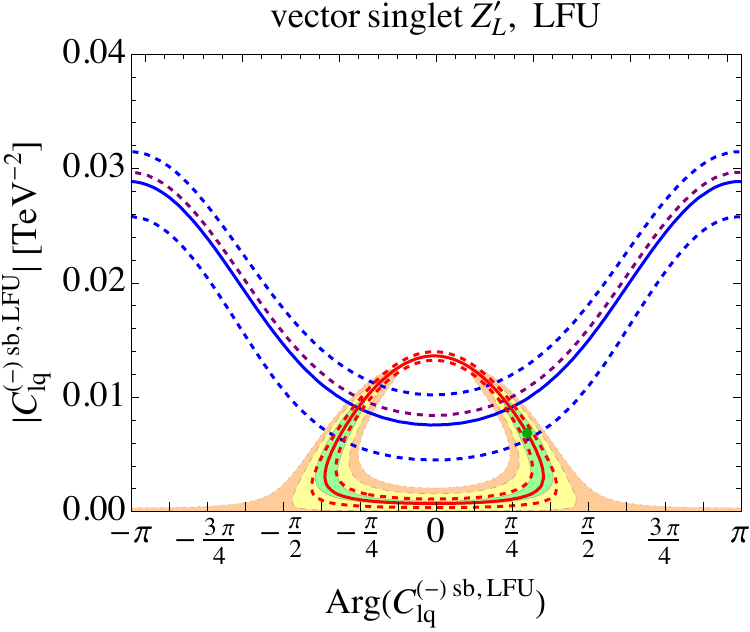}
	\quad
	\includegraphics[scale=0.45]{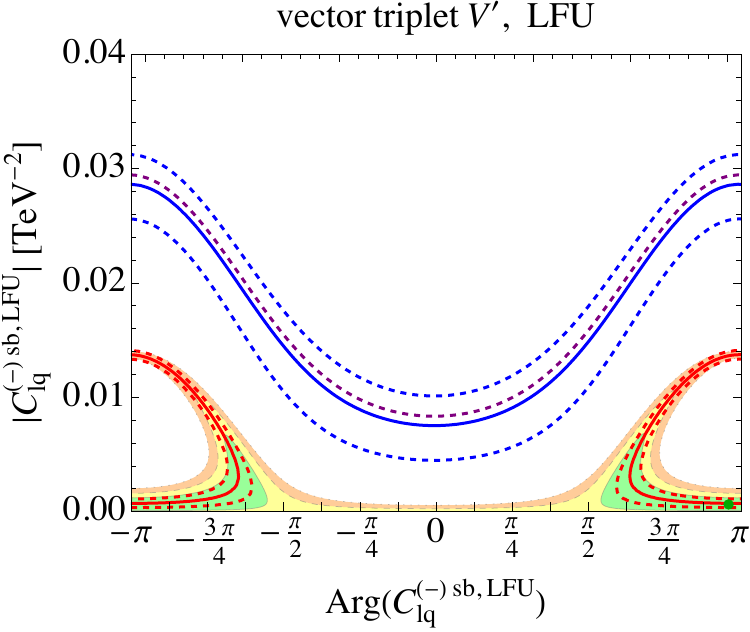}\\
    \vspace{4mm}
    \includegraphics[scale=0.54]{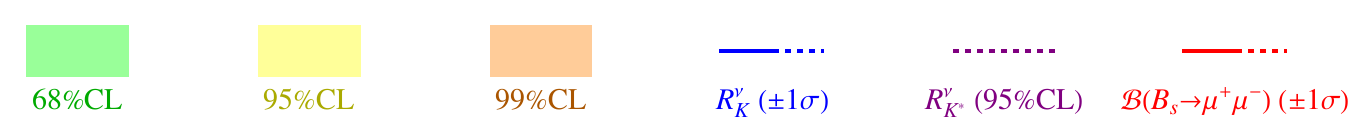}
	\caption{2D fit in the $\text{Arg}(C_{lq}^{(-)\alpha\beta sb})-|C_{lq}^{(-)\alpha\beta sb}|$ for $Z_{L}^\prime$ (on the left) and for $V^\prime$ (on the right) assuming LFU. The solid (dashed) blue line refers to the central value ($\pm1\sigma$ value) of $R_{K^+}^\nu$ observable. The solid (dashed) red lines refer to the central value ($\pm1\sigma$ contours) of the $B_s\to\mu^+\mu^-$ observable. The dashed purple line refers to the upper limit at 95\% of $R_{K^*}^\nu$ observable.}
	\label{fig:SMEFT_sb_Zp}
\end{figure}

We discussed in the main text the case of a vector triplet $V^\prime$ and the vector singlets $Z_{L,R}^{\prime}$ coupled to the third family of leptons, see Sec.~\ref{sec:models}.
Here, instead, we illustrate in Fig.~\ref{fig:SMEFT_sb_Zp} the 2D fits for a $V^\prime$ and a $Z_{L}^{\prime}$ that are universally coupled to all the lepton families. We do not show the $Z_{R}^{\prime}$, as this case is similar to $V^{\prime}$. Indeed, they generate a contribution of the same sign to the Wilson Coefficients in Eq.~\ref{eq:BrBll}, getting a strong contraint from the  $B_s \to\mu^+\mu^-$ decay. As a result, we find these UV states to be unable to address the Belle~II excess.
The LFU $Z_L^\prime$ could fit reasonably well the $R^\nu_K$ excess, with a large phase in the EFT coefficient. This phase implies constraints from $B_s$-mixing even stronger than those discussed in the main text for $V^\prime$ or $Z^\prime$s only coupled to tau doublet, therefore we do not consider it further.

\subsection*{Leptoquarks}

\begin{figure}
	\centering
	\includegraphics[scale=0.45]{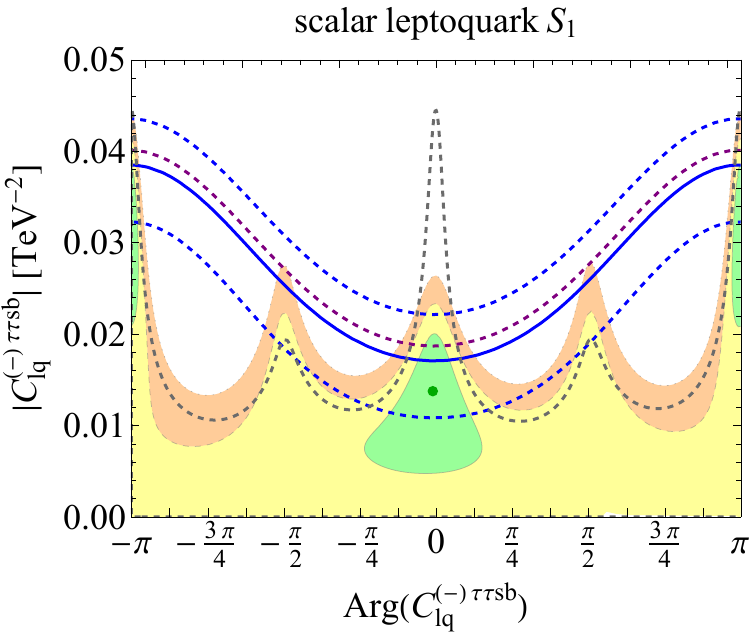}
	\quad
	\includegraphics[scale=0.45]{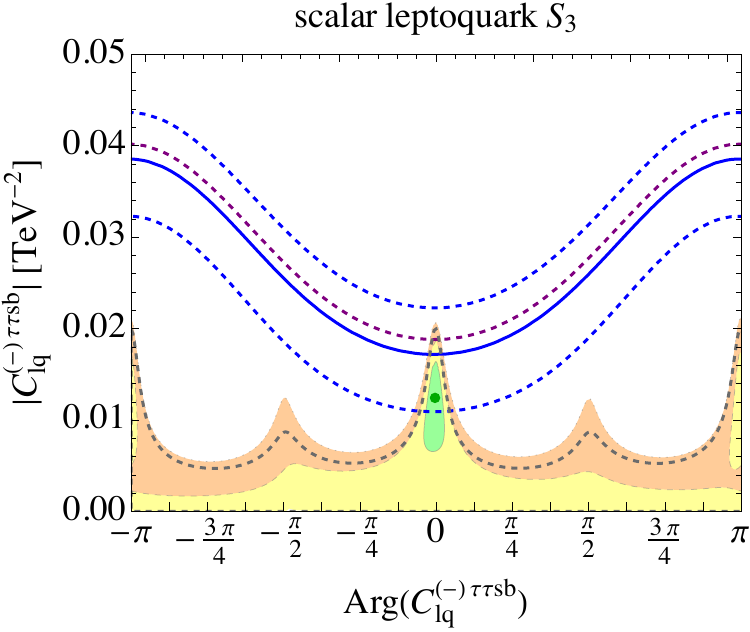}\\
    \vspace{4mm}
	\includegraphics[scale=0.45]{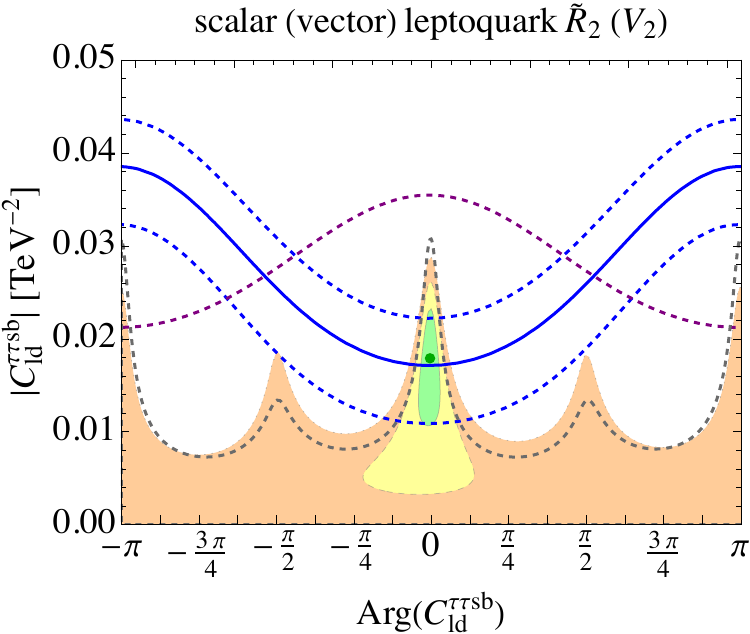}
	\quad
	\includegraphics[scale=0.45]{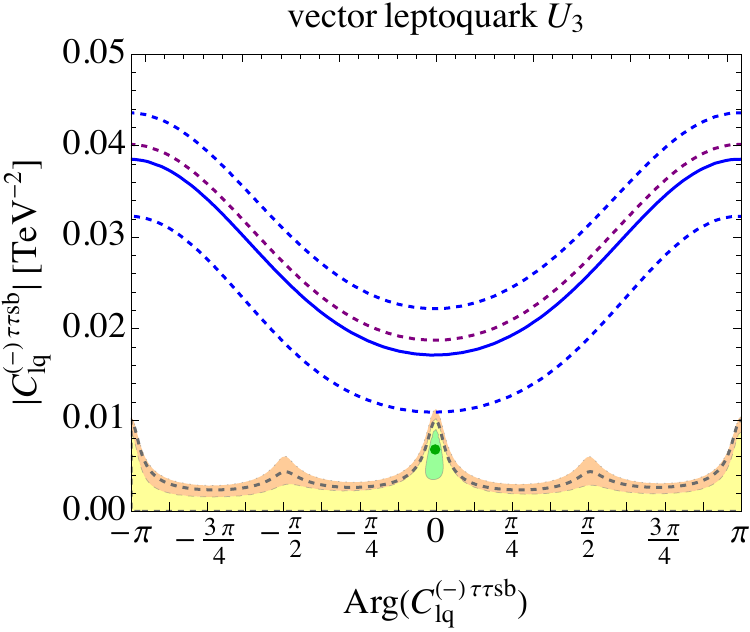}\\
    \vspace{4mm}
    \includegraphics[scale=0.54]{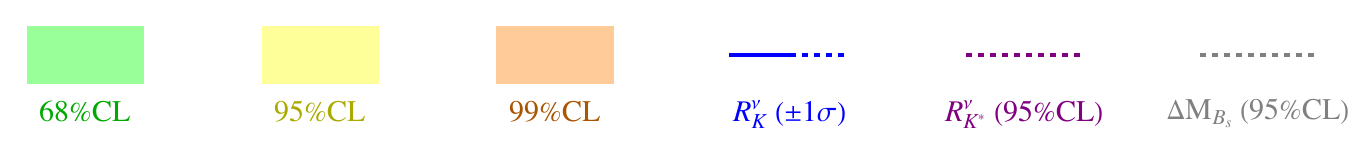}
	\caption{2D fit in the $\text{Arg}(C_{lq}^{(-)\tau \tau sb}(C_{ld}^{\tau\tau sb}))-|C_{lq}^{(-)\tau\tau sb}(C_{ld}^{\tau \tau sb})|$ plane assuming coupling only to tau sector. UV mediators that can give rise to these effective operators are outlined. The solid (dashed) blue lines refer to the central value ($\pm1\sigma$ contours) of the $R_{K^+}^\nu$ observable. The dashed purple line refers to the upper limit at 95\% of $R_{K^*}^\nu$ observable. The region above the dashed gray line is excluded at 95\% by $\Delta M_{B_s}$ observable.}
	\label{fig:SMEFT_sb_LQ}
\end{figure}

Regard the LQs, since in our analysis we choose to fix the mass (at 2 TeV), the SMEFT coefficient of the four-quark operator $\mathcal{O}_{qq}^{(1,3)\tau\tau sb}$ ($\mathcal{O}_{dd}^{\tau\tau sb}$), see App.~\ref{appendix:four_quark_ops}, turns out to be exactly proportional to the square of the semileptonic ones $\mathcal{O}_{lq}^{(1,3)\tau\tau sb}$ ($\mathcal{O}_{ld}^{\tau\tau sb}$), e.g.:
\be
    C_{lq}^{(1,3)\tau\tau sb} \propto \lambda_{b\tau}\lambda_{s\tau}^*~, \qquad
    C_{qq}^{(1,3)\tau\tau sb} \propto (\lambda_{b\tau}\lambda_{s\tau}^*)^2 \propto \left( C_{lq}^{(1,3)\tau\tau sb} \right)^2 ~.
\ee
Fixing the semileptonic operator to fit $R^\nu_K$ would therefore fix also the coefficient responsible for $B_s$ mixing.
Hence, we have to include $\Delta M_{B_s}$ among the observables of the 2D fits.
We plot in Fig.~\ref{fig:SMEFT_sb_LQ} the resulting $1\sigma$, $2\sigma$ and $3\sigma$ regions of the 2D fits for the LQs simplified models. Note that $\tilde{R}_2$ and $V_2$ generate the same semileptonic coefficient $C_{ld}^{\tau\tau sb}$ and thus same four-quark coefficient $C_{dd}^{\tau\tau sb}$ at fixed value of the mass, so they produce the same 2D fit.
We conclude that the best scenarios appear to be $\tilde{R}_{2}$ and $V_{2}$, slightly favoured over $S_1$ and $S_3$, while $U_3$ has to be discarded.
This result is in agreement with the results of the LEFT analysis where we observe a preference over NP coupled to right-handed quarks.
Hence, assuming for simplicity the coefficient to be real, we report the best-fit value of $\tilde{R}_{2}$ and $V_{2}$, together with $S_1$ as an example of LQ coupled to left-handed quarks, in the main text.

%%%%%%%%%%%%%%%%%%%%%%%%%%%%%%%%%%%
\section{LNV operators} 
\label{app:LNV}

The role of $\Delta L=2$ operators in the context of rare meson decays has been studied in several works: implications for kaon decays have been considered in Refs.~\cite{Li:2019fhz,Deppisch:2020oyx}, while Ref.~\cite{Felkl:2021uxi} discussed their impact on $B$-meson (semi)invisible modes. Recently, Ref.~\cite{Fridell:2023rtr} provided a comprehensive study of dim-7 $\Delta L=2$ SMEFT operators, considering both low-energy observables and high-energy collider limits. Following the analysis of \cite{Fridell:2023rtr}, we update the limits from $B\to K^{(*)} \nu \bar{\nu}$ in light of the Belle~II measurement.

At the level of LEFT, the effective (dim-6) operators relevant for $d_i\to d_j\nu\nu$ transitions are:
\begin{equation}\label{eq:LNVcoeff}
\begin{split}
&\mathcal{O}_{d\nu}^{S,LL}=(\overline{d_R}d_L)(\overline{\nu^C}\nu)+\mathrm{h.c.}\,,\\
&\mathcal{O}_{d\nu}^{S,LR}=(\overline{d_L}d_R)(\overline{\nu^C}\nu)+\mathrm{h.c.}\,,\\
&\mathcal{O}_{d\nu}^{T,LL}=(\overline{d_R}\sigma^{\mu\nu} d_L)(\overline{\nu^C}\sigma_{\mu\nu}\nu)+\mathrm{h.c.}\,,
\end{split}
\end{equation}
employing the basis of Ref.~\cite{Jenkins:2017dyc}. The dim-5 dipole operator $\overline{\nu^C}\sigma_{\mu\nu}\nu F^{\mu\nu}$is not considered, as it can only appear with the down quark dipole operator $\overline{d_L}\sigma^{\mu\nu}d_R F_{\mu\nu}$, bringing a double loop suppression. In addition, there are stringent bounds from neutrino magnetic moment searches. We also neglect the operator $\mathcal{O}_{d\nu}^{S,LR}$, since it is not generated at tree level by any dim-7 $\Delta L=2$ SMEFT operator. The tree level matching expressions for the remaining two coefficients are:
\begin{equation}
\begin{split}
    &L_{d\nu,prst}^{S,LL}=-\frac{\sqrt{2}v}{8}V_{xr}\left(C_{\bar{d}LQLH1}^{ptxs}+C_{\bar{d}LQLH1}^{psxt}\right)\,, \\
    &L_{d\nu,prst}^{T,LL}=-\frac{\sqrt{2}v}{32}V_{xr}\left(C_{\bar{d}LQLH1}^{ptxs}-C_{\bar{d}LQLH1}^{psxt}\right)\,,
\end{split}
\end{equation}
where the dim-7 SMEFT operator appearing in these expressions is 
\begin{equation}
O_{\bar{d}LQLH1}^{prst}=\epsilon_{ji}\epsilon_{mn}\left(\overline{d_p}L_r^i \right)\left(\overline{Q_s^{c}}^jL_t^m \right)H^n\,.
\end{equation} 

\begin{figure}
    \centering
    \includegraphics[width=0.46\textwidth]{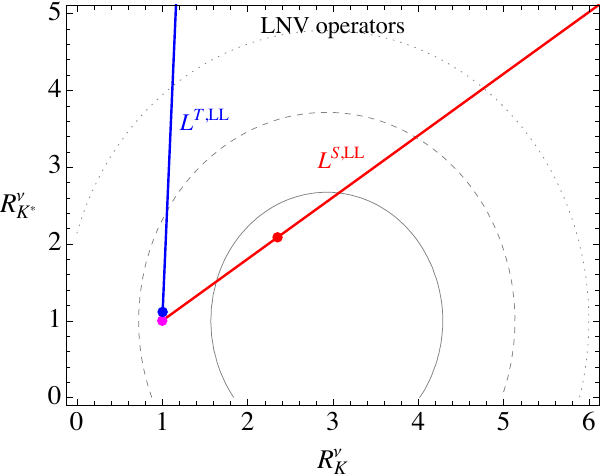}
    \caption{Fit of available data on all $B \to K^{(*)} \nu \bar\nu$ channels. Solid, dashed, and dotted gray lines represent 68, 95, and 99\% CL contours.}
    \label{fig:fit_BKnunu_data_LNV}
\end{figure}

The dependence of the $B \to K^{(*)} \nu \nu$ branching ratios on the LNV LEFT coefficients is:
\begin{equation}\begin{split}\label{eq:BKnunuLNV}
    \mathcal{B}(B^+ \to K^+ \nu_\alpha \nu_\beta) = \frac{10^{10} \, \text{GeV}^2}{1 + \delta_\alpha^\beta} &\left[ 4.01  \left(|L_{d\nu,s b \alpha \beta}^{S,LL} + L_{d\nu,s b \alpha \beta}^{S,LR}|^2 + \text{perm} \right) \right. \\
    & \left. 6.15 \left(|L_{d\nu,s b \alpha \beta }^{T,LL}|^2  + \text{perm} \right) \right]~, \\
    \mathcal{B}(B^0 \to K^{*0} \nu_\alpha \nu_\beta) = \frac{10^{10} \, \text{GeV}^2}{1 + \delta_\alpha^\beta} &\left[ 1.49  \left(|L_{d\nu,s b \alpha \beta}^{S,LL} - L_{d\nu,s b \alpha \beta}^{S,LR}|^2 + \text{perm} \right) \right. \\
    & \left. 72.3 \left(|L_{d\nu,s b \alpha \beta }^{T,LL}|^2  + \text{perm} \right) \right]~,
\end{split}\end{equation}
where the EFT coefficients are evaluated at the $m_b$ scale. The QCD RG evolution up to $m_Z$ is given by: $L_{d\nu}^{S,LL}(m_b) = 1.137 L_{d\nu}^{S,LL}(m_Z)$, $L_{d\nu}^{S,LR}(m_b) = 1.137 L_{d\nu}^{S,LR}(m_Z)$, and $L_{d\nu}^{T,LL}(m_b) = 0.900 L_{d\nu}^{T,LL}(m_Z)$.
The expressions in Eq.~\eqref{eq:BKnunuLNV} have been derived integrating the differential rate of Ref.~\cite{Fridell:2023rtr} in $q^2$. For the $K$ mode we employed the form factors fit from Ref.~\cite{Becirevic:2023aov}, while for the $K^*$ mode we used the results from Ref.~\cite{Bharucha:2015bzk}.

The dependence of the two modes on the LEFT coefficients is shown in Fig.~\ref{fig:fit_BKnunu_data_LNV}.
We see that the scalar operator allows a better fit than the tensor one, the preferred values being $L_{d\nu,sb\tau\tau}^{S,LL}(m_b) = (1.22 \pm 0.34) \times 10^{-8} \text{GeV}^{-2}$ (the best-fit point is shown as a red dot).
In terms of the dim-7 SMEFT operator, evaluated at the $m_Z$ scale, the best-fit point corresponds to a scale of $|C_{\bar{d}LQLH1}^{sb\tau\tau}|^{-1/3} \approx 1.6$~TeV.
The tensor operator (blue line and blue dot for the best-fit point) does not improve the fit over the Standard Model (magenta dot).




\end{appendices}


\bibliography{biblio}% common bib file


\begin{thebibliography}{73}
\ifx \bisbn   \undefined \def \bisbn  #1{ISBN #1}\fi
\ifx \binits  \undefined \def \binits#1{#1}\fi
\ifx \bauthor  \undefined \def \bauthor#1{#1}\fi
\ifx \batitle  \undefined \def \batitle#1{#1}\fi
\ifx \bjtitle  \undefined \def \bjtitle#1{#1}\fi
\ifx \bvolume  \undefined \def \bvolume#1{\textbf{#1}}\fi
\ifx \byear  \undefined \def \byear#1{#1}\fi
\ifx \bissue  \undefined \def \bissue#1{#1}\fi
\ifx \bfpage  \undefined \def \bfpage#1{#1}\fi
\ifx \blpage  \undefined \def \blpage #1{#1}\fi
\ifx \burl  \undefined \def \burl#1{\textsf{#1}}\fi
\ifx \doiurl  \undefined \def \doiurl#1{\url{https://doi.org/#1}}\fi
\ifx \betal  \undefined \def \betal{\textit{et al.}}\fi
\ifx \binstitute  \undefined \def \binstitute#1{#1}\fi
\ifx \binstitutionaled  \undefined \def \binstitutionaled#1{#1}\fi
\ifx \bctitle  \undefined \def \bctitle#1{#1}\fi
\ifx \beditor  \undefined \def \beditor#1{#1}\fi
\ifx \bpublisher  \undefined \def \bpublisher#1{#1}\fi
\ifx \bbtitle  \undefined \def \bbtitle#1{#1}\fi
\ifx \bedition  \undefined \def \bedition#1{#1}\fi
\ifx \bseriesno  \undefined \def \bseriesno#1{#1}\fi
\ifx \blocation  \undefined \def \blocation#1{#1}\fi
\ifx \bsertitle  \undefined \def \bsertitle#1{#1}\fi
\ifx \bsnm \undefined \def \bsnm#1{#1}\fi
\ifx \bsuffix \undefined \def \bsuffix#1{#1}\fi
\ifx \bparticle \undefined \def \bparticle#1{#1}\fi
\ifx \barticle \undefined \def \barticle#1{#1}\fi
\bibcommenthead
\ifx \bconfdate \undefined \def \bconfdate #1{#1}\fi
\ifx \botherref \undefined \def \botherref #1{#1}\fi
\ifx \url \undefined \def \url#1{\textsf{#1}}\fi
\ifx \bchapter \undefined \def \bchapter#1{#1}\fi
\ifx \bbook \undefined \def \bbook#1{#1}\fi
\ifx \bcomment \undefined \def \bcomment#1{#1}\fi
\ifx \oauthor \undefined \def \oauthor#1{#1}\fi
\ifx \citeauthoryear \undefined \def \citeauthoryear#1{#1}\fi
\ifx \endbibitem  \undefined \def \endbibitem {}\fi
\ifx \bconflocation  \undefined \def \bconflocation#1{#1}\fi
\ifx \arxivurl  \undefined \def \arxivurl#1{\textsf{#1}}\fi
\csname PreBibitemsHook\endcsname

%%% 1
\bibitem[\protect\citeauthoryear{Altmannshofer
  et~al.}{2009}]{Altmannshofer:2009ma}
\begin{barticle}
\bauthor{\bsnm{Altmannshofer}, \binits{W.}},
\bauthor{\bsnm{Buras}, \binits{A.J.}},
\bauthor{\bsnm{Straub}, \binits{D.M.}},
\bauthor{\bsnm{Wick}, \binits{M.}}:
\batitle{{New strategies for New Physics search in $B \to K^{*} \nu \bar{\nu}$,
  $B \to K \nu \bar{\nu}$ and $B \to X_{s} \nu \bar{\nu}$ decays}}.
\bjtitle{JHEP}
\bvolume{04},
\bfpage{022}
(\byear{2009})
\doiurl{10.1088/1126-6708/2009/04/022}
{\href{https://arxiv.org/abs/0902.0160}{{arXiv:0902.0160}}}
{[hep-ph]}
\end{barticle}
\endbibitem

%%% 2
\bibitem[\protect\citeauthoryear{Buras et~al.}{2015a}]{Buras:2014fpa}
\begin{barticle}
\bauthor{\bsnm{Buras}, \binits{A.J.}},
\bauthor{\bsnm{Girrbach-Noe}, \binits{J.}},
\bauthor{\bsnm{Niehoff}, \binits{C.}},
\bauthor{\bsnm{Straub}, \binits{D.M.}}:
\batitle{{$ B\to {K}^{\left(\ast \right)}\nu \overline{\nu} $ decays in the
  Standard Model and beyond}}.
\bjtitle{JHEP}
\bvolume{02},
\bfpage{184}
(\byear{2015})
\doiurl{10.1007/JHEP02(2015)184}
{\href{https://arxiv.org/abs/1409.4557}{{arXiv:1409.4557}}}
{[hep-ph]}
\end{barticle}
\endbibitem

%%% 3
\bibitem[\protect\citeauthoryear{Buras et~al.}{2015b}]{Buras:2015yca}
\begin{barticle}
\bauthor{\bsnm{Buras}, \binits{A.J.}},
\bauthor{\bsnm{Buttazzo}, \binits{D.}},
\bauthor{\bsnm{Knegjens}, \binits{R.}}:
\batitle{{$ K\to \pi \nu \overline{\nu} $ and
  \ensuremath{\varepsilon}'/\ensuremath{\varepsilon} in simplified new physics
  models}}.
\bjtitle{JHEP}
\bvolume{11},
\bfpage{166}
(\byear{2015})
\doiurl{10.1007/JHEP11(2015)166}
{\href{https://arxiv.org/abs/1507.08672}{{arXiv:1507.08672}}}
{[hep-ph]}
\end{barticle}
\endbibitem

%%% 4
\bibitem[\protect\citeauthoryear{Blake et~al.}{2017}]{Blake:2016olu}
\begin{barticle}
\bauthor{\bsnm{Blake}, \binits{T.}},
\bauthor{\bsnm{Lanfranchi}, \binits{G.}},
\bauthor{\bsnm{Straub}, \binits{D.M.}}:
\batitle{{Rare $B$ Decays as Tests of the Standard Model}}.
\bjtitle{Prog. Part. Nucl. Phys.}
\bvolume{92},
\bfpage{50}--\blpage{91}
(\byear{2017})
\doiurl{10.1016/j.ppnp.2016.10.001}
{\href{https://arxiv.org/abs/1606.00916}{{arXiv:1606.00916}}}
{[hep-ph]}
\end{barticle}
\endbibitem

%%% 5
\bibitem[\protect\citeauthoryear{Parrott et~al.}{2023}]{Parrott:2022zte}
\begin{barticle}
\bauthor{\bsnm{Parrott}, \binits{W.G.}},
\bauthor{\bsnm{Bouchard}, \binits{C.}},
\bauthor{\bsnm{Davies}, \binits{C.T.H.}}:
\batitle{{Standard Model predictions for
  B\textrightarrow{}K\ensuremath{\ell}+\ensuremath{\ell}-,
  B\textrightarrow{}K\ensuremath{\ell}1-\ensuremath{\ell}2+ and
  B\textrightarrow{}K\ensuremath{\nu}\ensuremath{\nu}\textasciimacron{} using
  form factors from Nf=2+1+1 lattice QCD}}.
\bjtitle{Phys. Rev. D}
\bvolume{107}(\bissue{1}),
\bfpage{014511}
(\byear{2023})
\doiurl{10.1103/PhysRevD.107.014511}
{\href{https://arxiv.org/abs/2207.13371}{{arXiv:2207.13371}}}
{[hep-ph]}.
\bcomment{[Erratum: Phys.Rev.D 107, 119903 (2023)]}
\end{barticle}
\endbibitem

%%% 6
\bibitem[\protect\citeauthoryear{Be\v{c}irevi\'c
  et~al.}{2023}]{Becirevic:2023aov}
\begin{barticle}
\bauthor{\bsnm{Be\v{c}irevi\'c}, \binits{D.}},
\bauthor{\bsnm{Piazza}, \binits{G.}},
\bauthor{\bsnm{Sumensari}, \binits{O.}}:
\batitle{{Revisiting $B\rightarrow K^{(*)} \nu {\bar{\nu }}$ decays in the
  Standard Model and beyond}}.
\bjtitle{Eur. Phys. J. C}
\bvolume{83}(\bissue{3}),
\bfpage{252}
(\byear{2023})
\doiurl{10.1140/epjc/s10052-023-11388-z}
{\href{https://arxiv.org/abs/2301.06990}{{arXiv:2301.06990}}}
{[hep-ph]}
\end{barticle}
\endbibitem

%%% 7
\bibitem[\protect\citeauthoryear{Grygier et~al.}{2017}]{Belle:2017oht}
\begin{barticle}
\bauthor{\bsnm{Grygier}, \binits{J.}}, \betal:
\batitle{{Search for $\boldsymbol{B\to h\nu\bar{\nu}}$ decays with semileptonic
  tagging at Belle}}.
\bjtitle{Phys. Rev. D}
\bvolume{96}(\bissue{9}),
\bfpage{091101}
(\byear{2017})
\doiurl{10.1103/PhysRevD.96.091101}
{\href{https://arxiv.org/abs/1702.03224}{{arXiv:1702.03224}}}
{[hep-ex]}.
\bcomment{[Addendum: Phys.Rev.D 97, 099902 (2018)]}
\end{barticle}
\endbibitem

%%% 8
\bibitem[\protect\citeauthoryear{Adachi et~al.}{2023}]{Belle-II:2023esi}
\begin{botherref}
\oauthor{\bsnm{Adachi}, \binits{I.}}, et al.:
{Evidence for $B^{+}\to K^{+}\nu\bar{\nu}$ Decays}
(2023)
{\href{https://arxiv.org/abs/2311.14647}{{arXiv:2311.14647}}}
{[hep-ex]}
\end{botherref}
\endbibitem

%%% 9
\bibitem[\protect\citeauthoryear{Altmannshofer et~al.}{2019}]{Belle-II:2018jsg}
\begin{barticle}
\bauthor{\bsnm{Altmannshofer}, \binits{W.}}, \betal:
\batitle{{The Belle II Physics Book}}.
\bjtitle{PTEP}
\bvolume{2019}(\bissue{12}),
\bfpage{123}--\blpage{01}
(\byear{2019})
\doiurl{10.1093/ptep/ptz106}
{\href{https://arxiv.org/abs/1808.10567}{{arXiv:1808.10567}}}
{[hep-ex]}.
\bcomment{[Erratum: PTEP 2020, 029201 (2020)]}
\end{barticle}
\endbibitem

%%% 10
\bibitem[\protect\citeauthoryear{Bause et~al.}{2024}]{Bause:2023mfe}
\begin{barticle}
\bauthor{\bsnm{Bause}, \binits{R.}},
\bauthor{\bsnm{Gisbert}, \binits{H.}},
\bauthor{\bsnm{Hiller}, \binits{G.}}:
\batitle{{Implications of an enhanced
  B\textrightarrow{}K\ensuremath{\nu}\ensuremath{\nu}\textasciimacron{}
  branching ratio}}.
\bjtitle{Phys. Rev. D}
\bvolume{109}(\bissue{1}),
\bfpage{015006}
(\byear{2024})
\doiurl{10.1103/PhysRevD.109.015006}
{\href{https://arxiv.org/abs/2309.00075}{{arXiv:2309.00075}}}
{[hep-ph]}
\end{barticle}
\endbibitem

%%% 11
\bibitem[\protect\citeauthoryear{Allwicher et~al.}{2024}]{Allwicher:2023xba}
\begin{barticle}
\bauthor{\bsnm{Allwicher}, \binits{L.}},
\bauthor{\bsnm{Becirevic}, \binits{D.}},
\bauthor{\bsnm{Piazza}, \binits{G.}},
\bauthor{\bsnm{Rosauro-Alcaraz}, \binits{S.}},
\bauthor{\bsnm{Sumensari}, \binits{O.}}:
\batitle{{Understanding the first measurement of
  B(B\textrightarrow{}K\ensuremath{\nu}\ensuremath{\nu}\textasciimacron{})}}.
\bjtitle{Phys. Lett. B}
\bvolume{848},
\bfpage{138411}
(\byear{2024})
\doiurl{10.1016/j.physletb.2023.138411}
{\href{https://arxiv.org/abs/2309.02246}{{arXiv:2309.02246}}}
{[hep-ph]}
\end{barticle}
\endbibitem

%%% 12
\bibitem[\protect\citeauthoryear{Athron et~al.}{2024}]{Athron:2023hmz}
\begin{barticle}
\bauthor{\bsnm{Athron}, \binits{P.}},
\bauthor{\bsnm{Martinez}, \binits{R.}},
\bauthor{\bsnm{Sierra}, \binits{C.}}:
\batitle{{B meson anomalies and large $ {B}^{+}\to {K}^{+}\nu \overline{\nu} $
  in non-universal U(1)$^{\prime}$ models}}.
\bjtitle{JHEP}
\bvolume{02},
\bfpage{121}
(\byear{2024})
\doiurl{10.1007/JHEP02(2024)121}
{\href{https://arxiv.org/abs/2308.13426}{{arXiv:2308.13426}}}
{[hep-ph]}
\end{barticle}
\endbibitem

%%% 13
\bibitem[\protect\citeauthoryear{Greljo and Thomsen}{2024}]{Greljo:2023bix}
\begin{barticle}
\bauthor{\bsnm{Greljo}, \binits{A.}},
\bauthor{\bsnm{Thomsen}, \binits{A.E.}}:
\batitle{{Rising Through the Ranks: Flavor Hierarchies from a Gauged \boldmath
  $ \mathrm{ SU(2)}$ Symmetry}}.
\bjtitle{Eur. Phys. J. C}
\bvolume{84}(\bissue{2}),
\bfpage{213}
(\byear{2024})
\doiurl{10.1140/epjc/s10052-024-12556-5}
{\href{https://arxiv.org/abs/2309.11547}{{arXiv:2309.11547}}}
{[hep-ph]}
\end{barticle}
\endbibitem

%%% 14
\bibitem[\protect\citeauthoryear{Chen et~al.}{2024}]{Chen:2024jlj}
\begin{botherref}
\oauthor{\bsnm{Chen}, \binits{F.-Z.}},
\oauthor{\bsnm{Wen}, \binits{Q.}},
\oauthor{\bsnm{Xu}, \binits{F.}}:
{Correlating $B\to K^{(\ast)} \nu\bar{\nu}$ and flavor anomalies in SMEFT}
(2024)
{\href{https://arxiv.org/abs/2401.11552}{{arXiv:2401.11552}}}
{[hep-ph]}
\end{botherref}
\endbibitem

%%% 15
\bibitem[\protect\citeauthoryear{D'Alise et~al.}{2024}]{DAlise:2024qmp}
\begin{botherref}
\oauthor{\bsnm{D'Alise}, \binits{A.}},
\oauthor{\bsnm{Fabiano}, \binits{G.}},
\oauthor{\bsnm{Frattulillo}, \binits{D.}},
\oauthor{\bsnm{Iacobacci}, \binits{D.}},
\oauthor{\bsnm{Sannino}, \binits{F.}},
\oauthor{\bsnm{Santorelli}, \binits{P.}},
\oauthor{\bsnm{Vignaroli}, \binits{N.}}:
{New Physics Pathways from B Processes}
(2024)
{\href{https://arxiv.org/abs/2403.17614}{{arXiv:2403.17614}}}
{[hep-ph]}
\end{botherref}
\endbibitem

%%% 16
\bibitem[\protect\citeauthoryear{Felkl et~al.}{2023}]{Felkl:2023ayn}
\begin{barticle}
\bauthor{\bsnm{Felkl}, \binits{T.}},
\bauthor{\bsnm{Giri}, \binits{A.}},
\bauthor{\bsnm{Mohanta}, \binits{R.}},
\bauthor{\bsnm{Schmidt}, \binits{M.A.}}:
\batitle{{When energy goes missing: new physics in $b\rightarrow s \nu \nu $
  with sterile neutrinos}}.
\bjtitle{Eur. Phys. J. C}
\bvolume{83}(\bissue{12}),
\bfpage{1135}
(\byear{2023})
\doiurl{10.1140/epjc/s10052-023-12326-9}
{\href{https://arxiv.org/abs/2309.02940}{{arXiv:2309.02940}}}
{[hep-ph]}
\end{barticle}
\endbibitem

%%% 17
\bibitem[\protect\citeauthoryear{He et~al.}{2023}]{He:2023bnk}
\begin{botherref}
\oauthor{\bsnm{He}, \binits{X.-G.}},
\oauthor{\bsnm{Ma}, \binits{X.-D.}},
\oauthor{\bsnm{Valencia}, \binits{G.}}:
{Revisiting models that enhance $B^+\to K^+ \nu\bar\nu$ in light of the new
  Belle II measurement}
(2023)
{\href{https://arxiv.org/abs/2309.12741}{{arXiv:2309.12741}}}
{[hep-ph]}
\end{botherref}
\endbibitem

%%% 18
\bibitem[\protect\citeauthoryear{Berezhnoy and
  Melikhov}{2024}]{Berezhnoy:2023rxx}
\begin{barticle}
\bauthor{\bsnm{Berezhnoy}, \binits{A.}},
\bauthor{\bsnm{Melikhov}, \binits{D.}}:
\batitle{{$B\to K^* M_X$ vs $B\to K M_X$ as a probe of a scalar-mediator dark
  matter scenario}}.
\bjtitle{EPL}
\bvolume{145}(\bissue{1}),
\bfpage{14001}
(\byear{2024})
\doiurl{10.1209/0295-5075/ad1d03}
{\href{https://arxiv.org/abs/2309.17191}{{arXiv:2309.17191}}}
{[hep-ph]}
\end{barticle}
\endbibitem

%%% 19
\bibitem[\protect\citeauthoryear{Altmannshofer
  et~al.}{2023}]{Altmannshofer:2023hkn}
\begin{botherref}
\oauthor{\bsnm{Altmannshofer}, \binits{W.}},
\oauthor{\bsnm{Crivellin}, \binits{A.}},
\oauthor{\bsnm{Haigh}, \binits{H.}},
\oauthor{\bsnm{Inguglia}, \binits{G.}},
\oauthor{\bsnm{Martin~Camalich}, \binits{J.}}:
{Light New Physics in $B\to K^{(*)}\nu\bar\nu$?}
(2023)
{\href{https://arxiv.org/abs/2311.14629}{{arXiv:2311.14629}}}
{[hep-ph]}
\end{botherref}
\endbibitem

%%% 20
\bibitem[\protect\citeauthoryear{McKeen et~al.}{2023}]{McKeen:2023uzo}
\begin{botherref}
\oauthor{\bsnm{McKeen}, \binits{D.}},
\oauthor{\bsnm{Ng}, \binits{J.N.}},
\oauthor{\bsnm{Tuckler}, \binits{D.}}:
{Higgs Portal Interpretation of the Belle II $B^+ \to K^+ \nu \nu$ Measurement}
(2023)
{\href{https://arxiv.org/abs/2312.00982}{{arXiv:2312.00982}}}
{[hep-ph]}
\end{botherref}
\endbibitem

%%% 21
\bibitem[\protect\citeauthoryear{Fridell et~al.}{2023}]{Fridell:2023ssf}
\begin{botherref}
\oauthor{\bsnm{Fridell}, \binits{K.}},
\oauthor{\bsnm{Ghosh}, \binits{M.}},
\oauthor{\bsnm{Okui}, \binits{T.}},
\oauthor{\bsnm{Tobioka}, \binits{K.}}:
{Decoding the $B \to K \nu \nu$ excess at Belle II: kinematics, operators, and
  masses}
(2023)
{\href{https://arxiv.org/abs/2312.12507}{{arXiv:2312.12507}}}
{[hep-ph]}
\end{botherref}
\endbibitem

%%% 22
\bibitem[\protect\citeauthoryear{Ho et~al.}{2024}]{Ho:2024cwk}
\begin{botherref}
\oauthor{\bsnm{Ho}, \binits{S.-Y.}},
\oauthor{\bsnm{Kim}, \binits{J.}},
\oauthor{\bsnm{Ko}, \binits{P.}}:
{Recent $B^+ \!\to K^+\nu\bar{\nu}$ Excess and Muon $g-2$ Illuminating Light
  Dark Sector with Higgs Portal}
(2024)
{\href{https://arxiv.org/abs/2401.10112}{{arXiv:2401.10112}}}
{[hep-ph]}
\end{botherref}
\endbibitem

%%% 23
\bibitem[\protect\citeauthoryear{Gabrielli et~al.}{2024}]{Gabrielli:2024wys}
\begin{botherref}
\oauthor{\bsnm{Gabrielli}, \binits{E.}},
\oauthor{\bsnm{Marzola}, \binits{L.}},
\oauthor{\bsnm{M\"u\"ursepp}, \binits{K.}},
\oauthor{\bsnm{Raidal}, \binits{M.}}:
{Explaining the $B^+\to K^+ \nu \bar{\nu}$ excess via a massless dark photon}
(2024)
{\href{https://arxiv.org/abs/2402.05901}{{arXiv:2402.05901}}}
{[hep-ph]}
\end{botherref}
\endbibitem

%%% 24
\bibitem[\protect\citeauthoryear{Hou et~al.}{2024}]{Hou:2024vyw}
\begin{botherref}
\oauthor{\bsnm{Hou}, \binits{B.-F.}},
\oauthor{\bsnm{Li}, \binits{X.-Q.}},
\oauthor{\bsnm{Shen}, \binits{M.}},
\oauthor{\bsnm{Yang}, \binits{Y.-D.}},
\oauthor{\bsnm{Yuan}, \binits{X.-B.}}:
{Deciphering the Belle II data on $B\to K \nu \bar\nu$ decay in the (dark)
  SMEFT with minimal flavour violation}
(2024)
{\href{https://arxiv.org/abs/2402.19208}{{arXiv:2402.19208}}}
{[hep-ph]}
\end{botherref}
\endbibitem

%%% 25
\bibitem[\protect\citeauthoryear{He et~al.}{2024}]{He:2024iju}
\begin{botherref}
\oauthor{\bsnm{He}, \binits{X.-G.}},
\oauthor{\bsnm{Ma}, \binits{X.-D.}},
\oauthor{\bsnm{Schmidt}, \binits{M.A.}},
\oauthor{\bsnm{Valencia}, \binits{G.}},
\oauthor{\bsnm{Volkas}, \binits{R.R.}}:
{Scalar dark matter explanation of the excess in the Belle II $B^+\to K^+ +
  \mbox{invisible}$ measurement}
(2024)
{\href{https://arxiv.org/abs/2403.12485}{{arXiv:2403.12485}}}
{[hep-ph]}
\end{botherref}
\endbibitem

%%% 26
\bibitem[\protect\citeauthoryear{Bolton et~al.}{2024}]{Bolton:2024egx}
\begin{botherref}
\oauthor{\bsnm{Bolton}, \binits{P.D.}},
\oauthor{\bsnm{Fajfer}, \binits{S.}},
\oauthor{\bsnm{Kamenik}, \binits{J.F.}},
\oauthor{\bsnm{Novoa-Brunet}, \binits{M.}}:
{Signatures of Light New Particles in $B\to K^{(*)} E_{\rm miss}$}
(2024)
{\href{https://arxiv.org/abs/2403.13887}{{arXiv:2403.13887}}}
{[hep-ph]}
\end{botherref}
\endbibitem

%%% 27
\bibitem[\protect\citeauthoryear{Datta et~al.}{2024}]{Datta:2023iln}
\begin{barticle}
\bauthor{\bsnm{Datta}, \binits{A.}},
\bauthor{\bsnm{Marfatia}, \binits{D.}},
\bauthor{\bsnm{Mukherjee}, \binits{L.}}:
\batitle{{B\textrightarrow{}K\ensuremath{\nu}\ensuremath{\nu}\textasciimacron{},
  MiniBooNE and muon g-2 anomalies from a dark sector}}.
\bjtitle{Phys. Rev. D}
\bvolume{109}(\bissue{3}),
\bfpage{031701}
(\byear{2024})
\doiurl{10.1103/PhysRevD.109.L031701}
{\href{https://arxiv.org/abs/2310.15136}{{arXiv:2310.15136}}}
{[hep-ph]}
\end{barticle}
\endbibitem

%%% 28
\bibitem[\protect\citeauthoryear{Gherardi et~al.}{2019}]{Gherardi:2019zil}
\begin{barticle}
\bauthor{\bsnm{Gherardi}, \binits{V.}},
\bauthor{\bsnm{Marzocca}, \binits{D.}},
\bauthor{\bsnm{Nardecchia}, \binits{M.}},
\bauthor{\bsnm{Romanino}, \binits{A.}}:
\batitle{{Rank-One Flavor Violation and B-meson anomalies}}.
\bjtitle{JHEP}
\bvolume{10},
\bfpage{112}
(\byear{2019})
\doiurl{10.1007/JHEP10(2019)112}
{\href{https://arxiv.org/abs/1903.10954}{{arXiv:1903.10954}}}
{[hep-ph]}
\end{barticle}
\endbibitem

%%% 29
\bibitem[\protect\citeauthoryear{Cortina~Gil et~al.}{2020}]{NA62:2020fhy}
\begin{barticle}
\bauthor{\bsnm{Cortina~Gil}, \binits{E.}}, \betal:
\batitle{{An investigation of the very rare $ {K}^{+}\to {\pi}^{+}\nu
  \overline{\nu} $ decay}}.
\bjtitle{JHEP}
\bvolume{11},
\bfpage{042}
(\byear{2020})
\doiurl{10.1007/JHEP11(2020)042}
{\href{https://arxiv.org/abs/2007.08218}{{arXiv:2007.08218}}}
{[hep-ex]}
\end{barticle}
\endbibitem

%%% 30
\bibitem[\protect\citeauthoryear{Cortina~Gil et~al.}{2021}]{NA62:2021zjw}
\begin{barticle}
\bauthor{\bsnm{Cortina~Gil}, \binits{E.}}, \betal:
\batitle{{Measurement of the very rare K$^{+}$\textrightarrow{}$ {\pi}^{+}\nu
  \overline{\nu} $ decay}}.
\bjtitle{JHEP}
\bvolume{06},
\bfpage{093}
(\byear{2021})
\doiurl{10.1007/JHEP06(2021)093}
{\href{https://arxiv.org/abs/2103.15389}{{arXiv:2103.15389}}}
{[hep-ex]}
\end{barticle}
\endbibitem

%%% 31
\bibitem[\protect\citeauthoryear{Ahn et~al.}{2019}]{KOTO:2018dsc}
\begin{barticle}
\bauthor{\bsnm{Ahn}, \binits{J.K.}}, \betal:
\batitle{{Search for the $K_L \!\to\! \pi^0 \nu \overline{\nu}$ and $K_L
  \!\to\! \pi^0 X^0$ decays at the J-PARC KOTO experiment}}.
\bjtitle{Phys. Rev. Lett.}
\bvolume{122}(\bissue{2}),
\bfpage{021802}
(\byear{2019})
\doiurl{10.1103/PhysRevLett.122.021802}
{\href{https://arxiv.org/abs/1810.09655}{{arXiv:1810.09655}}}
{[hep-ex]}
\end{barticle}
\endbibitem

%%% 32
\bibitem[\protect\citeauthoryear{{Joel Christopher Swallow (for the NA62
  collaboration)}}{2024}]{Kpiseminar}
\begin{botherref}
\oauthor{\bsnm{{Joel Christopher Swallow (for the NA62 collaboration)}}}:
{New measurement of the $K^+ \to \pi^+ \nu\bar\nu$ decay by the NA62
  Experiment}.
\url{https://indico.cern.ch/event/1447422/}
(2024)
\end{botherref}
\endbibitem

%%% 33
\bibitem[\protect\citeauthoryear{Piccini}{2020}]{Piccini:2020xgx}
\begin{barticle}
\bauthor{\bsnm{Piccini}, \binits{M.}}:
\batitle{{Status of the NA62 Experiment}}.
\bjtitle{EPJ Web Conf.}
\bvolume{234},
\bfpage{01012}
(\byear{2020})
\doiurl{10.1051/epjconf/202023401012}
\end{barticle}
\endbibitem

%%% 34
\bibitem[\protect\citeauthoryear{Ashraf et~al.}{2023}]{HIKE:2023ext}
\begin{botherref}
\oauthor{\bsnm{Ashraf}, \binits{M.U.}}, et al.:
{High Intensity Kaon Experiments (HIKE) at the CERN SPS Proposal for Phases 1
  and 2}
(2023)
{\href{https://arxiv.org/abs/2311.08231}{{arXiv:2311.08231}}}
{[hep-ex]}
\end{botherref}
\endbibitem

%%% 35
\bibitem[\protect\citeauthoryear{Aoki et~al.}{2021}]{Aoki:2021cqa}
\begin{botherref}
\oauthor{\bsnm{Aoki}, \binits{K.}}, et al.:
{Extension of the J-PARC Hadron Experimental Facility: Third White Paper}
(2021)
{\href{https://arxiv.org/abs/2110.04462}{{arXiv:2110.04462}}}
{[nucl-ex]}
\end{botherref}
\endbibitem

%%% 36
\bibitem[\protect\citeauthoryear{Falkowski and
  Straub}{2020}]{Falkowski:2019hvp}
\begin{barticle}
\bauthor{\bsnm{Falkowski}, \binits{A.}},
\bauthor{\bsnm{Straub}, \binits{D.}}:
\batitle{{Flavourful SMEFT likelihood for Higgs and electroweak data}}.
\bjtitle{JHEP}
\bvolume{04},
\bfpage{066}
(\byear{2020})
\doiurl{10.1007/JHEP04(2020)066}
{\href{https://arxiv.org/abs/1911.07866}{{arXiv:1911.07866}}}
{[hep-ph]}
\end{barticle}
\endbibitem

%%% 37
\bibitem[\protect\citeauthoryear{Allwicher et~al.}{2023}]{Allwicher:2022mcg}
\begin{barticle}
\bauthor{\bsnm{Allwicher}, \binits{L.}},
\bauthor{\bsnm{Faroughy}, \binits{D.A.}},
\bauthor{\bsnm{Jaffredo}, \binits{F.}},
\bauthor{\bsnm{Sumensari}, \binits{O.}},
\bauthor{\bsnm{Wilsch}, \binits{F.}}:
\batitle{{HighPT: A tool for~ high-$p_T$ Drell-Yan tails beyond the standard
  model}}.
\bjtitle{Comput. Phys. Commun.}
\bvolume{289},
\bfpage{108749}
(\byear{2023})
\doiurl{10.1016/j.cpc.2023.108749}
{\href{https://arxiv.org/abs/2207.10756}{{arXiv:2207.10756}}}
{[hep-ph]}
\end{barticle}
\endbibitem

%%% 38
\bibitem[\protect\citeauthoryear{Jenkins et~al.}{2014}]{Jenkins:2013wua}
\begin{barticle}
\bauthor{\bsnm{Jenkins}, \binits{E.E.}},
\bauthor{\bsnm{Manohar}, \binits{A.V.}},
\bauthor{\bsnm{Trott}, \binits{M.}}:
\batitle{{Renormalization Group Evolution of the Standard Model Dimension Six
  Operators II: Yukawa Dependence}}.
\bjtitle{JHEP}
\bvolume{01},
\bfpage{035}
(\byear{2014})
\doiurl{10.1007/JHEP01(2014)035}
{\href{https://arxiv.org/abs/1310.4838}{{arXiv:1310.4838}}}
{[hep-ph]}
\end{barticle}
\endbibitem

%%% 39
\bibitem[\protect\citeauthoryear{Jenkins et~al.}{2013}]{Jenkins:2013zja}
\begin{barticle}
\bauthor{\bsnm{Jenkins}, \binits{E.E.}},
\bauthor{\bsnm{Manohar}, \binits{A.V.}},
\bauthor{\bsnm{Trott}, \binits{M.}}:
\batitle{{Renormalization Group Evolution of the Standard Model Dimension Six
  Operators I: Formalism and lambda Dependence}}.
\bjtitle{JHEP}
\bvolume{10},
\bfpage{087}
(\byear{2013})
\doiurl{10.1007/JHEP10(2013)087}
{\href{https://arxiv.org/abs/1308.2627}{{arXiv:1308.2627}}}
{[hep-ph]}
\end{barticle}
\endbibitem

%%% 40
\bibitem[\protect\citeauthoryear{Alonso et~al.}{2014}]{Alonso:2013hga}
\begin{barticle}
\bauthor{\bsnm{Alonso}, \binits{R.}},
\bauthor{\bsnm{Jenkins}, \binits{E.E.}},
\bauthor{\bsnm{Manohar}, \binits{A.V.}},
\bauthor{\bsnm{Trott}, \binits{M.}}:
\batitle{{Renormalization Group Evolution of the Standard Model Dimension Six
  Operators III: Gauge Coupling Dependence and Phenomenology}}.
\bjtitle{JHEP}
\bvolume{04},
\bfpage{159}
(\byear{2014})
\doiurl{10.1007/JHEP04(2014)159}
{\href{https://arxiv.org/abs/1312.2014}{{arXiv:1312.2014}}}
{[hep-ph]}
\end{barticle}
\endbibitem

%%% 41
\bibitem[\protect\citeauthoryear{Jenkins et~al.}{2018a}]{Jenkins:2017jig}
\begin{barticle}
\bauthor{\bsnm{Jenkins}, \binits{E.E.}},
\bauthor{\bsnm{Manohar}, \binits{A.V.}},
\bauthor{\bsnm{Stoffer}, \binits{P.}}:
\batitle{{Low-Energy Effective Field Theory below the Electroweak Scale:
  Operators and Matching}}.
\bjtitle{JHEP}
\bvolume{03},
\bfpage{016}
(\byear{2018})
\doiurl{10.1007/JHEP03(2018)016}
{\href{https://arxiv.org/abs/1709.04486}{{arXiv:1709.04486}}}
{[hep-ph]}.
\bcomment{[Erratum: JHEP 12, 043 (2023)]}
\end{barticle}
\endbibitem

%%% 42
\bibitem[\protect\citeauthoryear{Jenkins et~al.}{2018b}]{Jenkins:2017dyc}
\begin{barticle}
\bauthor{\bsnm{Jenkins}, \binits{E.E.}},
\bauthor{\bsnm{Manohar}, \binits{A.V.}},
\bauthor{\bsnm{Stoffer}, \binits{P.}}:
\batitle{{Low-Energy Effective Field Theory below the Electroweak Scale:
  Anomalous Dimensions}}.
\bjtitle{JHEP}
\bvolume{01},
\bfpage{084}
(\byear{2018})
\doiurl{10.1007/JHEP01(2018)084}
{\href{https://arxiv.org/abs/1711.05270}{{arXiv:1711.05270}}}
{[hep-ph]}
\end{barticle}
\endbibitem

%%% 43
\bibitem[\protect\citeauthoryear{Fuentes-Martin
  et~al.}{2021}]{Fuentes-Martin:2020zaz}
\begin{barticle}
\bauthor{\bsnm{Fuentes-Martin}, \binits{J.}},
\bauthor{\bsnm{Ruiz-Femenia}, \binits{P.}},
\bauthor{\bsnm{Vicente}, \binits{A.}},
\bauthor{\bsnm{Virto}, \binits{J.}}:
\batitle{{DsixTools 2.0: The Effective Field Theory Toolkit}}.
\bjtitle{Eur. Phys. J. C}
\bvolume{81}(\bissue{2}),
\bfpage{167}
(\byear{2021})
\doiurl{10.1140/epjc/s10052-020-08778-y}
{\href{https://arxiv.org/abs/2010.16341}{{arXiv:2010.16341}}}
{[hep-ph]}
\end{barticle}
\endbibitem

%%% 44
\bibitem[\protect\citeauthoryear{}{}]{UTfit}
\begin{botherref}
New Physics Fit results: Summer 2023.
\url{http://utfit.org/UTfit/ResultsSummer2023NP}
\end{botherref}
\endbibitem

%%% 45
\bibitem[\protect\citeauthoryear{Bona et~al.}{2023}]{UTfit:2022hsi}
\begin{barticle}
\bauthor{\bsnm{Bona}, \binits{M.}}, \betal:
\batitle{{New UTfit Analysis of the Unitarity Triangle in the
  Cabibbo-Kobayashi-Maskawa scheme}}.
\bjtitle{Rend. Lincei Sci. Fis. Nat.}
\bvolume{34},
\bfpage{37}--\blpage{57}
(\byear{2023})
\doiurl{10.1007/s12210-023-01137-5}
{\href{https://arxiv.org/abs/2212.03894}{{arXiv:2212.03894}}}
{[hep-ph]}
\end{barticle}
\endbibitem

%%% 46
\bibitem[\protect\citeauthoryear{Barducci et~al.}{2023}]{Barducci:2023lqx}
\begin{barticle}
\bauthor{\bsnm{Barducci}, \binits{D.}},
\bauthor{\bsnm{Nardecchia}, \binits{M.}},
\bauthor{\bsnm{Toni}, \binits{C.}}:
\batitle{{Perturbative unitarity constraints on generic vector interactions}}.
\bjtitle{JHEP}
\bvolume{09},
\bfpage{134}
(\byear{2023})
\doiurl{10.1007/JHEP09(2023)134}
{\href{https://arxiv.org/abs/2306.11533}{{arXiv:2306.11533}}}
{[hep-ph]}
\end{barticle}
\endbibitem

%%% 47
\bibitem[\protect\citeauthoryear{Aad et~al.}{2015}]{ATLAS:2015rbx}
\begin{barticle}
\bauthor{\bsnm{Aad}, \binits{G.}}, \betal:
\batitle{{A search for high-mass resonances decaying to $\tau^{+}\tau^{-}$ in
  $pp$ collisions at $\sqrt{s}=8$ TeV with the ATLAS detector}}.
\bjtitle{JHEP}
\bvolume{07},
\bfpage{157}
(\byear{2015})
\doiurl{10.1007/JHEP07(2015)157}
{\href{https://arxiv.org/abs/1502.07177}{{arXiv:1502.07177}}}
{[hep-ex]}
\end{barticle}
\endbibitem

%%% 48
\bibitem[\protect\citeauthoryear{Aad et~al.}{2020}]{ATLAS:2020zms}
\begin{barticle}
\bauthor{\bsnm{Aad}, \binits{G.}}, \betal:
\batitle{{Search for heavy Higgs bosons decaying into two tau leptons with the
  ATLAS detector using $pp$ collisions at $\sqrt{s}=13$ TeV}}.
\bjtitle{Phys. Rev. Lett.}
\bvolume{125}(\bissue{5}),
\bfpage{051801}
(\byear{2020})
\doiurl{10.1103/PhysRevLett.125.051801}
{\href{https://arxiv.org/abs/2002.12223}{{arXiv:2002.12223}}}
{[hep-ex]}
\end{barticle}
\endbibitem

%%% 49
\bibitem[\protect\citeauthoryear{Aaboud et~al.}{2018}]{ATLAS:2017qwn}
\begin{barticle}
\bauthor{\bsnm{Aaboud}, \binits{M.}}, \betal:
\batitle{{Search for the direct production of charginos and neutralinos in
  final states with tau leptons in $\sqrt{s} = $ 13 TeV $pp$ collisions with
  the ATLAS detector}}.
\bjtitle{Eur. Phys. J. C}
\bvolume{78}(\bissue{2}),
\bfpage{154}
(\byear{2018})
\doiurl{10.1140/epjc/s10052-018-5583-9}
{\href{https://arxiv.org/abs/1708.07875}{{arXiv:1708.07875}}}
{[hep-ex]}
\end{barticle}
\endbibitem

%%% 50
\bibitem[\protect\citeauthoryear{Aad et~al.}{2021}]{ATLAS:2021jyv}
\begin{barticle}
\bauthor{\bsnm{Aad}, \binits{G.}}, \betal:
\batitle{{Search for new phenomena in $pp$ collisions in final states with tau
  leptons, b-jets, and missing transverse momentum with the ATLAS detector}}.
\bjtitle{Phys. Rev. D}
\bvolume{104}(\bissue{11}),
\bfpage{112005}
(\byear{2021})
\doiurl{10.1103/PhysRevD.104.112005}
{\href{https://arxiv.org/abs/2108.07665}{{arXiv:2108.07665}}}
{[hep-ex]}
\end{barticle}
\endbibitem

%%% 51
\bibitem[\protect\citeauthoryear{Allwicher et~al.}{2021}]{Allwicher:2021rtd}
\begin{barticle}
\bauthor{\bsnm{Allwicher}, \binits{L.}},
\bauthor{\bsnm{Arnan}, \binits{P.}},
\bauthor{\bsnm{Barducci}, \binits{D.}},
\bauthor{\bsnm{Nardecchia}, \binits{M.}}:
\batitle{{Perturbative unitarity constraints on generic Yukawa interactions}}.
\bjtitle{JHEP}
\bvolume{10},
\bfpage{129}
(\byear{2021})
\doiurl{10.1007/JHEP10(2021)129}
{\href{https://arxiv.org/abs/2108.00013}{{arXiv:2108.00013}}}
{[hep-ph]}
\end{barticle}
\endbibitem

%%% 52
\bibitem[\protect\citeauthoryear{Marzocca et~al.}{2022}]{Marzocca:2021miv}
\begin{barticle}
\bauthor{\bsnm{Marzocca}, \binits{D.}},
\bauthor{\bsnm{Trifinopoulos}, \binits{S.}},
\bauthor{\bsnm{Venturini}, \binits{E.}}:
\batitle{{From B-meson anomalies to Kaon physics with scalar leptoquarks}}.
\bjtitle{Eur. Phys. J. C}
\bvolume{82}(\bissue{4}),
\bfpage{320}
(\byear{2022})
\doiurl{10.1140/epjc/s10052-022-10271-7}
{\href{https://arxiv.org/abs/2106.15630}{{arXiv:2106.15630}}}
{[hep-ph]}
\end{barticle}
\endbibitem

%%% 53
\bibitem[\protect\citeauthoryear{Buras et~al.}{2015}]{Buras:2015qea}
\begin{barticle}
\bauthor{\bsnm{Buras}, \binits{A.J.}},
\bauthor{\bsnm{Buttazzo}, \binits{D.}},
\bauthor{\bsnm{Girrbach-Noe}, \binits{J.}},
\bauthor{\bsnm{Knegjens}, \binits{R.}}:
\batitle{{$ {K}^{+}\to {\pi}^{+}\nu \overline{\nu} $ and $ {K}_L\to {\pi}^0\nu
  \overline{\nu} $ in the Standard Model: status and perspectives}}.
\bjtitle{JHEP}
\bvolume{11},
\bfpage{033}
(\byear{2015})
\doiurl{10.1007/JHEP11(2015)033}
{\href{https://arxiv.org/abs/1503.02693}{{arXiv:1503.02693}}}
{[hep-ph]}
\end{barticle}
\endbibitem

%%% 54
\bibitem[\protect\citeauthoryear{Bause et~al.}{2021}]{Bause:2021cna}
\begin{barticle}
\bauthor{\bsnm{Bause}, \binits{R.}},
\bauthor{\bsnm{Gisbert}, \binits{H.}},
\bauthor{\bsnm{Golz}, \binits{M.}},
\bauthor{\bsnm{Hiller}, \binits{G.}}:
\batitle{{Interplay of dineutrino modes with semileptonic rare B-decays}}.
\bjtitle{JHEP}
\bvolume{12},
\bfpage{061}
(\byear{2021})
\doiurl{10.1007/JHEP12(2021)061}
{\href{https://arxiv.org/abs/2109.01675}{{arXiv:2109.01675}}}
{[hep-ph]}
\end{barticle}
\endbibitem

%%% 55
\bibitem[\protect\citeauthoryear{Buras}{1998}]{Buras:1998raa}
\begin{bchapter}
\bauthor{\bsnm{Buras}, \binits{A.J.}}:
\bctitle{{Weak Hamiltonian, CP violation and rare decays}}.
In: \bbtitle{{Les Houches Summer School in Theoretical Physics, Session 68:
  Probing the Standard Model of Particle Interactions}},
pp. \bfpage{281}--\blpage{539}
(\byear{1998})
\end{bchapter}
\endbibitem

%%% 56
\bibitem[\protect\citeauthoryear{Be\v{c}irevi\'c
  et~al.}{2016}]{Becirevic:2016zri}
\begin{barticle}
\bauthor{\bsnm{Be\v{c}irevi\'c}, \binits{D.}},
\bauthor{\bsnm{Sumensari}, \binits{O.}},
\bauthor{\bsnm{Zukanovich~Funchal}, \binits{R.}}:
\batitle{{Lepton flavor violation in exclusive $b\rightarrow s$ decays}}.
\bjtitle{Eur. Phys. J. C}
\bvolume{76}(\bissue{3}),
\bfpage{134}
(\byear{2016})
\doiurl{10.1140/epjc/s10052-016-3985-0}
{\href{https://arxiv.org/abs/1602.00881}{{arXiv:1602.00881}}}
{[hep-ph]}
\end{barticle}
\endbibitem

%%% 57
\bibitem[\protect\citeauthoryear{De~Bruyn et~al.}{2012a}]{DeBruyn:2012wj}
\begin{barticle}
\bauthor{\bsnm{De~Bruyn}, \binits{K.}},
\bauthor{\bsnm{Fleischer}, \binits{R.}},
\bauthor{\bsnm{Knegjens}, \binits{R.}},
\bauthor{\bsnm{Koppenburg}, \binits{P.}},
\bauthor{\bsnm{Merk}, \binits{M.}},
\bauthor{\bsnm{Tuning}, \binits{N.}}:
\batitle{{Branching Ratio Measurements of $B_s$ Decays}}.
\bjtitle{Phys. Rev. D}
\bvolume{86},
\bfpage{014027}
(\byear{2012})
\doiurl{10.1103/PhysRevD.86.014027}
{\href{https://arxiv.org/abs/1204.1735}{{arXiv:1204.1735}}}
{[hep-ph]}
\end{barticle}
\endbibitem

%%% 58
\bibitem[\protect\citeauthoryear{De~Bruyn et~al.}{2012b}]{DeBruyn:2012wk}
\begin{barticle}
\bauthor{\bsnm{De~Bruyn}, \binits{K.}},
\bauthor{\bsnm{Fleischer}, \binits{R.}},
\bauthor{\bsnm{Knegjens}, \binits{R.}},
\bauthor{\bsnm{Koppenburg}, \binits{P.}},
\bauthor{\bsnm{Merk}, \binits{M.}},
\bauthor{\bsnm{Pellegrino}, \binits{A.}},
\bauthor{\bsnm{Tuning}, \binits{N.}}:
\batitle{{Probing New Physics via the $B^0_s\to \mu^+\mu^-$ Effective
  Lifetime}}.
\bjtitle{Phys. Rev. Lett.}
\bvolume{109},
\bfpage{041801}
(\byear{2012})
\doiurl{10.1103/PhysRevLett.109.041801}
{\href{https://arxiv.org/abs/1204.1737}{{arXiv:1204.1737}}}
{[hep-ph]}
\end{barticle}
\endbibitem

%%% 59
\bibitem[\protect\citeauthoryear{Workman and Others}{2022}]{Workman:2022ynf}
\begin{barticle}
\bauthor{\bsnm{Workman}, \binits{R.L.}},
\bauthor{\bsnm{Others}}:
\batitle{{Review of Particle Physics}}.
\bjtitle{PTEP}
\bvolume{2022},
\bfpage{083}--\blpage{01}
(\byear{2022})
\doiurl{10.1093/ptep/ptac097}
\end{barticle}
\endbibitem

%%% 60
\bibitem[\protect\citeauthoryear{D'Ambrosio et~al.}{2022}]{DAmbrosio:2022kvb}
\begin{barticle}
\bauthor{\bsnm{D'Ambrosio}, \binits{G.}},
\bauthor{\bsnm{Iyer}, \binits{A.M.}},
\bauthor{\bsnm{Mahmoudi}, \binits{F.}},
\bauthor{\bsnm{Neshatpour}, \binits{S.}}:
\batitle{{Anatomy of kaon decays and prospects for lepton flavour universality
  violation}}.
\bjtitle{JHEP}
\bvolume{09},
\bfpage{148}
(\byear{2022})
\doiurl{10.1007/JHEP09(2022)148}
{\href{https://arxiv.org/abs/2206.14748}{{arXiv:2206.14748}}}
{[hep-ph]}
\end{barticle}
\endbibitem

%%% 61
\bibitem[\protect\citeauthoryear{Isidori and
  Unterdorfer}{2004}]{Isidori:2003ts}
\begin{barticle}
\bauthor{\bsnm{Isidori}, \binits{G.}},
\bauthor{\bsnm{Unterdorfer}, \binits{R.}}:
\batitle{{On the short distance constraints from K(L,S) ---\ensuremath{>} mu+
  mu-}}.
\bjtitle{JHEP}
\bvolume{01},
\bfpage{009}
(\byear{2004})
\doiurl{10.1088/1126-6708/2004/01/009}
{\href{https://arxiv.org/abs/hep-ph/0311084}{{arXiv:hep-ph/0311084}}}
\end{barticle}
\endbibitem

%%% 62
\bibitem[\protect\citeauthoryear{Aaij et~al.}{2020}]{LHCb:2020ycd}
\begin{barticle}
\bauthor{\bsnm{Aaij}, \binits{R.}}, \betal:
\batitle{{Constraints on the $K^0_S \rightarrow \mu^+ \mu^-$ Branching
  Fraction}}.
\bjtitle{Phys. Rev. Lett.}
\bvolume{125}(\bissue{23}),
\bfpage{231801}
(\byear{2020})
\doiurl{10.1103/PhysRevLett.125.231801}
{\href{https://arxiv.org/abs/2001.10354}{{arXiv:2001.10354}}}
{[hep-ex]}
\end{barticle}
\endbibitem

%%% 63
\bibitem[\protect\citeauthoryear{}{}]{hflav}
\begin{botherref}
Preliminary average of {R(D)} and {R(D*)} for Winter 2023.
\url{https://hflav-eos.web.cern.ch/hflav-eos/semi/winter23_prel/html/RDsDsstar/RDRDs.html}
\end{botherref}
\endbibitem

%%% 64
\bibitem[\protect\citeauthoryear{Dor\v{s}ner et~al.}{2016}]{Dorsner:2016wpm}
\begin{barticle}
\bauthor{\bsnm{Dor\v{s}ner}, \binits{I.}},
\bauthor{\bsnm{Fajfer}, \binits{S.}},
\bauthor{\bsnm{Greljo}, \binits{A.}},
\bauthor{\bsnm{Kamenik}, \binits{J.F.}},
\bauthor{\bsnm{Ko\v{s}nik}, \binits{N.}}:
\batitle{{Physics of leptoquarks in precision experiments and at particle
  colliders}}.
\bjtitle{Phys. Rept.}
\bvolume{641},
\bfpage{1}--\blpage{68}
(\byear{2016})
\doiurl{10.1016/j.physrep.2016.06.001}
{\href{https://arxiv.org/abs/1603.04993}{{arXiv:1603.04993}}}
{[hep-ph]}
\end{barticle}
\endbibitem

%%% 65
\bibitem[\protect\citeauthoryear{Gherardi et~al.}{2020}]{Gherardi:2020det}
\begin{barticle}
\bauthor{\bsnm{Gherardi}, \binits{V.}},
\bauthor{\bsnm{Marzocca}, \binits{D.}},
\bauthor{\bsnm{Venturini}, \binits{E.}}:
\batitle{{Matching scalar leptoquarks to the SMEFT at one loop}}.
\bjtitle{JHEP}
\bvolume{07},
\bfpage{225}
(\byear{2020})
\doiurl{10.1007/JHEP07(2020)225}
{\href{https://arxiv.org/abs/2003.12525}{{arXiv:2003.12525}}}
{[hep-ph]}.
\bcomment{[Erratum: JHEP 01, 006 (2021)]}
\end{barticle}
\endbibitem

%%% 66
\bibitem[\protect\citeauthoryear{Bobeth and Buras}{2018}]{Bobeth:2017ecx}
\begin{barticle}
\bauthor{\bsnm{Bobeth}, \binits{C.}},
\bauthor{\bsnm{Buras}, \binits{A.J.}}:
\batitle{{Leptoquarks meet $\varepsilon'/\varepsilon$ and rare Kaon
  processes}}.
\bjtitle{JHEP}
\bvolume{02},
\bfpage{101}
(\byear{2018})
\doiurl{10.1007/JHEP02(2018)101}
{\href{https://arxiv.org/abs/1712.01295}{{arXiv:1712.01295}}}
{[hep-ph]}
\end{barticle}
\endbibitem

%%% 67
\bibitem[\protect\citeauthoryear{Davidson et~al.}{1994}]{Davidson:1993qk}
\begin{barticle}
\bauthor{\bsnm{Davidson}, \binits{S.}},
\bauthor{\bsnm{Bailey}, \binits{D.C.}},
\bauthor{\bsnm{Campbell}, \binits{B.A.}}:
\batitle{{Model independent constraints on leptoquarks from rare processes}}.
\bjtitle{Z. Phys. C}
\bvolume{61},
\bfpage{613}--\blpage{644}
(\byear{1994})
\doiurl{10.1007/BF01552629}
{\href{https://arxiv.org/abs/hep-ph/9309310}{{arXiv:hep-ph/9309310}}}
\end{barticle}
\endbibitem

%%% 68
\bibitem[\protect\citeauthoryear{Aebischer et~al.}{2020}]{Aebischer:2020dsw}
\begin{barticle}
\bauthor{\bsnm{Aebischer}, \binits{J.}},
\bauthor{\bsnm{Bobeth}, \binits{C.}},
\bauthor{\bsnm{Buras}, \binits{A.J.}},
\bauthor{\bsnm{Kumar}, \binits{J.}}:
\batitle{{SMEFT ATLAS of $\Delta$F = 2 transitions}}.
\bjtitle{JHEP}
\bvolume{12},
\bfpage{187}
(\byear{2020})
\doiurl{10.1007/JHEP12(2020)187}
{\href{https://arxiv.org/abs/2009.07276}{{arXiv:2009.07276}}}
{[hep-ph]}
\end{barticle}
\endbibitem

%%% 69
\bibitem[\protect\citeauthoryear{Li et~al.}{2020}]{Li:2019fhz}
\begin{barticle}
\bauthor{\bsnm{Li}, \binits{T.}},
\bauthor{\bsnm{Ma}, \binits{X.-D.}},
\bauthor{\bsnm{Schmidt}, \binits{M.A.}}:
\batitle{{Implication of $K\to \pi \nu \bar{\nu}$ for generic neutrino
  interactions in effective field theories}}.
\bjtitle{Phys. Rev. D}
\bvolume{101}(\bissue{5}),
\bfpage{055019}
(\byear{2020})
\doiurl{10.1103/PhysRevD.101.055019}
{\href{https://arxiv.org/abs/1912.10433}{{arXiv:1912.10433}}}
{[hep-ph]}
\end{barticle}
\endbibitem

%%% 70
\bibitem[\protect\citeauthoryear{Deppisch et~al.}{2020}]{Deppisch:2020oyx}
\begin{barticle}
\bauthor{\bsnm{Deppisch}, \binits{F.F.}},
\bauthor{\bsnm{Fridell}, \binits{K.}},
\bauthor{\bsnm{Harz}, \binits{J.}}:
\batitle{{Constraining lepton number violating interactions in rare kaon
  decays}}.
\bjtitle{JHEP}
\bvolume{12},
\bfpage{186}
(\byear{2020})
\doiurl{10.1007/JHEP12(2020)186}
{\href{https://arxiv.org/abs/2009.04494}{{arXiv:2009.04494}}}
{[hep-ph]}
\end{barticle}
\endbibitem

%%% 71
\bibitem[\protect\citeauthoryear{Felkl et~al.}{2021}]{Felkl:2021uxi}
\begin{barticle}
\bauthor{\bsnm{Felkl}, \binits{T.}},
\bauthor{\bsnm{Li}, \binits{S.L.}},
\bauthor{\bsnm{Schmidt}, \binits{M.A.}}:
\batitle{{A tale of invisibility: constraints on new physics in b
  \textrightarrow{} s\ensuremath{\nu}\ensuremath{\nu}}}.
\bjtitle{JHEP}
\bvolume{12},
\bfpage{118}
(\byear{2021})
\doiurl{10.1007/JHEP12(2021)118}
{\href{https://arxiv.org/abs/2111.04327}{{arXiv:2111.04327}}}
{[hep-ph]}
\end{barticle}
\endbibitem

%%% 72
\bibitem[\protect\citeauthoryear{Fridell et~al.}{2023}]{Fridell:2023rtr}
\begin{botherref}
\oauthor{\bsnm{Fridell}, \binits{K.}},
\oauthor{\bsnm{Gr\'af}, \binits{L.}},
\oauthor{\bsnm{Harz}, \binits{J.}},
\oauthor{\bsnm{Hati}, \binits{C.}}:
{Probing Lepton Number Violation: A Comprehensive Survey of Dimension-7 SMEFT}
(2023)
{\href{https://arxiv.org/abs/2306.08709}{{arXiv:2306.08709}}}
{[hep-ph]}
\end{botherref}
\endbibitem

%%% 73
\bibitem[\protect\citeauthoryear{Bharucha et~al.}{2016}]{Bharucha:2015bzk}
\begin{barticle}
\bauthor{\bsnm{Bharucha}, \binits{A.}},
\bauthor{\bsnm{Straub}, \binits{D.M.}},
\bauthor{\bsnm{Zwicky}, \binits{R.}}:
\batitle{{$B\to V\ell^+\ell^-$ in the Standard Model from light-cone sum
  rules}}.
\bjtitle{JHEP}
\bvolume{08},
\bfpage{098}
(\byear{2016})
\doiurl{10.1007/JHEP08(2016)098}
{\href{https://arxiv.org/abs/1503.05534}{{arXiv:1503.05534}}}
{[hep-ph]}
\end{barticle}
\endbibitem

\end{thebibliography}

\end{document}